\documentclass[a4paper,fleqn,usenatbib]{mnras}

\usepackage{color}
\usepackage{xcolor}
\usepackage{graphicx,url,amssymb,amsmath,longtable,rotating,units,wasysym,listings,bm}
\usepackage{amsmath}	
\usepackage{amssymb}	
\usepackage{mathptmx}
\usepackage{ulem}
\usepackage{natbib}
\usepackage{subfig}

\newcommand{\simlt}{\lower.5ex\hbox{$\; \buildrel < \over \sim \;$}}
\newcommand{\simgt}{\lower.5ex\hbox{$\; \buildrel > \over \sim \;$}}

\title[]{Monte Carlo simulations of relativistic shock breakout from a stellar wind}
\author[Ito]{Hirotaka Ito $^{1,2}$\thanks{Email: hirotaka.ito@riken.jp}, Amir Levinson$^{3,4}$, Ehud Nakar$^{3}$, Shigehiro Nagataki$^{1,2,5}$
\\
$1$ Astrophysical Big Bang Laboratory (ABBL), RIKEN Pioneering Research Institute (PRI), 2-1 Hirosawa, Wako, Saitama 351-0198, Japan\\
$2$ RIKEN Center for Interdisciplinary Theoretical and Mathematical Sciences (iTHEMS), 2-1 Hirosawa, Wako, Saitama 351-0198, Japan\\
$3$ School of Physics \& Astronomy, Tel Aviv University, Tel Aviv 69978, Israel\\
$4$ Yukawa Institute for Theoretical Physics, Kyoto University, Oiwake-cho, Kitashirakawa, Sakyo-ku, Kyoto 606-8502, Japan\\
$5$ Astrophysical Big Bang Group (ABBG), Okinawa Institute of Science and Technology (OIST)
1919-1 Tancha, Onna-son, Kunigami-gun, Okinawa 904-0495, Japan\\
}

\begin{document}
\maketitle

\begin{abstract}
  We present Monte Carlo simulations of relativistic radiation-mediated shocks (RRMS) in the photon-starved regime, incorporating photon escape from the upstream region—characterized by the escape fraction, $f_{\rm esc}$—under a steady-state assumption. These simulations, performed for shock Lorentz factors $\Gamma_u = 2$, $3.5$, $6$, $10$, and $15$, are applicable to RRMS breakouts 
   in shallowly declining density profiles such as stellar winds.
  We find that vigorous pair production acts as a thermostat, regulating the downstream temperature to $\sim 100$-$200~{\rm keV}$, largely independent of $f_{\rm esc}$. A subshock forms and strengthens with increasing $f_{\rm esc}$. The escaping spectra 
   peak at $E_p \approx 300$-$600~{\rm keV}$ in the shock frame and deviate from a Wien distribution, exhibiting low-energy flattening ($f_\nu \propto \nu^{0}$) due to free-free emission and high-energy extensions caused by inverse Compton scattering from subshock-heated pairs.
  While an earlier analytical model
   reproduces the velocity structure well at $\Gamma_u = 2$, it significantly overestimates the shock width at higher Lorentz factors, particularly for $f_{\rm esc} \gtrsim$ a few $\%$. Based on this finding, we provide updated predictions for breakout observables in wind environments for $\Gamma_u \gtrsim 6$. Notably, the duration of the relativistic breakout becomes largely insensitive to the explosion energy and ejecta mass, typically exceeding analytical predictions by orders of magnitude and capable of producing a $\sim$300 
    s flash of MeV photons with a radiated energy of $\sim 10^{50}$
    erg for an energetic explosion yielding $\Gamma_{bo} \sim 6$. We also discuss limitations of our modelling assumptions and their implications for the predicted breakout observables.

\end{abstract}



\section{Introduction}
\label{sec:Intro}

Shock breakout governs the initial electromagnetic emission from various cosmic explosions, including supernovae and gamma-ray bursts. During breakout, photons that were initially trapped within the transition region of a radiation-mediated shock (RMS) are released \citep[see, e.g.,][]{katz2017,Levinson_Nakar2020}. A detailed understanding of RMS structure is therefore essential for interpreting the properties of breakout emission.

The shock velocity at the breakout can range from
sub-relativistic to mildly relativistic, and in extreme cases even ultra-
relativistic, as anticipated, e.g., in energetic explosions of a compact progenitor, or highly collimated explosions such as low-luminosity gamma-ray bursts (LLGRBs). 
  When the velocity of RMS is slow ($\beta_u \lesssim 0.05$), where $\beta_u$ denotes the shock velocity normalized by the speed of light $c$, thermal equilibrium is achieved throughout the shock. Additionally, the diffusion approximation can be applied to solve radiation transfer within the shock.
 These aspects significantly simplify the problem, and the physics of slow RMSs has been extensively studied and is well established in the literature, particularly in the regime well before breakout where the shock resides in a highly optically thick region ($\tau \gg \beta_u^{-1}$), and thus the steady-state assumption holds \citep[e.g.,][]{Zeldovich1967,Weaver1976,Blandford1981}.

On the other hand, in fast Newtonian shocks ($0.05 \lesssim \beta_u  \lesssim 0.5$), the system drops out of the thermal equilibrium, necessitating the computation of thermal temperature by considering photon production \citep{Weaver1976,Katz2010}. 
When the shock velocity becomes relativistic ($\beta_u \gtrsim 0.5$), additional complexities arise: the diffusion approximation breaks down, and effects such as pair production and Klein-Nishina modifications to the scattering cross-section must be considered. 
Due to these challenges, even steady-state calculations applicable to shocks formed in highly optically thick regions remain scarce for faster shocks, especially in the relativistic regime, compared to the extensive studies available for slow shocks.
The first ab-initio study of relativistic RMS (RRMS) was carried out by \citet{Budnik2010}, followed by further updates covering the transition from fast Newtonian to relativistic velocity conducted by \citet{Ito2020a} (hereafter ILN20a).
These studies focused on RRMSs in the photon-starved regime, applicable to shocks propagating through stellar envelopes or extended envelopes, where the number of photons advected from upstream is negligible compared to those produced within the shock. It is worth noting that complementary studies have also investigated photon-rich RRMSs, which are more relevant to breakouts in high-entropy environments such as gamma-ray burst (GRB) outflows, where upstream photon advection dominates over in-shock photon production \citep{Beloborodov2017, Lundman2018, Ito2018}.

The studies outlined above focus on the steady-state profiles of shocks in which photons are trapped. While these studies offer valuable insights, they cannot be directly applied to the breakout phase, during which photons escape from the shock. In particular, shock breakout occurring in a region with a sharp density drop, such as at the edge of a stellar envelope or the outermost layers of an extended envelope, is inherently an abrupt phenomenon. Accurately capturing the resulting emission requires time-dependent modeling.
In the case of slow and fast shocks, substantial progress has been made through both analytical approaches \citep[e.g.,][]{Chevelier1992, Nakar_Sari2010, Katz2012, Irwin_Hotokezaka2024} and radiation hydrodynamics simulations, which adopt the diffusion approximation for radiative transfer with varying simplifications across studies \citep[e.g.,][]{Klein_Chevelier1978, Ensman_Burrows1992, Blinnikov2000, Tolstov2013, Sapir2013, Sapir2014}. These radiation-hydrodynamics-based simulations have explored the evolution of the shock structure during breakout and the resulting emission.
In contrast, studies of relativistic shock breakout remain limited and are largely restricted to analytical treatments \citep[e.g.,][]{Nakar_Sari2012, Faran_Sari2023}. These studies suggest that the breakout flash is triggered when the shocked ejecta expands sufficiently and its temperature decreases to approximately $50~{\rm keV}$. At this point, previously generated pairs annihilate, allowing photons that had been trapped by pair opacity to escape and produce the observable breakout emission. However, a fully self-consistent calculation of this process has not yet been conducted. \footnote{The only ab initio simulation of relativistic RMS breakout to date is by \citet{Lundman2021}, which focuses on the photon-rich regime but neglects pair production, limiting its applicability.}

Shock breakout can also occur in shallowly decaying density regions, such as the optically thick stellar wind surrounding a progenitor star \citep{Ofek2014, Gal-Yam2014}, representing an important class of breakout events alongside those occurring at sharp density drop regions.
Numerous studies have investigated shock breakout in stellar winds, particularly in the sub-relativistic regime \citep[e.g.,][]{Balberg_Loeb2011, Chevalier_Irwin2011, Svirski2012, Svirski2014, Khatami_Kasen2024}.
A key difference, also highlighted in the present study, is that in shallowly decaying density regions, the breakout proceeds relatively gradually, provided that the shock-crossing time is comparable to (not significantly shorter than) the timescale over which the upstream conditions evolve during breakout.
This is in contrast to breakouts at the stellar edge, where the upstream density drops steeply and the breakout is more abrupt.
Taking advantage of this more gradual transition, photon-starved shock breakout in winds has been modeled both analytically and numerically in recent years.
\footnote{However, a recent high-resolution radiation hydrodynamics study by \citet{Wasserman2025} suggests that steady-state treatments may be invalid under these conditions.}
\citet{Ioka2019} developed a simple analytic model for RMS breakout in the fast Newtonian regime ($\beta_u \lesssim 0.3$), incorporating radiative losses under the diffusion approximation.
This was followed by numerical simulations by \citet{Ito2020b} (hereafter ILN20b), which also assumed steady-state conditions but employed full radiative transfer. These simulations showed good agreement with the analytic model for shock velocities of $\beta_u = 0.1$ and $\beta_u \lesssim 0.25$.
At higher velocities, approaching the mildly relativistic regime ($\beta_u \approx 0.5$), the simulations revealed vigorous pair production and the formation of a collisionless subshock, both of which substantially alter the shock structure and emergent spectrum.

Similarly, under the steady-state assumption, wind breakout in the highly relativistic regime has been analytically investigated by \citet{granot2018} (hereafter GNL18). 
Their analysis indicates that vigorous pair production within the shock transition layer generates sufficient self-generated opacity to sustain the shock to be radiation-mediated during the breakout phase.
Unlike relativistic breakouts at sharp edges, where the breakout flash originates abruptly from a shocked layer with optical depth near unity, the breakout emission arises from the gradual leakage of photons from the shock front, which advances over orders of magnitude in radius as the fraction of photons escaping from the shock increases.
The model introduced in GNL18, however, has some important limitations.  
First, it is based on assumptions that are applicable only in highly relativistic regimes; hence, the shock profile cannot be properly extended to the downstream region.  
 Second, as a result, the model was expected to fail in describing mildly relativistic shocks.  
Third, the model cannot be used to compute the spectrum.  
  Given these limitations, a more comprehensive study is required to deepen our understanding of relativistic RMS, particularly in the presence of photon escape.  
  To achieve this, we employ a numerical approach developed in \citet{Ito2020b}, which allows us to extend the analysis beyond the highly relativistic regime and investigate shocks with mildly relativistic velocities. Specifically, we examine cases with $\Gamma_u = 2$, $3.5$ $6$, $10$, and $15$.  
  Our numerical approach enables us to construct a complete shock structure, spanning from the upstream (escaping) boundary to the far downstream region, and to evaluate the spectral properties of the emitted radiation.  
Contrary to naive expectations, our results show that mildly relativistic shocks with $\Gamma_u = 2$ can be well reproduced by the analytical model, regardless of the amount of photon escape from the shock. In contrast, for higher Lorentz factors ($\Gamma_u \geq 3.5$),
while a shock without photon escape shows excellent agreement with the analytical model, the model breaks down as photon escape from the shock becomes significant. This behavior is discussed in detail in this paper.
We note that both the GNL18 model and our numerical study adopt a steady-state, planar treatment of the shock structure; a definitive validation of this approximation would require fully time-dependent simulations, which are beyond the scope of the present work.

This paper is organized as follows. 
In Section \ref{sec:numerical}, we outline the numerical methods and the simulation setup. Section \ref{sec:structure} presents the calculated structure of RMS, while Section \ref{sec:spectrum} discusses the spectral characteristics of photons escaping from RMS. 
The comparison between our numerical results and the analytical model of GNL18 is presented in Section~\ref{sec:comparison}, and its implications for the observable properties of relativistic shock breakout from a stellar wind are discussed in Section~\ref{sec:Breakout}.
We discuss the limitations of our modelling assumptions, including the steady-state planar treatment, in Section~\ref{sec:limitations}.
Finally, we conclude in Section \ref{sec:summary}.
Throughout the paper, the subscript $u$ and $d$  refer to the physical quantities at the far upstream and far downstream  regions of the shock, respectively.

\section{Monte-Carlo simulations of leaking RRMS}
\label{sec:numerical}

The numerical method used in the current study is outlined in ILN20b, and we provide a brief summary of the  methodology here. 
The code is an updated version of the one used in ILN20a, which focused on the steady-state structure of RMS with an optically thick upstream that prevents photons from escaping from the shock (referred to hereafter as 'infinite shock').
The updated code incorporates radiative losses through the upstream boundary, enabling simulations of RRMS with finite optical depths upstream, as anticipated during the breakout phase.
Similar modifications have been applied previously to simulations of fast Newtonian ($\beta_u = 0.1$ and $0.25$) and mildly relativistic ($\beta_u = 0.5$) 
shock breakouts, as reported in ILN20b.
In this study, we extend our analysis to the relativistic regime, specifically considering cases with $\Gamma_u =2$, $3.5$, $6$, $10$, and $15$.

In the simulations, the pair-loaded optical depth upstream of the subshock, which is defined as $\tau_{u*} = \int_0^{z_{u}} \Gamma (n + n_{\pm}) \sigma_T dz$, serves as an input parameter that dictates the amount of photons escaping the upstream boundary.
Here, $\sigma_T$ denotes the Thomson cross-section.
The term $n$ refers to the density of the electrons advected from upstream, while $n_{\pm}$ represents the density of electron-positron pairs produced via photon-photon collisions. 
In this study, we focus on RMS formed in a proton-electron plasma; therefore, the density of advected electrons is equal to that of the baryons (protons). Throughout the paper, we use $n$ to represent the density of both the advected electrons and protons.
The range of integration extends from the subshock at $z=0$ to the upstream boundary at $z_u$ ($< 0$).\footnote{Since the spatial coordinate $z$ is aligned with the flow direction, $\tau_{*u}$ is defined as a negative value.}
Note that, as found in ILN20a for infinite shocks, the formation of a subshock is an inherent feature in relativistic shocks, also in cases of finite shocks.

Our numerical code employs an iterative method that seeks a self-consistent, steady shock structure in  planar geometry which conserves the energy-momentum flux throughout the flow. In the infinite shock calculations conducted in ILN20a, a large imposed value of $\tau_{u*}$ prevented the photons produced in the shock from diffusing back to the upstream boundary.
In contrast, in the current simulations, the smaller values of $\tau_{u*}$ invoked allow a fraction of these diffusing-back photons, which propagate counter to the flow direction into the upstream region (hereafter, counterstreaming photons), to escape through the upstream boundary of the simulation box, resulting in a decrease (increase) in the net energy (momentum) flux compared to the infinite shock.
Therefore, unlike in the infinite shock cases where the energy-momentum flux is determined by the upstream conditions, these quantities cannot be predetermined in finite shocks. Hence, while the code adjusts only the flow parameters (i.e., $\beta$, $T$, and $n_{\pm}$) in the infinite shock simulations, it additionally adjusts the energy and momentum fluxes in the finite shock simulations.

Following ILN20b, 
we employ the ratio of the energy flux carried by escaping photons to the incoming energy flux of the far-upstream baryons as an indicator of the amount of photons escaping the shock:
\begin{eqnarray}
\label{eq:fraction}
f_{esc} = - \frac{F_{esc}}{F_{b}}.
\end{eqnarray}
Here, $F_{esc}<0$ denotes the net energy flux of the photons at the upstream boundary of the simulation box\footnote{$F_{esc}$ is a negative quantity since we define positive energy flux in the direction along the flow.}  and
$F_{b} = \Gamma_{\rm u}(\Gamma_{\rm u} - 1)n_{\rm u}m_p \beta_{\rm u} c^3$ represents the energy flux of the baryons, with $m_p$ being the proton rest mass. 
By performing a series of simulation runs for various values of $\tau_{u*}$, we can probe the structure and radiation signature of finite shocks with differing escape fractions.

In all cases, we set the baryon density at the upstream boundary to be $n_u = 10^{15}~{\rm cm^{-3}}$.\footnote{The results of the calculations exhibit a relatively weak dependence on $n_u$, which is particularly evident in relativistic shocks where temperature regulation by pair production plays a significant role.}
The plasma, consisting of protons and electron/positron pairs, is modeled as a single fluid in which all constituents share a common bulk velocity and follow a Maxwellian distribution at the same temperature.
This implies that we are assuming tight coupling over distances much shorter than the shock deceleration scale.
While the assumption holds for single-ion, sub-relativistic shocks, it may not be valid for relativistic shocks where vigorous pairs are produced, as indicated by \citet{Levinson2020,Vanthieghem2022}, although even mild magnetization of the upstream plasma can lead to a strong coupling \citep{Mahlmann2023}. 
We intend to further investigate this aspect in our future work.

As in ILN20a and ILN20b, we inject a small number of thermal photons at the upstream boundary, characterized by an energy distribution in the form of Wien spectra with a temperature equal to that of the plasma.
Specifically, the injected photons have a photon-to-baryon number ratio of $n_{\gamma u}/n_u = 10^{-2}$, and have a temperature set at $T_u = (3k n_{\gamma u} a_{rad})^{1/3}$, where $k$ and $a_{rad}$ are the Boltzmann constant and the radiation constant, respectively.
 This step is taken merely for numerical convenience and has no noticeable impact on the results, given that the number and temperature of the injected photons are much lower than those produced within the shock.

\section{THE STRUCTURE OF RRMS WITH ESCAPE}
\label{sec:structure}

\begin{figure}
  \centering
  {\includegraphics[width=1\columnwidth]{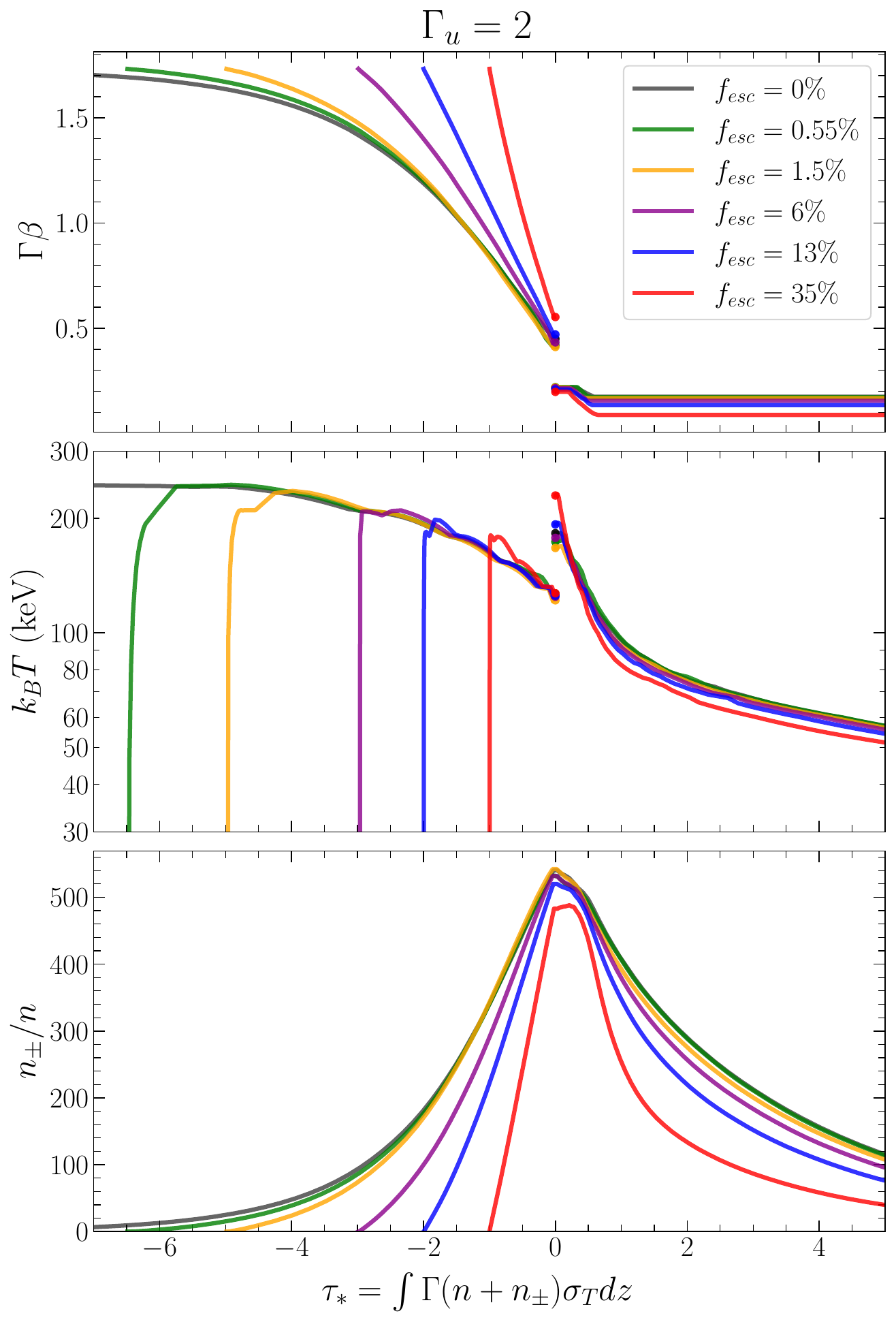}}
  \caption{Profiles of 4-velocity ({\it top}), temperature  ({\it middle}),  and pair-to-baryon ratio ({\it bottom}) as functions of the pair-loaded Thomson optical depth $\tau_* = \int \Gamma (n + n_{\pm})\sigma_T dz$ for the simulations with an upstream Lorentz factor of $\Gamma_u = 2$. 
  The profiles are represented by solid lines in green, orange, purple, blue, and red, each indicating the results from finite shocks with different escape fractions as detailed in the legend. For comparative purposes, the structure of an infinite shock is depicted with a black line. It is noted that for infinite shocks, the upstream boundary extends beyond the left edge of the figure, whereas the leftmost points of the finite shock profiles mark their respective upstream boundaries. A discontinuous jump in both the 4-velocity and temperature profiles at $\tau_{*}=0$ reflects the presence of a subshock. 
  }
  \label{fig:profileGu2}
\end{figure}

\begin{figure}
  \centering
  {\includegraphics[width=1\columnwidth]{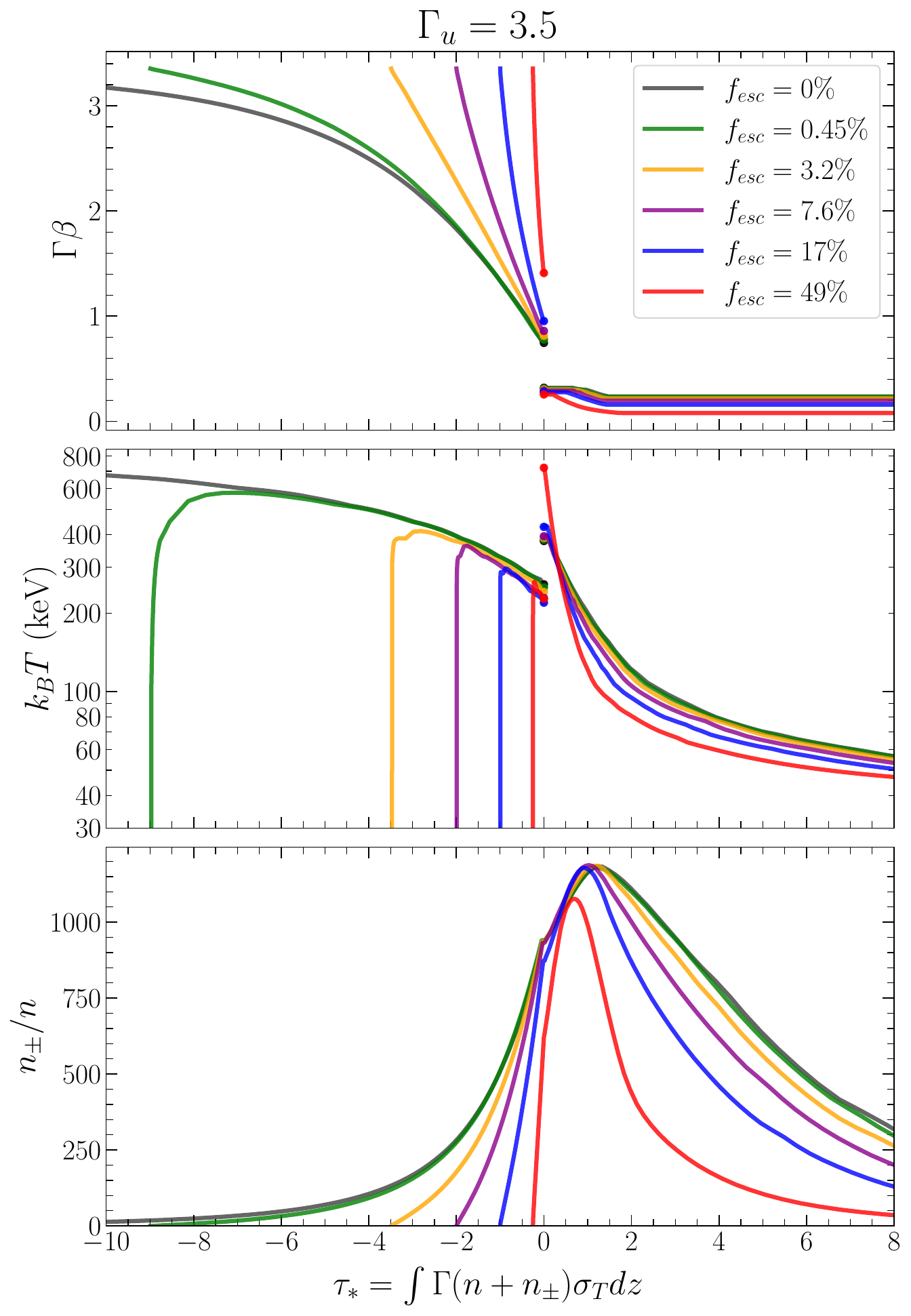}}
  \caption{Same as Figure \ref{fig:profileGu2}, but for $\Gamma_u = 3.5$.}
  \label{fig:profileGu35}
\end{figure}

\begin{figure}
  \centering
  {\includegraphics[width=1\columnwidth]{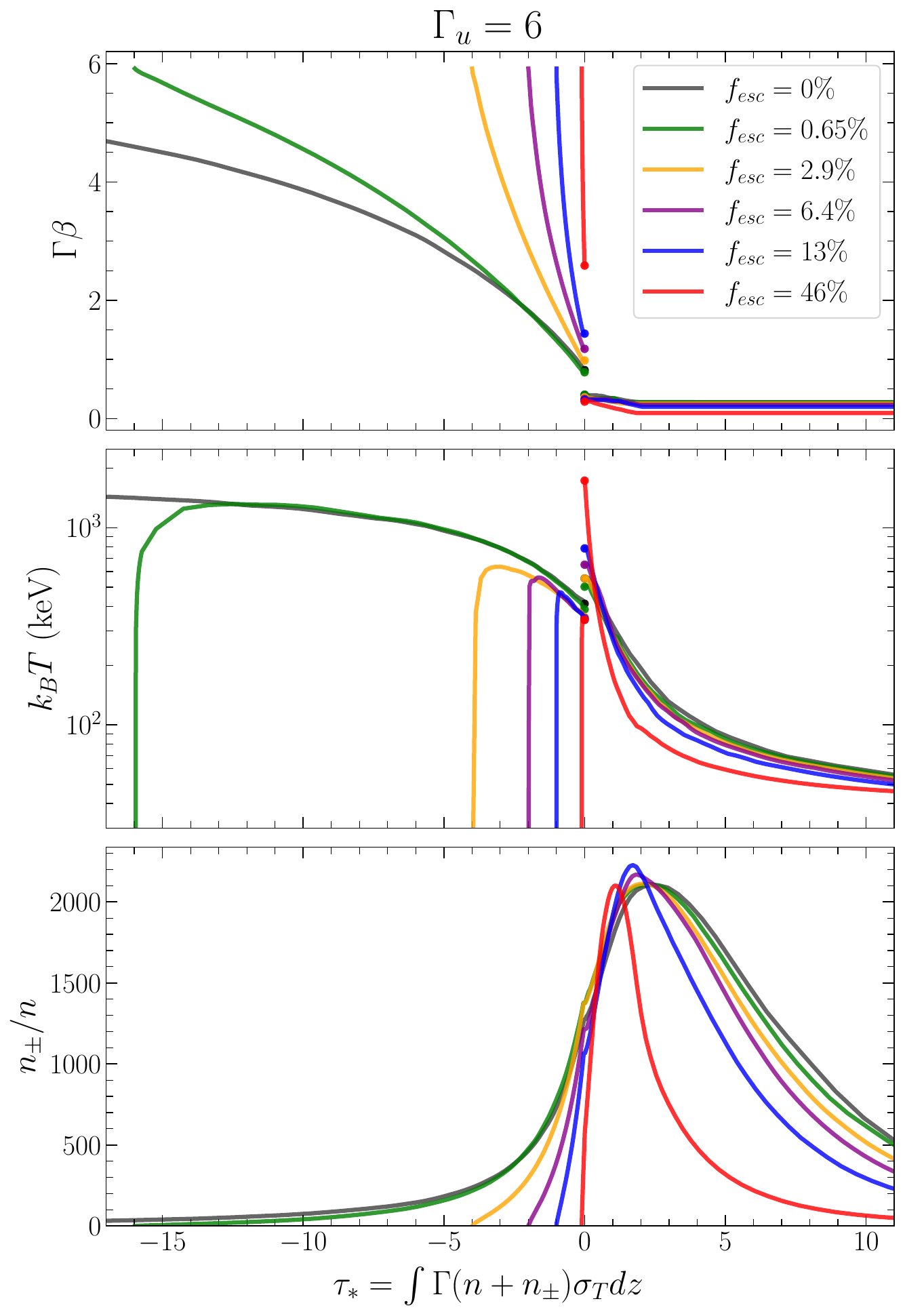}}
  \caption{Same as Figure \ref{fig:profileGu2}, but for $\Gamma_u = 6$.}
  \label{fig:profileGu6}
\end{figure}

\begin{figure}
  \centering
  {\includegraphics[width=1\columnwidth]{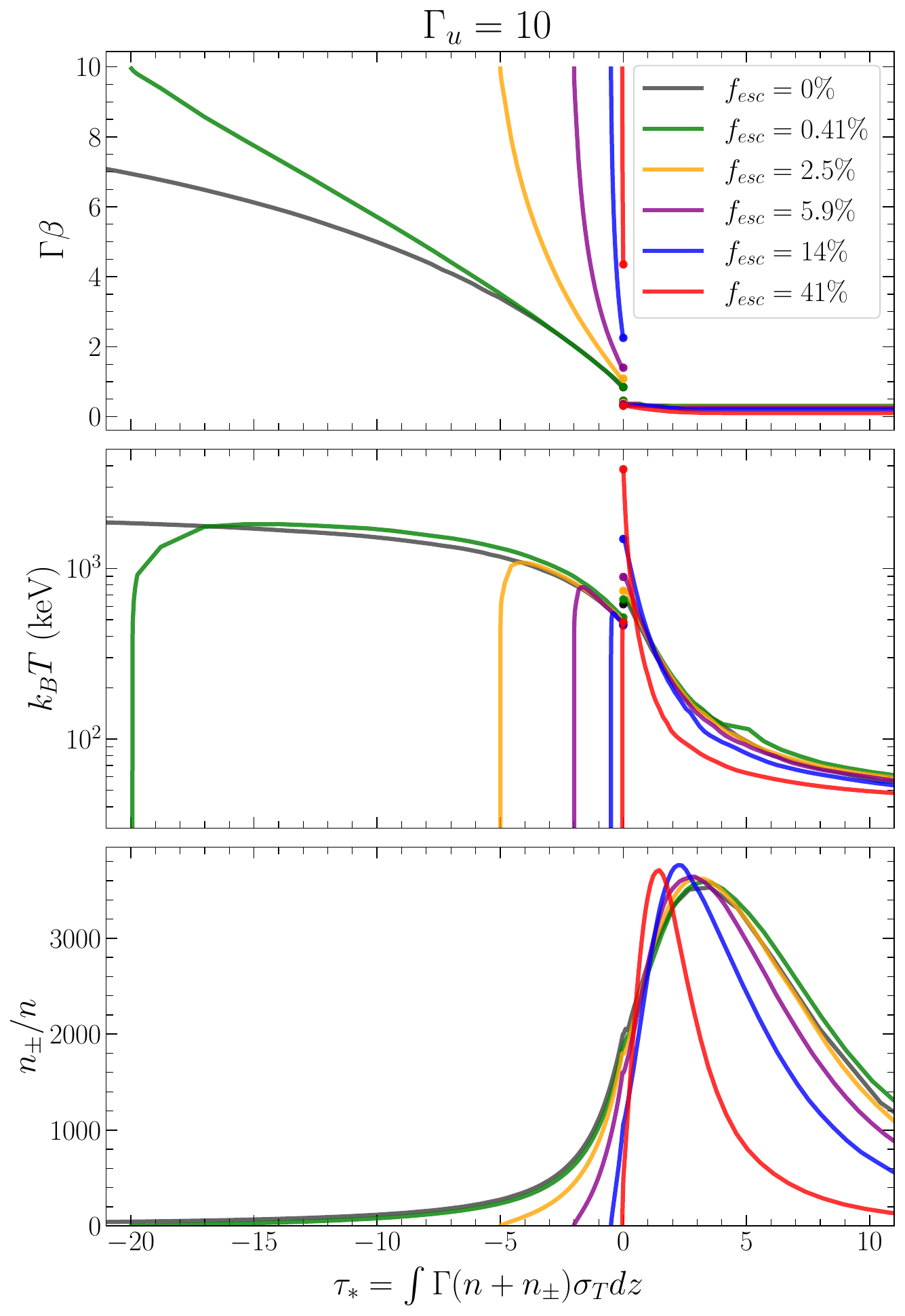}}
  \caption{Same as Figure \ref{fig:profileGu2}, but for $\Gamma_u = 10$.}
  \label{fig:profileGu10}
\end{figure}

Figures \ref{fig:profileGu2} - \ref{fig:profileGu10} display the shock structures derived from simulations for upstream Lorentz factors $\Gamma_u = 2$, $3.5$, $6$, and $10$ across five values of the escape fraction $f_{esc}$ spanning $0$ to $\lesssim 0.5$.
We omit a description of the $\Gamma_u=15$ shock structure here and the corresponding spectra in Section~\ref{sec:spectrum} because we performed only two runs at $f_{esc}=0.14$ and $0.4$, which do not cover the $f_{\rm esc}$ dependence from $0$ to high values. We note, however, that these runs show no qualitative differences from the high-$f_{esc}$ cases at $\Gamma_u = 10$.
As described in Section~\ref{sec:comparison}, we use these two runs to examine how the shock width depends on $\Gamma_u$ at high $f_{\rm esc}$.
As depicted in the figure, the shock width significantly reduces with increasing $f_{esc}$.
This result contrasts with the sub-relativistic shocks \citep[][ILN20b]{Ioka2019}, where the shock width is mostly consistent with the diffusion length $\tau_* \sim \beta_u^{-1}$, regardless of the value of $f_{esc}$. 
This is due to the fact that relativistic shocks have an ability to adjust its opacity by generating pairs (GNL18).
The current simulation finds that a minute amount of escape $f_{esc}\lesssim 0.01$ is sufficient to appreciably narrow the shock width, particularly when the Lorentz factor of the shock is high.
This feature is consistent with the analytical model of GNL18, which suggests that finite shocks deviate markedly from the infinite shock configuration when the escape fraction exceeds $\sim 1/\Gamma_u^2$. 
However, it should be noted that our results also reveal quantitative differences in the shock structure from that of the analytical model at high escape fractions, as will be discussed in Section \ref{sec:comparison}.

As mentioned earlier, the subshock, characterized by a discontinuous jump in the velocity and temperature, is an inherent feature of relativistic shocks regardless of the escape fraction.\footnote{As reported in ILN20b, a subshock is not present in the sub-relativistic regime ($\beta_u = 0.1$ and $0.25$). In the mildly relativistic regime, where $\beta_u = 0.5$, the subshock is absent in the case of an infinite shock; however, it manifests in finite shocks ($f_{esc} > 0$).}
In our simulation, the jumps between pre- and post-subshock velocity and temperature are connected by the Rankine-Hugoniot conditions, under the assumption that the bulk plasma is isolated from the radiation \citep{Ito2018}.
The strength of the subshock increases with $f_{esc}$, as observed in ILN20b for mildly relativistic shocks with $\beta_u = 0.5$.
\footnote{To be precise, in our iterative calculations, the subshock strength does not converge perfectly but fluctuates within a range of roughly $10$-$20\%$ in the jump of 4-velocity $\delta (\Gamma \beta)$. Consequently, the subshock might occasionally appear slightly stronger for smaller escape fractions. Specifically, the sequence of subshock strength depicted in Figure \ref{fig:profileGu2} ($\Gamma_u = 2$) is not directly proportional to the escape fraction in the range of $f_{esc} = 0\%$ to $6\%$. The fluctuations observed in the current simulations might suggest that a perfectly steady profile is unattainable, indicating that the subshock feature could be inherently dynamical. However, exploring this aspect further is beyond the scope of the current study.}
For instance, in shocks with $\Gamma_u = 10$, the 4-velocity at the immediate upstream of the subshock, $\Gamma_{u, sub} \beta_{u, sub}$, is approximately $0.84$, $0.85$, $1.08$, $1.4$, $2.25$, and $4.35$ for $f_{esc} \simeq 0$, $0.41$, $2.5$, $5.9$, $14$, and $41~\%$, respectively.
 This results in a fraction of the total kinetic energy being dissipated at the subshock, which we evaluate as the reduction in kinetic energy flux of the plasma across the subshock, normalized by the incoming kinetic energy flux at the upstream boundary expressed as $\delta F_{kin,sub}/F_b  = [1+(n_{\pm}/n)_{sub} m_e/m_p](\Gamma_{u, sub} - \Gamma_{d,sub})/(\Gamma_u - 1)$, to increase as  $4.8$, $4.8$, $8.8$, $14$, $25$, and $48\%$ for the corresponding values of $f_{esc}$.\footnote{For $\Gamma_u =2$ ($6$), the reduction of kinetic energy across the subshock $\delta F_{kin,sub}/F_b$  is roughly $15\%$ ($45\%$) for the simulation with the highest escape fraction of $f_{esc} = 35\%$ ($46\%$).}
Here, $\Gamma_{d, sub}$ denotes the Lorentz factor immediately downstream of the subshock, and $(n_{\pm}/n)_{sub}$ indicates the pair-to-baryon number density ratio at the subshock.
The result implies that the role of radiation mediation in the shock weakens as $f_{esc}$ increases, and a significant amount of energy is dissipated by the plasma interactions at the subshock in finite shocks with high escape fractions. 
This is in contrast with the sub-relativistic shocks, where subshocks do not appear even at a high escape fraction ($f_{esc} \sim 70\%$ for $\beta \leq 0.25$; ILN20b).

In all cases, the post-subshock velocity is close to that of the far downstream, implying that most of the shock deceleration is accomplished at this point.
Following the subshock, the velocity exhibits a slight and gradual decline, ultimately reaching a far downstream value that is lower for higher $f_{esc}$, due to increased radiative losses.

As mentioned above, the subshock also leads to a jump in the temperature, due to the dissipation of plasma kinetic energy, manifesting as a discontinuous spike in Figures \ref{fig:profileGu2} - \ref{fig:profileGu10}. 
Due to the proportionality of the subshock strength with the escape fraction, simulations with larger escape fractions exhibit a more pronounced temperature spike.
While the temperature in most regions is predominantly determined by the Compton temperature, where the net heating and cooling of the pairs balance, this discontinuous jump causes the temperature to deviate temporarily from equilibrium.
 However, the plasma quickly cools via inverse Compton scattering and rejoins the Compton equilibrium temperature within a pair-loaded optical depth of a few.
The equilibrium temperature subsequent to this spike falls within the range of $k T \sim 100$-$200$ keV, irrespective of the values of $f_{esc}$ or $\Gamma_u$. 
This regulation of temperature is attributed to the vigorous pair production, which functions as a thermostat, preventing the temperature from substantially exceeding the pair production threshold, a characteristic property of RRMS well-discussed in the literature \citep[see][and references therein]{Levinson_Nakar2020}.
As a result, except for the initial rise at the upstream boundary and the spike at the subshock, the temperature exhibits a similar profile among different escape fractions.

Regarding the pair density profile, current simulations confirm the ability of RRMS to self-generate its opacity.
As $f_{esc}$ increases, rapid pair production initiating from the upstream boundary is enhanced. This acceleration in pair production enables the shock to maintain its radiation-mediated nature, despite the pair-unloaded optical depth $\tilde{\tau}$ ($\ll 1$) decreasing by several orders of magnitude with increasing $f_{esc}$. 
Subsequent to this rapid increase, a similar peak level of pair density is observed across various $f_{esc}$ values in the immediate downstream region.

\section{THE SPECTRUM OF ESCAPING RADIATION}
\label{sec:spectrum}

\begin{figure*}
  \centering

  \setlength{\columnsep}{0.1em}

  \newlength{\panelw}
  \setlength{\panelw}{\dimexpr0.5\textwidth - 0.5\columnsep\relax}

  \begin{minipage}[t]{\panelw}\centering
    \includegraphics[width=\linewidth,keepaspectratio]{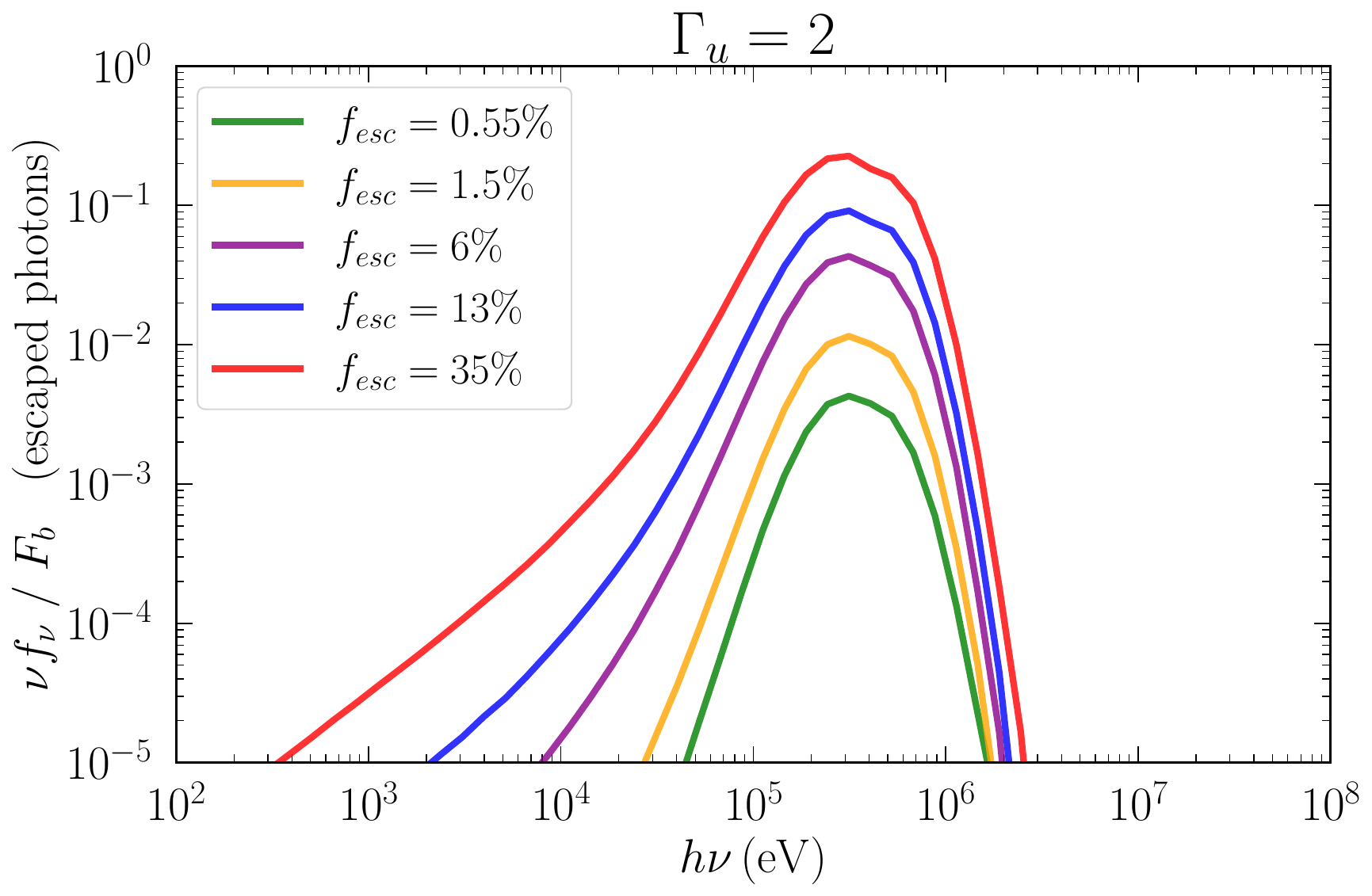}
  \end{minipage}\hspace{\columnsep}%
  \begin{minipage}[t]{\panelw}\centering
    \includegraphics[width=\linewidth,keepaspectratio]{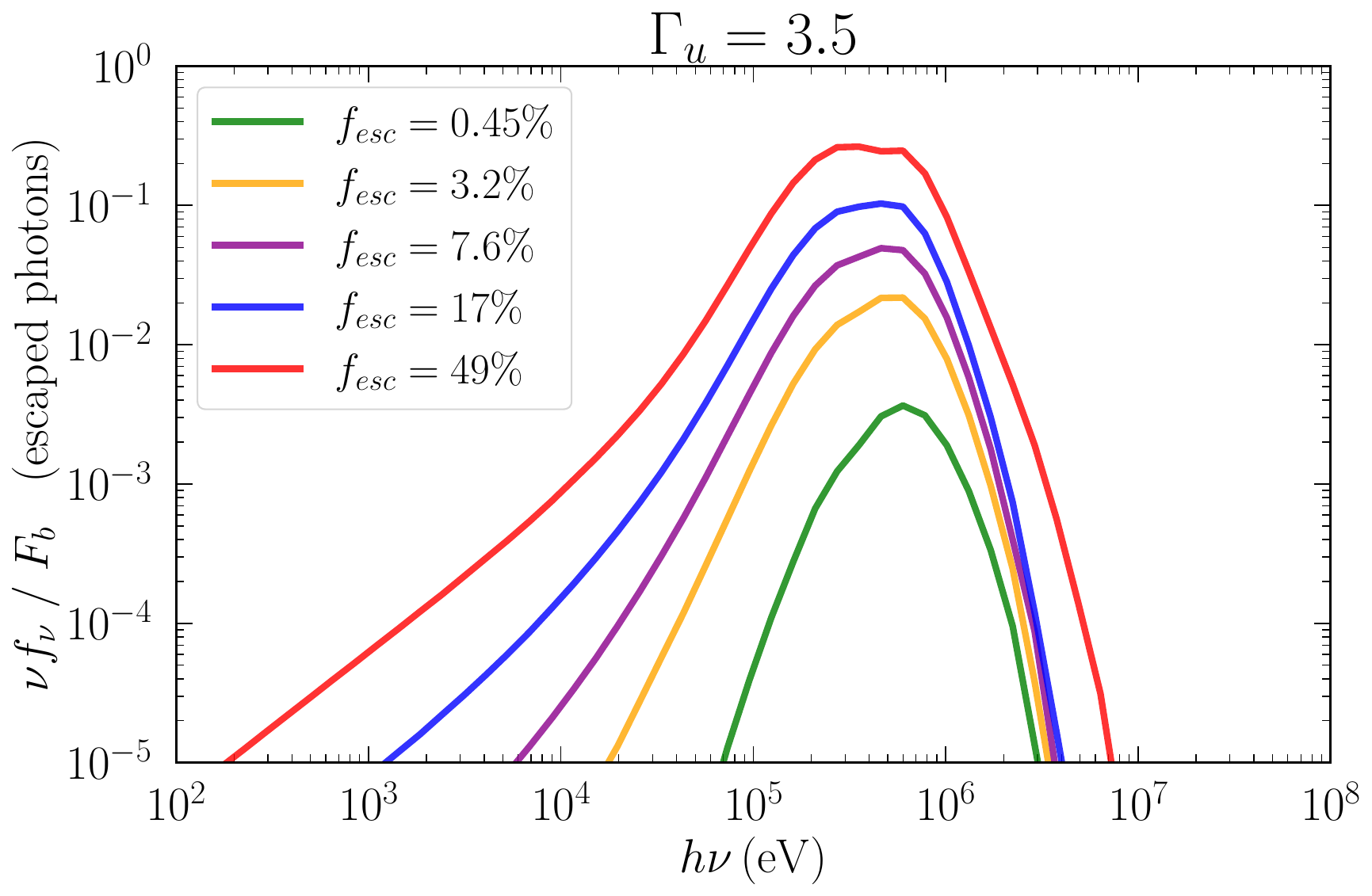}
  \end{minipage}

  \par\medskip %

   \begin{minipage}[t]{\panelw}\centering
    \includegraphics[width=\linewidth,keepaspectratio]{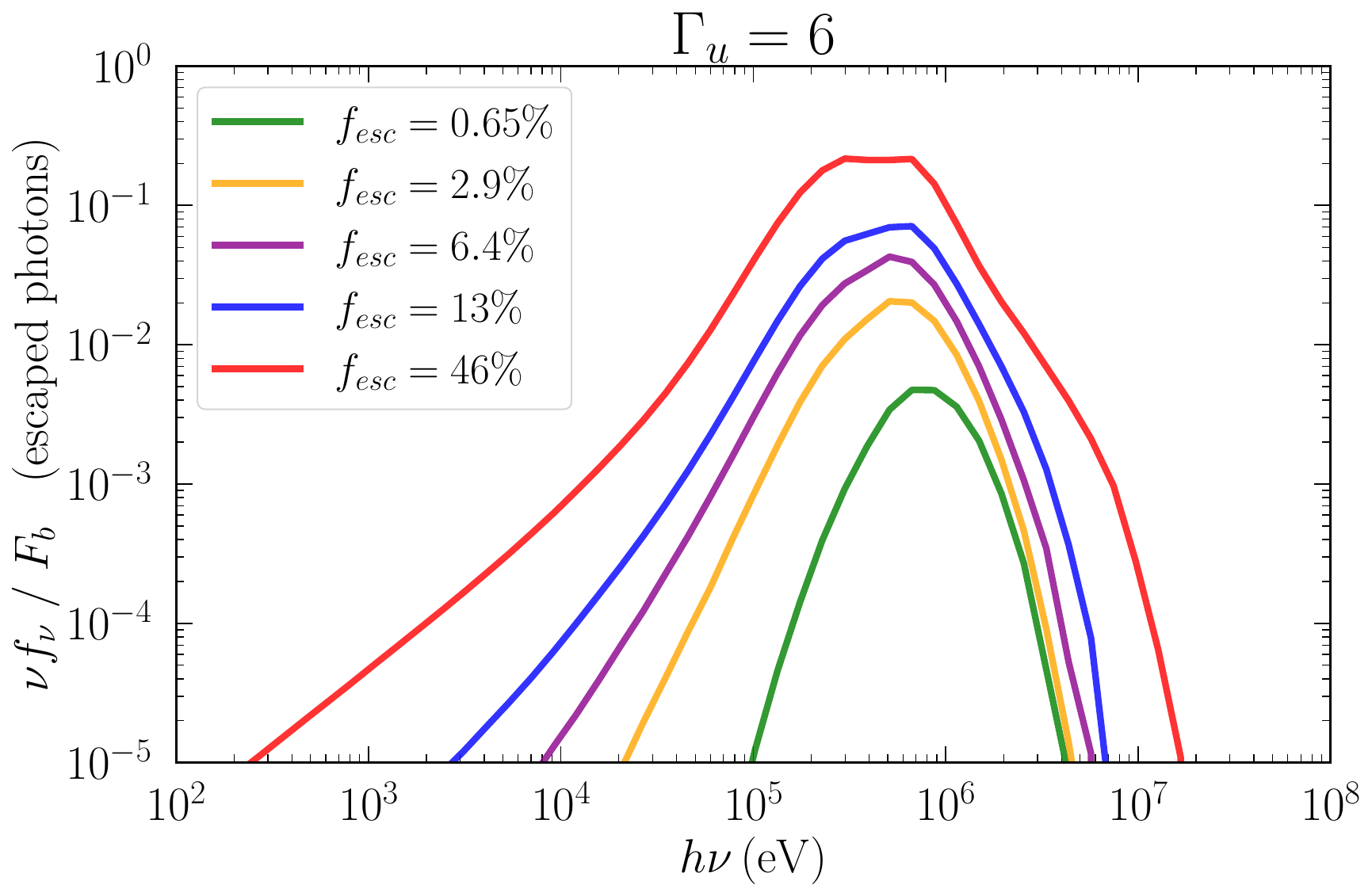}
  \end{minipage}\hspace{\columnsep}%
  \begin{minipage}[t]{\panelw}\centering
    \includegraphics[width=\linewidth,keepaspectratio]{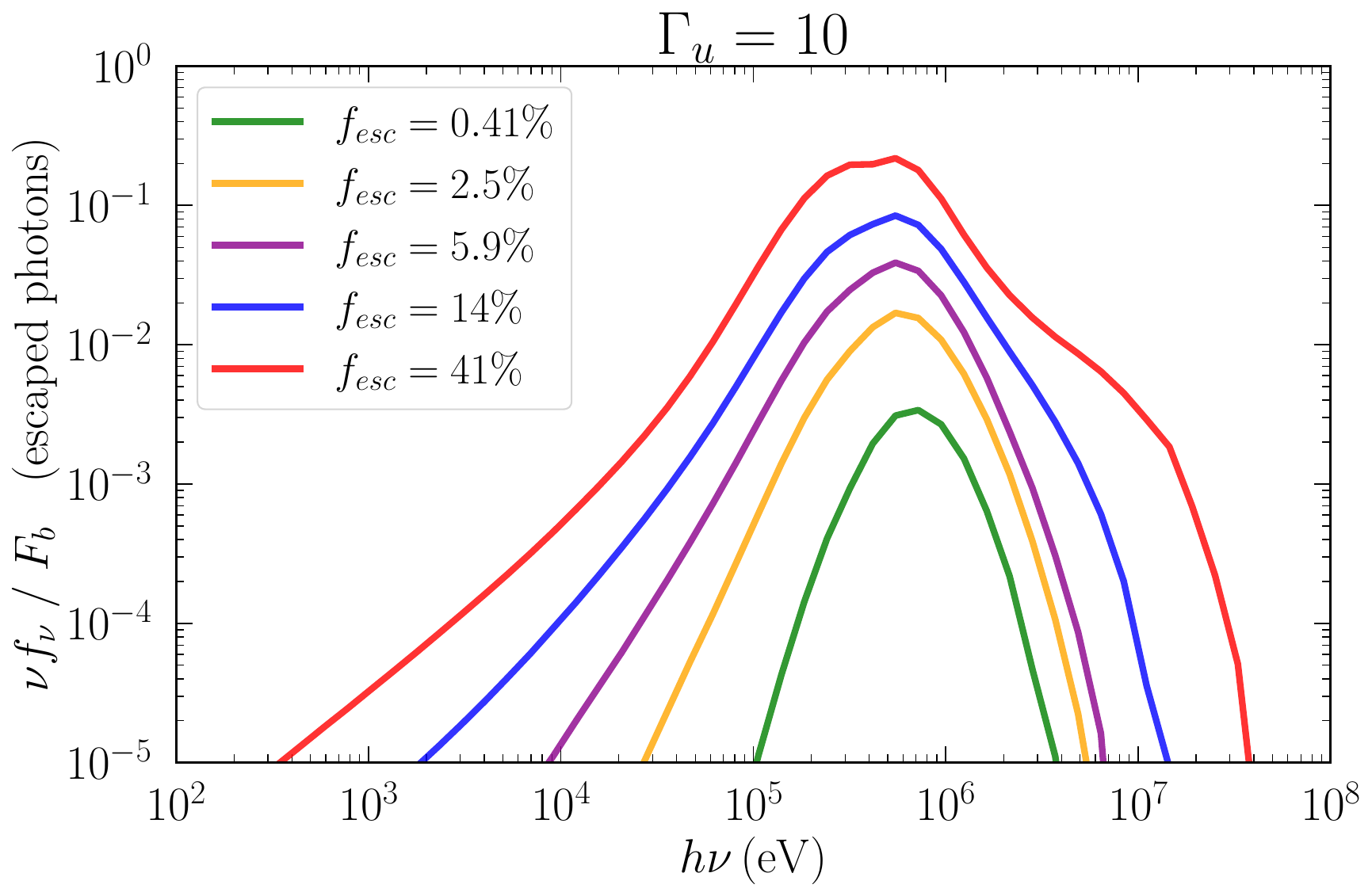}
  \end{minipage}

  \caption{Shock-frame, $\nu f_\nu$ flux of escaping photons normalized by the total kinetic energy flux of baryons at the upstream boundary, $F_b = \Gamma_{u} (\Gamma_{u}-1) n_{u} m_p c^3 \beta_{u}$. Each panel shows results for $\Gamma_{u}=2$ (top left), $3.5$ (top right), $6$ (bottom left), and $10$ (bottom right).} Lines indicate different escape fractions (see legends).
  \label{fig:nufnu}
\end{figure*}

In Figure \ref{fig:nufnu}, we display the resulting spectral energy distribution of the photons escaping from the upstream boundary, presented in the form of $\nu f_\nu$ normalized by the incoming energy flux $F_b$. Here, $\nu$ represents the frequency of the photon, and $f_\nu \equiv \frac{dF_{esc}}{d\nu}$ denotes the energy flux per unit frequency.
In all cases, the energy at the peak of $\nu f_\nu$ spectra falls within the range of $E_{p} \approx 300$ to $600~{\rm keV}$, irrespective of $\Gamma_u$ and $f_{esc}$. This stability reflects the downstream temperature's stabilization due to pair production ($E_p \sim 3 k T$).
Each line in the figure, corresponding to a specific value of $f_{esc}$ as indicated, represents the instantaneous spectrum emitted during the gradual shock breakout from a wind-like medium at the radius where the optical depth to infinity ahead of the shock roughly equals the local shock width (smaller $f_{esc}$ corresponds to earlier phase of the breakout).
It should be noted, however, that to model the actual breakout signal, the entire spectra, including $E_p$, must be boosted to the upstream (observer's) rest frame, resulting in the spectra being shifted to higher energies by a factor of $\sim \Gamma_u$.


  As found in ILN20b for sub- and mildly relativistic shocks ($\beta_u = 0.1$ - $0.5$), the overall shape of the spectra is substantially broader than that of the Wien spectrum ($f_\nu \propto \nu^3 {\rm exp}(-h\nu / kT)$).
  Notably, below the pronounced peak at $E_p$, produced by the thermal Comptonization in the downstream region, the spectra exhibit a soft extension.
  This extension is characterized by a slope that asymptotically approaches $f_{\nu} \propto \nu^0$, which is attributed to free-free emission.
  It extends down to energies below which the spectra show a break, resulting from the low-energy cutoff of the free-free emission due to the Coulomb screening effect.
  Though energetically less dominant in the spectra, the prominence of the soft tail (the spectral region where $f_{\nu} \propto \nu^{0}$) becomes more pronounced as the escape fraction increases.
  To clearly illustrate the feature of this soft extended tail, the spectra in the $f_{\nu}$ form are provided in Appendix \ref{App:fnu} (see Figure \ref{fig:fnu}).
  This characteristic feature,  aligning with the observations for both sub-relativistic and mildly relativistic shocks as detailed in ILN20b, suggests that it can be considered a distinctive signature of shock breakout, independent of shock velocity.

  A deviation from the thermal-like spectra is also observed above $E_p$, specifically, a non-exponential cut-off feature. While this feature is modest for $\Gamma_u = 2$, a clear deviation becomes evident for $\Gamma_u \geq 3.5$, 
  especially at higher escape fractions. 
  For example, a significant excess over the exponential cutoff is observed at
  $f_{esc} =  49\%$ for $\Gamma_u = 3.5$, and at $f_{esc} \gtrsim  10\%$ $\Gamma_u = 6$ and $10$. 
  This excess is generated by the effect of the subshock, wherein high-temperature pairs at the post-subshock region up-scatter a fraction of photons, immediately transferring their energy. Consequently, the stretch of the high-energy spectra becomes more pronounced for higher $f_{esc}$, since the post-subshock temperature increases due to the enhanced strength of the subshock as discussed in Section \ref{sec:structure}. For instance,  in the simulation for $\Gamma_u = 10$ with the highest escape fraction ($f_{esc} = 41\%$), the post-subshock temperature reaches $kT \sim 4~{\rm MeV}$. This aligns with the observed concave curvature observed at $h\nu > E_p$ in the spectra, which begins to exhibit a downturn around $h\nu \sim 3 kT \sim 10~{\rm MeV}$ (as illustrated by the red line in the bottom panel of Figure \ref{fig:nufnu}).

  This high-energy excess is a characteristic feature of relativistic shocks, where a significant fraction of energy is dissipated at the subshock as $f_{esc}$ increases. 
  This stands in contrast to the sub-relativistic ($\beta_u = 0.1$ and $0.25$) and mildly relativistic shocks ($\beta_u = 0.5$) explored in ILN20b, where no deviation from the exponential cutoff is observed. The absence (for $\beta_u = 0.1$ and $0.25$ up to $f_{esc} \sim 70\%$) or relative weakness (for $\beta_u = 0.5$ up to $f_{esc} \sim 45\%$) of a subshock prevents the emergence of such an excess.
  It should be noted that the excess observed in the present simulations has not yet reached its maximum and is expected to become more pronounced as $f_{esc}$ increases further. This is due to the anticipated continuous strengthening of the subshock with increasing $f_{esc}$, along with a reduction in $\gamma-\gamma$ attenuation caused by the decreasing optical depth upstream of the subshock.
  
  Regarding $\gamma-\gamma$ attenuation, we note that high-energy photons ($h\nu > E_p$) undergo significant absorption before reaching the upstream boundary. 
  For instance, in the simulation with $\Gamma_u = 10$ and $f_{esc} = 41\%$, the energy flux of counterstreaming photons at $h\nu \approx 10~{\rm MeV}$ at the escaping boundary ($\tau_{*} = \tau_{u} = 4 \times 10^{-2}$) decreases by approximately a factor of 10 compared to that at the immediate post-subshock ($\tau_{*} = 0$). 
  It is worth noting that, as will be discussed in Section~\ref{sec:comparison}, this $\gamma-\gamma$ attenuation of high-energy photons is not only crucial for shaping the resulting spectrum but also a dominant factor in the deceleration process of finite shocks with $\Gamma_u = 3.5$, $6$ and $10$, particularly when $f_{esc}$ is large.
  Additionally, particle acceleration at the subshock—--omitted in our present simulations—--may influence the high-energy end of the spectrum. We briefly address this point in Section~\ref{sec:paracc}.

  To summarize, our results indicate that while the spectral peak remains stable at $E_p \approx 300 - 600~{\rm keV}$ in relativistic shocks, spectral distortions from the thermal-like spectrum, i.e., Wien spectrum, become more pronounced as the escape fraction increases. This suggests that as the breakout progresses, the spectrum broadens both below and above $E_p$. 

 \section{Comparison with The Analytical Model}\label{sec:comparison}

 In this section, we compare our numerical simulation with the analytical model proposed by GNL18. Before presenting the comparison, we first provide a brief overview of the analytical model. A more detailed explanation, including the shock profile derived from the model, is given in Appendix \ref{App:Anamodel} and \ref{App:CompINF}.
 The model describes the deceleration of the relativistic flow in the shock transition region by modeling its interaction with the counterstreaming photons emanating from the immediate downstream region. 
  The fraction of the counterstreaming photons that reach the upstream boundary, which is equivalent to $f_{esc}$ defined in Equation (\ref{eq:fraction}), is an input parameter of the model. For given values of $\Gamma_u$ and $f_{esc}$, the model construct a structure of RRMS.
 In addition to these two input parameters, the model contains free parameters: $\eta$, which characterizes the thermal energy added to the flow per each interaction with counterstreaming photons; and $a$, which characterizes the average energy of counterstreaming photons as observed by thermal pairs. 
 For the given parameters, the analytical model derives a smooth flow structure that extends from the upstream boundary, where $\Gamma = \Gamma_u$, up to the immediate downstream region, a position where deceleration by the counterstreaming photons is completed ($\Gamma \approx 1$).

\begin{figure*}
  \centering

  \setlength{\columnsep}{0.1em}

  \setlength{\panelw}{\dimexpr0.5\textwidth - 0.5\columnsep\relax}

  \begin{minipage}[t]{\panelw}\centering
    \includegraphics[width=\linewidth,keepaspectratio]{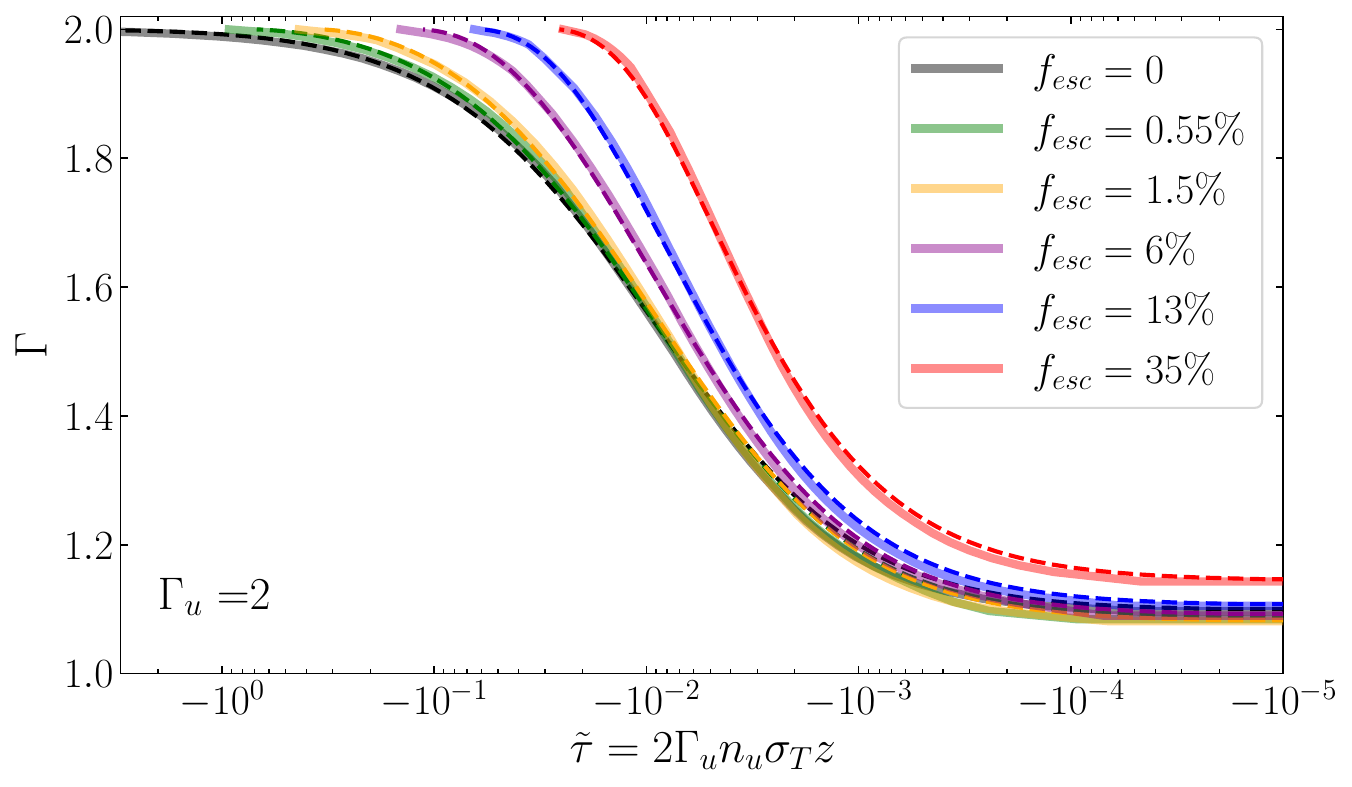}
  \end{minipage}\hspace{\columnsep}%
  \begin{minipage}[t]{\panelw}\centering
    \includegraphics[width=\linewidth,keepaspectratio]{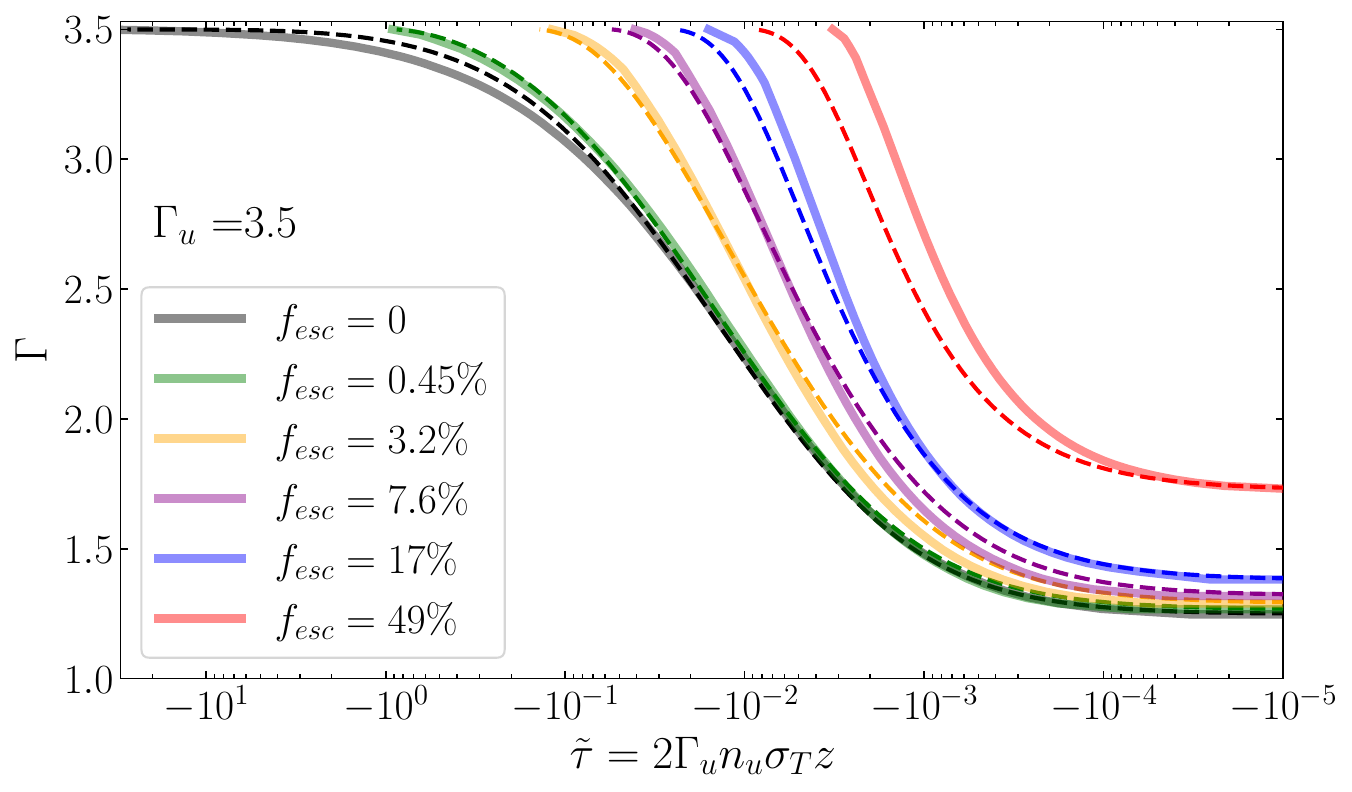}
  \end{minipage}

  \par\medskip %

   \begin{minipage}[t]{\panelw}\centering
    \includegraphics[width=\linewidth,keepaspectratio]{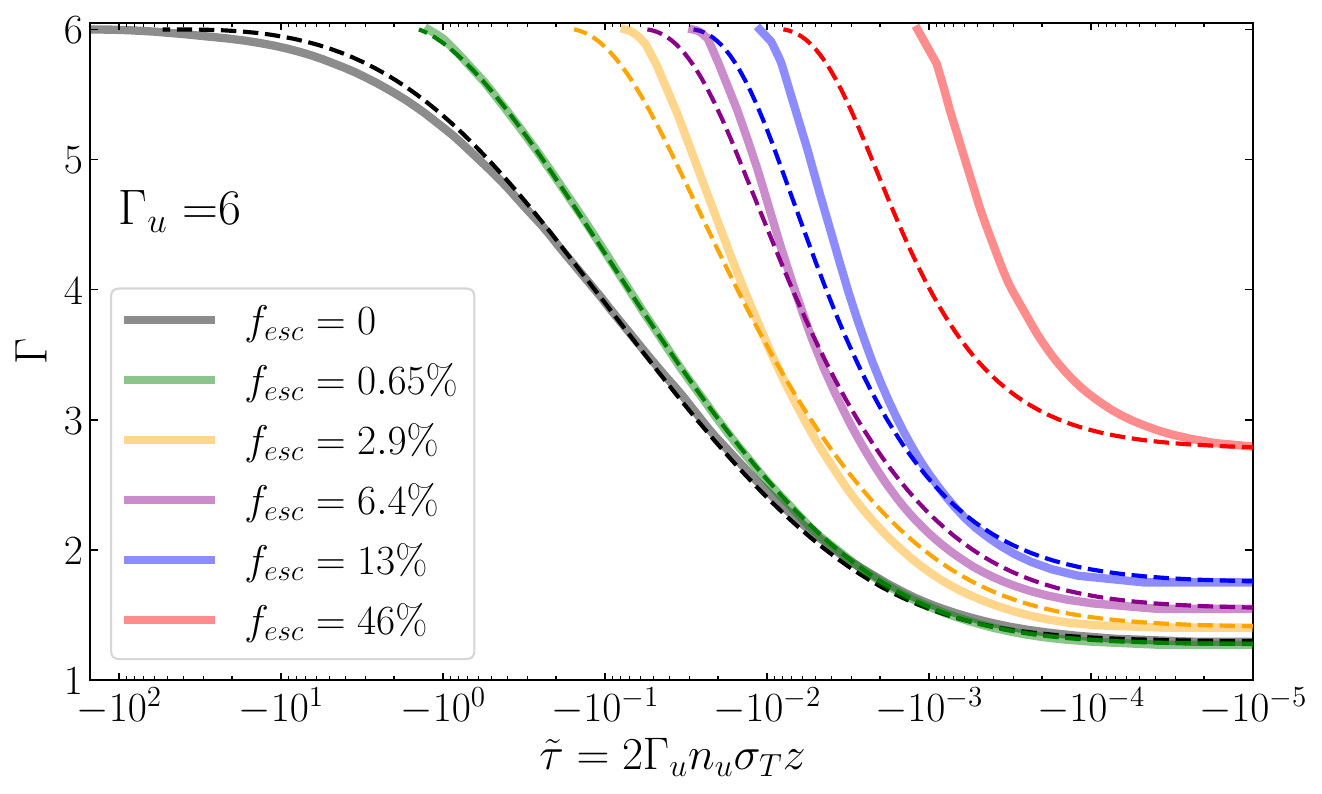}
  \end{minipage}\hspace{\columnsep}%
  \begin{minipage}[t]{\panelw}\centering
    \includegraphics[width=\linewidth,keepaspectratio]{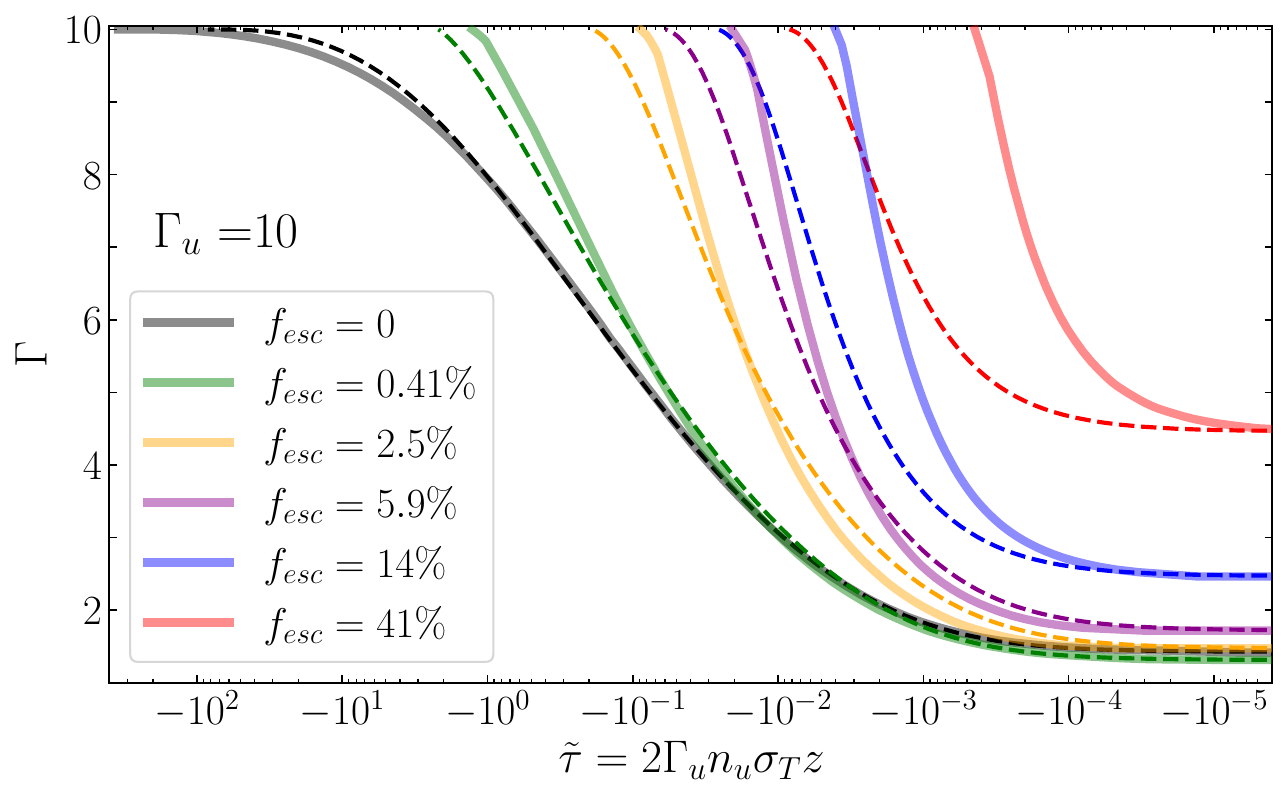}
  \end{minipage}

  \caption{
    Comparison of the Lorentz factor profile as a function of pair-unloaded optical depth, $\tilde{\tau} = 2\Gamma_u n_u \sigma_T z$, with the analytical model of GNL18 for $\Gamma_u = 2$ (top left), $3.5$ (top right), $6$ (bottom left), and $10$ (bottom right).
For the shock with $\Gamma_u = 2$, the analytical model reproduces the numerical results, with 
the optimal parameter pairs $(a, \eta)$ for each escape fraction $f_{esc}$ are determined as follows: at $f_{esc} = 0\%$, $a = 0.55$ and $\eta = 0.4$;
at $0.55~\%$, $a = 0.6$ and $\eta = 0.4$; at $f_{esc} = 1.5~\%$, $a = 0.65$ and $\eta = 0.4$; at $f_{esc} = 6~\%$ $a = 0.7$ and $\eta = 0.4$; at $f_{esc} = 13~\%$ $a = 0.8$ and $\eta = 0.4$; at $f_{esc} = 35~\%$ $a = 1.95$ and $\eta = 0.35$. Although the best fit values of the free parameters vary among the different escape fractions, a single pair, e.g., $a = 0.5$ and $\eta = 0.3$ exhibits a good agreement with the numerical results.
On the other hand, for the shock with $\Gamma_u=3.5$, the optimal parameter pairs $(a,\eta)$ for each escape fraction $f_{esc}$ are as follows: at $f_{esc}=0\%$, $a=0.5$ and $\eta=0.45$; at $0.45\%$, $a=0.85$ and $\eta=0.4$; at $3.2\%$, $a=0.75$ and $\eta=0.3$; at $7.6\%$, $a=0.75$ and $\eta=0.25$; at $17\%$, $a=0.5$ and $\eta=0.25$; and at $49\%$, $a=0.5$ and $\eta=0.25$. For the high–escape-fraction cases at $\Gamma_u=3.5$ ($f_{esc}\ge 0.17$), both parameters sit at the lower bounds of the fitting range ($a=0.5$, $\eta=0.25$), indicating the analytical model’s limitations in capturing the simulation results at high $f_{esc}$.
For the shock with $\Gamma_u=6$, the optimal pairs are: at $f_{esc}=0\%$, $a=2.4$ and $\eta=0.6$; at $0.65\%$, $a=0.7$ and $\eta=0.75$; at $2.9\%$, $a=0.5$ and $\eta=0.35$; and for $6.8\%$, $13\%$, and $46\%$, both parameters reach the lower bounds ($a=0.5$, $\eta=0.25$).
For the shock with $\Gamma_u=10$, the optimal pairs are: at $f_{esc}=0\%$, $a=1.75$ and $\eta=0.6$; at $0.41\%$, $a=0.5$ and $\eta=0.8$; at $2.5\%$, $a=0.5$ and $\eta=0.35$; and for $5.9\%$, $14\%$, and $41\%$, both parameters again take the lower bounds ($a=0.5$, $\eta=0.25$).
Taken together, the high-$f_{esc}$ points for $\Gamma_u=3.5$, $6$, and $10$ saturate at the lower bounds $(a=0.5,\ \eta=0.25)$, consistent with the above indication of the analytical model’s limitations in this regime.
}
   \label{fig:comparison}
\end{figure*}

 Although the model is designed to describe flows with highly relativistic velocities ($\Gamma \gg 1)$, we perform comparisons across all models, including mildly relativistic shocks with $\Gamma_u = 2$. 
 The comparison is conducted by fitting the Lorentz factor profile, expressed as a function of the pair-unloaded Thomson optical depth $\tilde{\tau} = 2\Gamma_u n_u \sigma_T z$, derived from the analytical model, to the corresponding profile from the simulation through the optimization of the parameters $a$ and $\eta$.\footnote{To facilitate the comparison, the optical depth used in the analytical model must be translated into $\tilde{\tau}$. The method for this translation, which is essentially identical to that adopted in GNL18 with a minor modification, is provided in Appendix \ref{App:CompINF}.}
 These values depend on the energy and angular distributions of pairs and photons and are expected to be on the order of unity. 
 Accordingly, we conduct the fitting procedure within predefined ranges, with $a$ set to vary between 0.5 and 3, and $\eta$ from 0.25 to 1.0.\footnote{Even if we succeed in obtaining a good fit with values far beyond the predefined range, the solution should be considered unphysical.}
 To determine the best-fit values, we systematically scanned these ranges in increments of $0.05$ for both parameters.
 The comparison is displayed in Figure \ref{fig:comparison}.
 It is worth noting that $\tilde{\tau}$ represents the pair-unloaded optical depth measured in the upstream rest frame.
 Therefore, when a shock propagates in a wind-like environment, the fraction of shock energy released at a given optical depth can be determined by identifying $f_{esc}$ that yields a shock structure with a depth $\tilde{\tau_u} (= 2\Gamma_u \sigma_T z_u)$ comparable to the optical depth.

 While the analytical solution derives a flow profile in which continuous deceleration due to counterstreaming photons continues until $\Gamma \approx 1$ regardless of $f_{esc}$, our numerical solution reveals an existence of a subshock where the Lorentz factor, $\Gamma_{u, sub}$, significantly deviates from 1.
 Therefore, in each fitting procedure, we identify a position of $\tau$ in the analytical solution that matches $\Gamma = \Gamma_{u, sub}$. 
 Starting from this position, we conducted a comparison of the Lorentz factor profiles up to the upstream boundary with the numerical results. 
 Specifically, we adjusted the position of $\tau = 0$ in the analytical solution from the initial position where $\Gamma \approx 1$ to a new position where $\Gamma = \Gamma_{u, sub}$, effectively shifting it to align with the subshock location in the numerical solution.

\subsection{Comparison of Shock Profiles}  

In the case of infinite shocks, we find that, within a reasonable range of $\eta$ and $a$, the analytical model reproduces the numerical results quite well in all cases ($\Gamma_u = 2$, $3.5$, $6$, and $10$). This agreement was previously reported in ILN20a for $\Gamma_u = 6$ and $10$ (as well as for $\Gamma_u = 20$, which is not included in the present paper), confirming the model's validity for highly relativistic shocks. Here, we extend this finding to mildly relativistic shocks ($\Gamma_u < 6$), showing that even for $\Gamma_u = 2$, the analytical model exhibits good agreement with the numerical results, demonstrating its applicability to lower-velocity shocks than previously expected.
A more comprehensive comparison between the analytical model and the numerical results for infinite shocks is presented in Appendix \ref{App:CompINF}.
For mildly relativistic shocks with $\Gamma_u = 2$, we also find a good match in the case of finite shocks, regardless of the value of $f_{esc}$. 
By contrast, for higher Lorentz factor shocks ($\Gamma_u \geq 3.5$), good agreement persists at small $f_{\rm esc}$, whereas a notable discrepancy emerges and grows as $f_{\rm esc}$ increases. This discrepancy cannot be reconciled by varying the parameters $\eta$ and $a$ within physically reasonable ranges; it becomes larger at higher $\Gamma_u$. This trend of increasing discrepancy with $\Gamma_u$ is also confirmed by our additional two $\Gamma_u = 15$ runs at $f_{\rm esc} = 0.14$ and $0.4$, which are not shown in Figure~\ref{fig:comparison}.

The result is counterintuitive, as one would naively expect highly relativistic shocks to match better with the analytical model, given that it is constructed under the assumption of $\Gamma \gg 1$. 
This discrepancy can be understood as a consequence of the role of the subshock in highly relativistic shocks, which influences the spectral shape of counterstreaming photons. As shown in Section \ref{sec:structure}, the subshock produces a high-temperature spike in the immediate downstream region, where relativistically heated pairs scatter photons to higher energies. While the post-subshock temperature remains in the range $kT_{d,sub} \lesssim 0.5 m_e c^2$ for $\Gamma_u = 2$ (at least up to $f_{esc} \leq 35\%$), it 
 exceeds $\sim m_e c^2$ and continues to rise as the escape fraction increases for $\Gamma_u \geq 3.5$.
%
As a result of this high temperature, a high-energy photon component  ($h\nu \sim  3 k T_{d, sub} \gtrsim 3 m_e c^2$) emerges in addition to the bulk of photons with energy $\sim m_e c^2$, as described in Section \ref{sec:spectrum}. The high-energy extension observed in the escaping photon spectrum above the spectral peak $E_p \sim m_e c^2$ (Figure \ref{fig:nufnu}) corresponds to the fraction of these high-energy photons that counterstream up to the upstream boundary. 
As mentioned in Section \ref{sec:spectrum}, these high-energy photons are efficiently absorbed via $\gamma$-$\gamma$ attenuation before reaching the upstream boundary. This absorption of high-energy counterstreaming photons is responsible for the deviation of the numerical results from the analytical model, which can be explained as follows.

The key point is that the absorption of counterstreaming high-energy photons primarily occurs via collisions with the bulk population of lower-energy photons that are likewise counterstreaming upstream from the immediate downstream region, rather than with photons advecting with the flow (i.e., backscattered photons).
This occurs because the high-energy photons possess sufficient energy to exceed the pair production threshold necessary for interaction with the bulk of the counterstreaming photon population.
As a result, the effective target photon population for the high-energy photons is significantly more abundant than the backscattered photons assumed in the analytical model.
As the escape fraction increases, the subshock strengthens, and this channel of $\gamma - \gamma$ pair production becomes the primary mechanism governing the shock deceleration process.
Consequently, as $f_{\rm esc}$ increases,  the assumption in the analytical model that shock dissipation is governed by the interaction of counterstreaming photons ($h\nu \sim m_e c^2$) with pairs and backscattered photons begins to break down, and the shock deceleration width becomes significantly shorter due to the high opacity arising from $\gamma$-$\gamma$ interactions among counterstreaming photons.
As noted earlier, for $\Gamma_u \geq 3.5$ the deviation from the analytical model increases with $\Gamma_u$—--a trend we confirm up to $\Gamma_u=15$ in the present study. The larger deviations at higher $\Gamma_u$ are due to the increasingly pronounced subshock temperature spike, $kT_{d,{\rm sub}}$.

This breakdown does not occur for $\Gamma_u = 2$, simply because the high-energy extension is modest, and $\gamma - \gamma$ interactions between counterstreaming photons do not play a significant role.
Note that, in the case of $\Gamma_u = 2$, the channel of $\gamma - \gamma$ pair production occurring via collisions between counterstreaming photons and backscattered photons also appears to play a negligible role in the deceleration process, with Compton scattering being the dominant mechanism, as described in Section \ref{App:CompINF}.
Nevertheless, although this qualitative difference is reflected in a moderate systematic deviation in the temperature profile and the total density of advected photons and pairs (see Appendix \ref{App:CompINF}), our results indicate that the analytical model is capable of reproducing the Lorentz factor profile for $\Gamma_u = 2$ with good accuracy, regardless of the value of $f_{esc}$.

\subsection{Comparison of Shock Widths}

\begin{figure*}
  \centering

  \setlength{\columnsep}{0.1em}

  \setlength{\panelw}{\dimexpr0.5\textwidth - 0.5\columnsep\relax}

  \begin{minipage}[t]{\panelw}\centering
    \includegraphics[width=\linewidth,keepaspectratio]{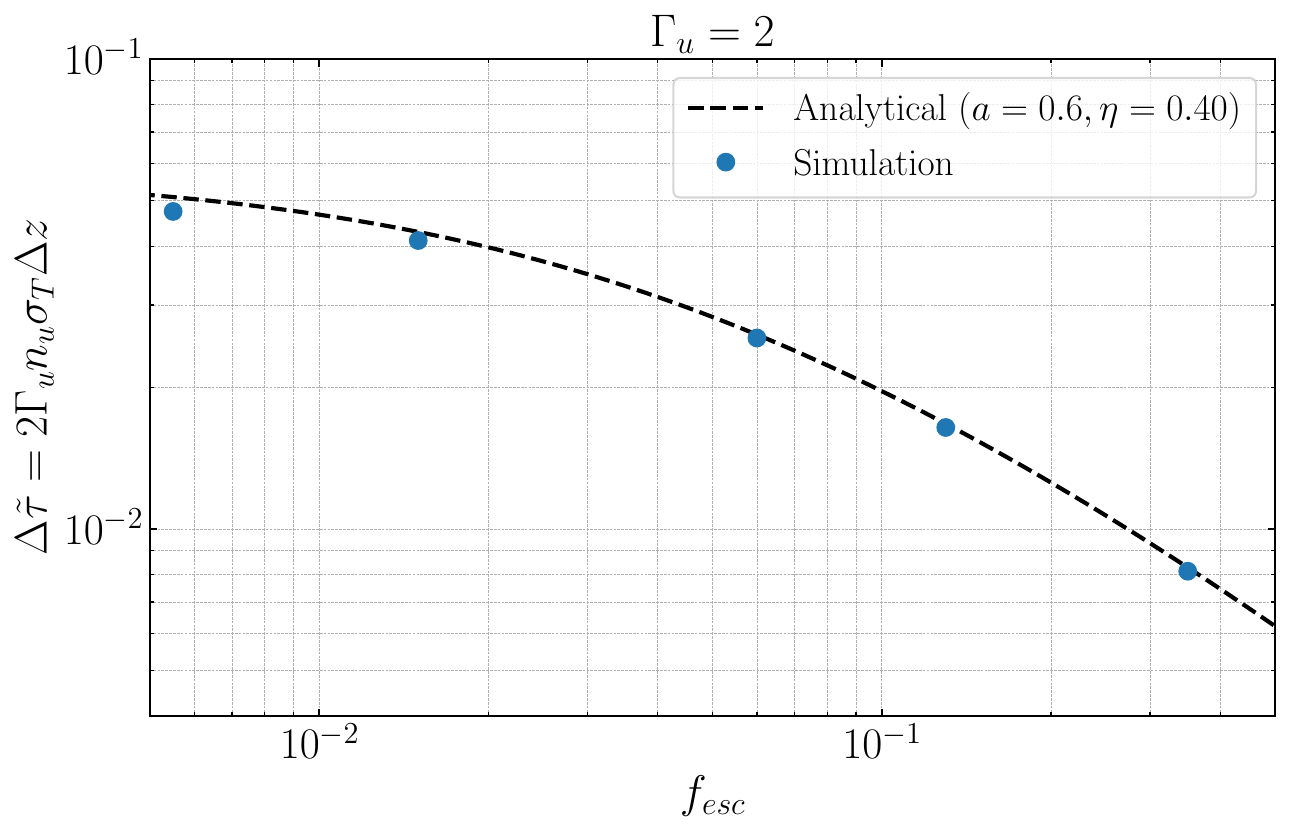}
  \end{minipage}\hspace{\columnsep}%
  \begin{minipage}[t]{\panelw}\centering
    \includegraphics[width=\linewidth,keepaspectratio]{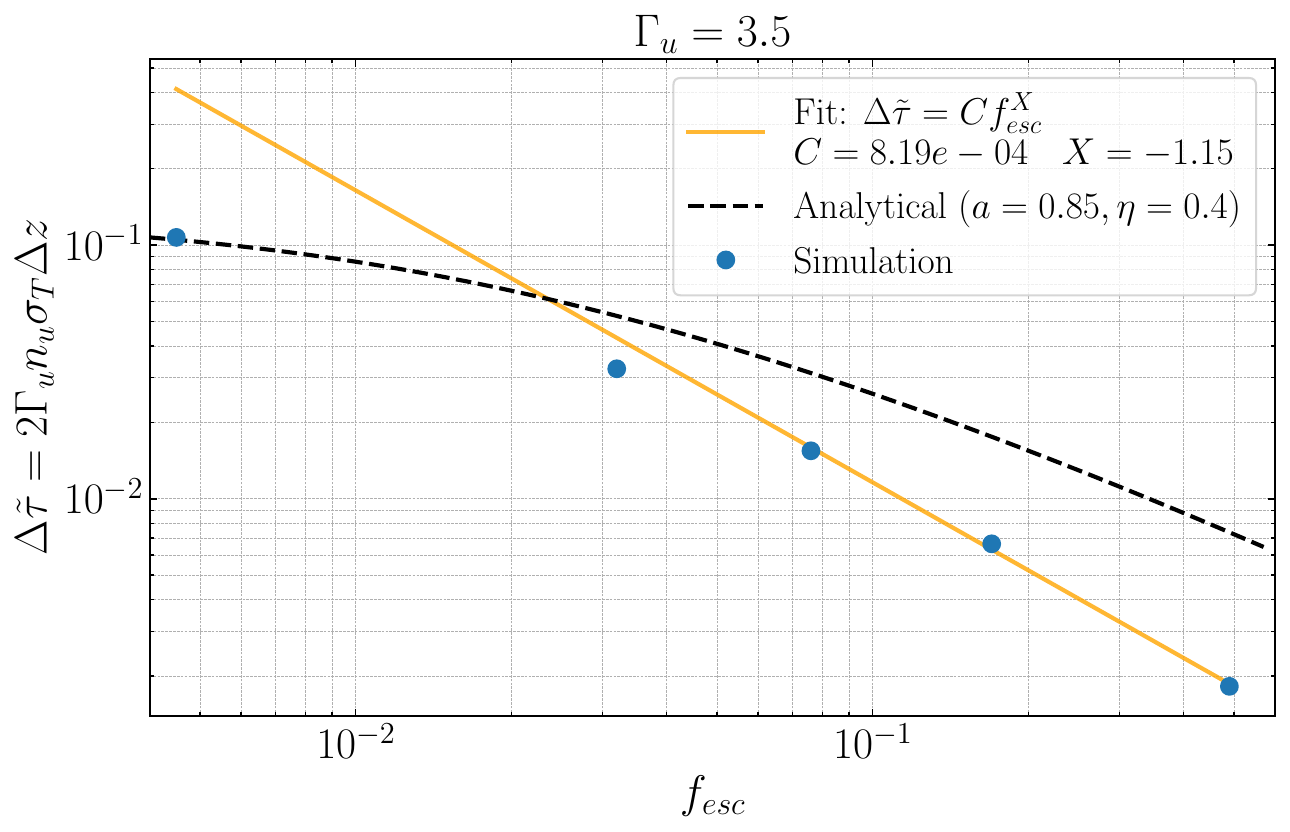}
  \end{minipage}

  \par\medskip %

   \begin{minipage}[t]{\panelw}\centering
    \includegraphics[width=\linewidth,keepaspectratio]{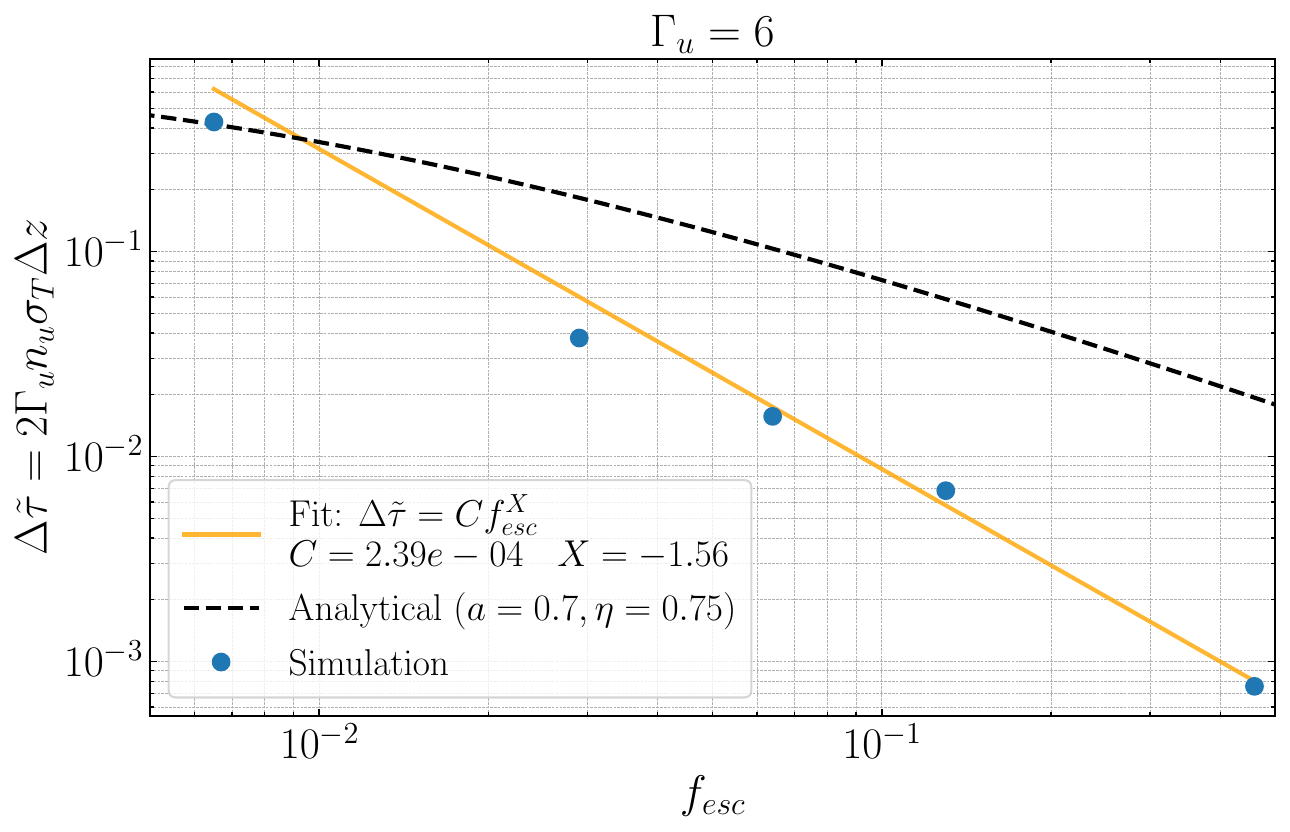}
  \end{minipage}\hspace{\columnsep}%
  \begin{minipage}[t]{\panelw}\centering
    \includegraphics[width=\linewidth,keepaspectratio]{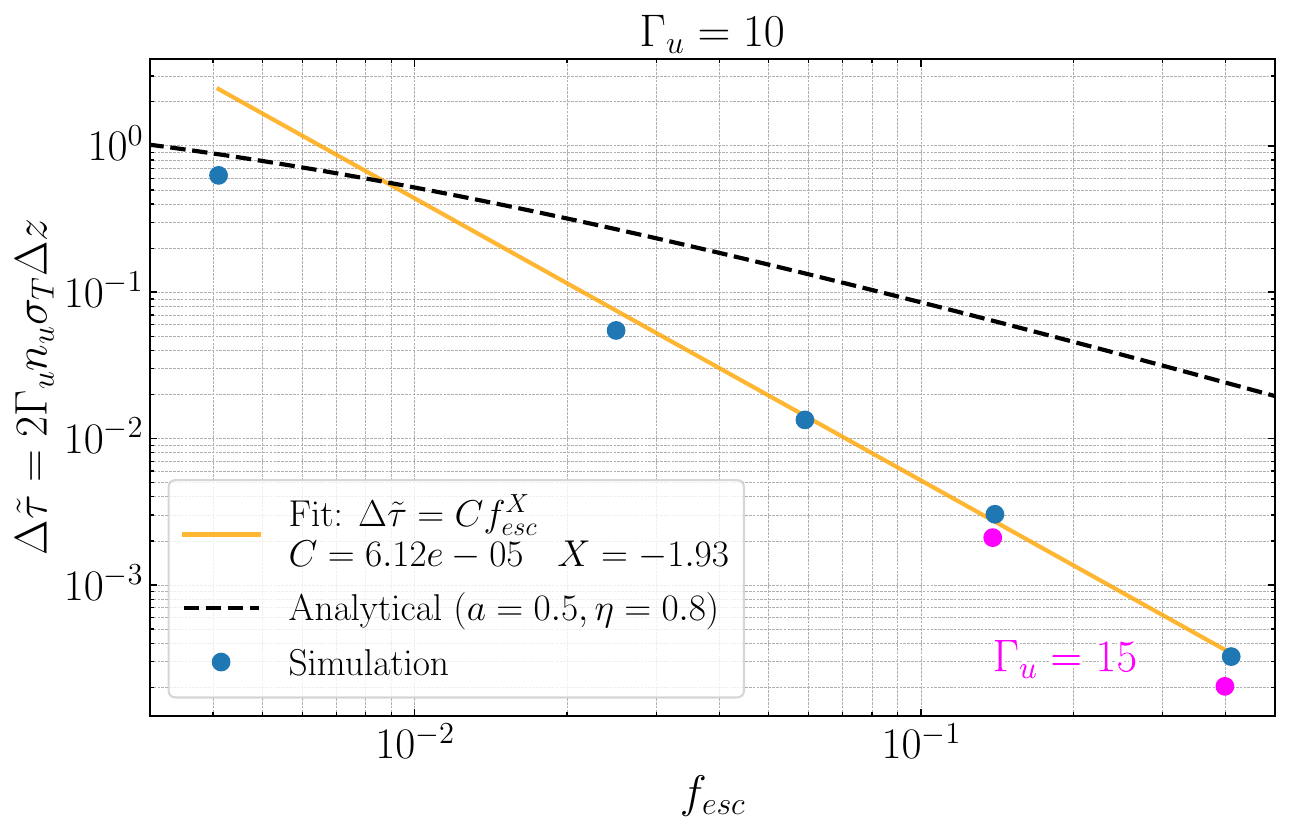}
  \end{minipage}

  \caption{ 
    Comparison of the shock width, $\Delta \tilde{\tau}$, obtained from the simulation ({\it blue circles}) and the analytical model ({\it black dashed lines}) as a function of $f_{esc}$ for $\Gamma_u = 2$ (top left),  $3.5$ (top right), $6$ (bottom left), and $\Gamma_u = 10$ (bottom right).
    The shock width, $\Delta \tilde{\tau}$, is defined as the optical depth measured from the subshock to the upstream position where the flow's four-velocity reaches 90\% of its far-upstream value ($\Gamma \beta = 0.9 \Gamma_u \beta_u$).
For the analytical model, the parameter pair $(a,\eta)$ is fixed to the values that best fit the lowest-$f_{esc}$ simulation in each panel.
Orange solid lines (shown for $\Gamma_u=3.5, 6,$ and $10$) indicate the best-fit power law $\Delta \tilde{\tau} = C f_{\rm esc}^{X}$ to the simulation data in the corresponding panel, obtained by excluding the single point with $f_{esc}<0.05$.
Magenta circles in the bottom right panel denote reference points from the $\Gamma_u=15$ simulations (shown for comparison only and excluded from the fits).
  }
  \label{fig:comparison2}
\end{figure*}

In Figure \ref{fig:comparison2}, we present the deceleration length of finite shocks, $\Delta \tilde{\tau}$, as a function of $f_{esc}$, compared with predictions from the analytical model. 
For the free parameters of the analytical model, we adopt the pair of $a$ and $\eta$ that best fit the $\Gamma - \tilde{\tau}$ profile of the numerical results for the lowest $f_{esc}$ cases (profiles shown by green lines in Figure \ref{fig:comparison}). 
As a reference, we also show two magenta points from simulations with $\Gamma_u=15$ in the $\Gamma_u=10$ panel.
It is worth noting that although the analytical model considered in this study is modified from the original by shifting the reference position of $\tilde{\tau} = 0$ to the location where $\Gamma = \Gamma_{u, sub}$---rather than $\Gamma = 1$, as in the original analytical solution---this adjustment, while shortening the shock width by removing part of the upstream region, has a negligible impact on its overall value. This is because the shock width is primarily determined by the far upstream region, where $\Gamma \sim \Gamma_u$, as evident from Figure \ref{fig:comparison}.


Since the shock profile is well reproduced by the analytical model, the dependence of the shock width on $f_{esc}$ observed in the simulation is also well captured by the analytical model for $\Gamma_u = 2$.
On the other hand, reflecting the breakdown of the analytical model at high escape fractions, the $\tilde{\tau} - f_{esc}$ dependence observed in the simulation cannot be reproduced by the analytical model for $\Gamma_u \geq 3.5$. 
Note that varying the free parameters does not alleviate this discrepancy, as evidenced by the fact that large values of $f_{esc}$ cannot be described within the framework of the analytical model, as mentioned above. 
At large escape fractions ($f_{esc} \gg \Gamma_u^{-2}$), the analytical model predicts a shock width approximately given by  
$
\Delta \tilde{\tau} \sim \frac{m_e}{m_p} \frac{\Gamma_u}{f_{esc}} = 5.4 \times 10^{-3} \Gamma_{u, 1} f_{esc}^{-1}.
$
In contrast, power-law fits of the form $\Delta \tilde{\tau} \propto f_{esc}^{X}$ at fixed $\Gamma_u$ yield $X\simeq -1.15$, $-1.56$, and $-1.93$ for $\Gamma_u=3.5$, $6$, and $10$, respectively, when using the three points with $f_{esc}>0.05$; for $\Gamma_u=15$, the two available points with $f_{esc}>0.1$ give $X\simeq -2.20$. This indicates a systematically steeper dependence on $f_{esc}$, with the power-law index becoming more negative as $\Gamma_u$ increases (i.e., $|X|$ grows).
Thus, the simulation exhibits a much narrower shock width at large $f_{esc}$, with the gap between the simulation and the analytical model exceeding an order of magnitude at the highest escape fractions.

Another notable discrepancy between the simulation and the analytical model regarding shock width lies in its dependence on the shock Lorentz factor. As mentioned above, the analytical model predicts a scaling of $\Delta \tilde{\tau} \propto \Gamma_u$ for $f_{esc} \gg \Gamma_u^{-2}$. 
Qualitatively, the tendency for larger $\Gamma_u$ to result in a wider shock can be understood as a consequence of the Klein-Nishina suppression of the cross section, which becomes more significant for faster shocks. 
Indeed, for infinite shocks, a clear trend of increasing shock width with Lorentz factor, $\Delta \tilde{\tau} \propto \Gamma_u^3$, is found both in simulations (ILN20a) and in the analytical study (GNL18). While this scaling breaks down in the mildly relativistic regime, the simulations (ILN20a) still indicate a positive correlation between $\Delta \tilde{\tau}$ and $\Gamma_u$, at least down to $\Gamma_u = 2$.
In contrast, although the analytical model also predicts a positive dependence of shock width on Lorentz factor for finite shocks---specifically, $\Delta \tilde{\tau} \propto \Gamma_u$ when $f_{esc} \gg \Gamma_u^{-2}$---our current simulations of finite shocks ($f_{esc} > 0$) exhibit the opposite trend: the shock width decreases with increasing $\Gamma_u$.\footnote{
Since the analytical model reproduces the numerical results well at $\Gamma_u=2$, a positive dependence of the shock width on Lorentz factor ($\Delta\tilde{\tau}\propto\Gamma_u$) may still hold within a quite narrow Lorentz-factor range around $\Gamma_u\simeq2$ (for $\Gamma_u<3.5$).
}
For example, for the simulations with highest $f_{esc} (= 0.35 - 49)$, the shock width is found to be approximately $\Delta \tilde{\tau} \sim 6.7 \times 10^{-3}$, $1.8\times 10^{-3}$, $7.6 \times 10^{-4}$, $3.2 \times 10^{-4}$,  and $1.6 \times 10^{-4}$ for $\Gamma_u = 2$, $3.5$, $6$, $10$, and $15$ respectively, indicating an inverse correlation with the Lorentz factor.  
Thus, an important implication of our simulation results is that higher Lorentz factor shocks propagating in a wind-like environment tend to break out at larger radii (i.e., at smaller optical depths), contrary to previous expectation by GNL18 which finds opposite trend.
We will revisit this point in more detail in Section~\ref{sec:SHbreakoutModel}.
This discrepancy can again be understood as a consequence of the breakdown of the analytical model discussed above. Since higher Lorentz factors lead to the emergence of more prominent subshock-generated high-energy photons, deceleration via $\gamma - \gamma$ attenuation between counterstreaming photons becomes more efficient, resulting in faster deceleration.

\section{Implications for the properties of relativistic shock breakout from a stellar wind}
\label{sec:Breakout}

The discrepancy between our simulation results and the analytical model of GNL18 has important implications for shock breakout predictions based on that model.
As noted earlier, the analytical model predicts the shock width to scale approximately as $\Delta \tilde{\tau} \sim (m_e / m_p) \Gamma_u f_{esc}^{-1}$.
On the other hand, our simulations show that, for shocks with $\Gamma_u \gtrsim 3.5$, the optical depth $\Delta \tilde{\tau}$ departs from the analytical model.
Focusing on the high-Lorentz-factor shocks ($\Gamma_u=6$, $10$, and $15$), where this departure is most evident, and restricting to the regime of appreciable escape ($f_{esc}\gtrsim 0.1$), we fit the simulation data with  an analytical function of the form
$\Delta \tilde{\tau}=C\,\Gamma_u^a f_{\rm esc}^b$ and obtain the following rough best-fit scaling:
\begin{eqnarray}
\label{eq:SHwidth}
\Delta \tilde{\tau} \approx 4.0\times10^{-3}\Gamma_u^{-1.8} f_{\rm esc}^{-1.9} \quad (6 \lesssim \Gamma_u \lesssim 15).
\end{eqnarray}
We do not include our $\Gamma_u=3.5$ simulations in this fit, because this Lorentz factor lies in a transitional regime where systematic deviations from the analytical model begin to emerge. Nevertheless, it is worth noting that the above scaling remains consistent with the $\Gamma_u=3.5$, $f_{esc}=0.49$ point.
As described earlier, this scaling indicates a narrower shock width for a given $\Gamma_u$ and $f_{esc}$ than predicted by the analytical model, at least in the relativistic regime.

Assuming that this revised scaling remains valid up to $f_{\rm esc} \approx 1$, the optical depth at which breakout occurs can be estimated as
\begin{eqnarray}
\label{eq:taubo}
  \tau_{bo} \approx 4.0\times10^{-3} \Gamma_{bo}^{-1.8} \quad (6 \lesssim \Gamma_{bo} \lesssim 15).
\end{eqnarray}
where $\Gamma_{\rm bo}$ is the Lorentz factor at breakout.
This is in stark contrast with the analytical prediction of $\tau_{ bo, {\rm GNL18}} \approx (m_e / m_p) \Gamma_{\rm bo}$, suggesting that our numerical results predict a much smaller breakout optical depth.
The ratio between the two can be expressed as
$\tau_{\rm bo}/\tau_{ bo, {\rm GNL18}} = 1.2 \times 10^{-2}\Gamma_{ bo,1}^{-2.8}$, where $\Gamma_{ bo,1} = \Gamma_{ bo}/10$, indicating that breakout occurs at a much larger radius in our model than in the analytical estimate for the same $\Gamma_{ bo}$.
In the following sections (\S\ref{sec:Closure_relation} and \S\ref{sec:SHbreakoutModel}), we revisit the breakout emission analysis originally presented in GNL18, adopting the revised shock width scaling up to $f_{esc} \approx 1$ for relativistic shocks ($\Gamma_u \gtrsim 6$), and highlight how the resulting observables differ from those predicted by the analytical model.

\subsection{Closure relation}
\label{sec:Closure_relation}

As discussed in previous works, three observables: the breakout duration, $t_{bo}$, the observed temperature, $T_{bo}$, and the total emitted energy, $E_{bo}$, are expected to satisfy a closure relation, as they are determined by two physical quantities: the breakout Lorentz factor, $\Gamma_{bo}$, and the breakout radius, $r_{bo}$.
Since the emitted spectrum peaks around $E_p \sim m_e c^2$ in the shock frame, the observed temperature is estimated as
\begin{eqnarray}
  \label{eq:Tobs}
  k T_{obs} = \frac{\Gamma_{bo} m_e c^2}{3} \approx 1.7 \Gamma_{bo, 1}~{\rm MeV}.
\end{eqnarray}
The breakout duration is given by
\begin{eqnarray}
  \label{eq:tbo}
  t_{bo} = \frac{r_{bo}}{2\Gamma_{bo}^2 c} \approx 1.7 r_{bo, 13} \Gamma_{bo,1}^{-2}~{\rm s},
\end{eqnarray}
where $r_{bo, 13} = r_{bo} / 10^{13}~{\rm cm}$.

While the expressions for $T_{obs}$ and $t_{bo}$ as functions of $\Gamma_{bo}$ and $r_{bo}$ given in the above equations remain unchanged from the previous study, the expression for the breakout energy is modified due to the updated breakout optical depth, $\tau_{bo}$.
The breakout energy can be estimated from the swept-up wind mass, $M_{bo}$, as $E_{bo} \approx \Gamma_{bo}^2 M_{bo} c^2$. Assuming a wind-like density profile, $\rho \propto r^{-2}$, the swept-up mass is given by $M_{bo} = 4\pi \tau_{bo} r_{bo}^2 / \kappa$, where $\kappa$ is the opacity.
Hence,  the breakout energy is estimated as
$E_{bo} \approx 4\pi r_{bo}^2 \Gamma_{bo}^3 c^2 m_e / (m_p \kappa)$ in GNL18.
In contrast, adopting the revised breakout optical depth from Equation~(\ref{eq:taubo}), we obtain the breakout energy as
\begin{eqnarray}
  \label{eq:Ebo0}
  E_{bo} = \frac{4\pi \Gamma_{bo}^2 \tau_{bo} r_{bo}^2 c^2}{\kappa} \approx 3.6 \times 10^{46} r_{bo, 13}^2 \Gamma_{bo,1}^{0.2} \kappa_{0.2}^{-1} ~{\rm erg},
\end{eqnarray}
where $\kappa_{0.2} = \kappa / (0.2~{\rm cm}^2{g}^{-1})$. 
Notably, unlike the earlier prediction $E_{bo} \propto \Gamma_{bo}^3$, our result shows very weak dependence on $\Gamma_{bo}$, reflecting the revised scaling $\tau_{bo} \propto \Gamma_{bo}^{-1.8}$.

From Equations (\ref{eq:Tobs}), (\ref{eq:tbo}), and (\ref{eq:Ebo0}) we derive a modified closure relation:
\begin{eqnarray}
  \label{eq:closure}
    E_{bo} \approx 3.6\times10^{46} \kappa_{0.2}^{-1} \left(\frac{t_{bo}}{1.7\, \mathrm{s}}\right)^2 \left(\frac{kT_{obs}}{1.7\,\mathrm{MeV}}\right)^{4.2}\quad\text{erg}.
\end{eqnarray}
This differs from the previous prediction of $E_{bo} \propto T_{obs}^7$, reducing the temperature dependence to $E_{bo} \propto T_{obs}^{4.2}$.
 However, an important caveat of this rough scaling relation is that it has only been examined over a limited range of Lorentz factor,
$6 \lesssim \Gamma_{\rm bo} \lesssim 15$. 
Extending the analysis to a broader range of $\Gamma_u$ is necessary to determine whether this scaling remains valid beyond this regime.
On the other hand, for mildly relativistic shocks with Lorentz factors close to $\Gamma_{bo} \sim 2$, the agreement between the analytical model and numerical results at $\Gamma_u \sim 2$ suggests that the original closure relation recover its validity in a limited range, $2 \lesssim \Gamma_{bo} < 3.5$. Again, further analysis is required to precisely determine the range of validity, which is beyond the scope of this study.

\subsection{Properties wind shock breakout based on a spherical explosion model}
\label{sec:SHbreakoutModel}

Here, focusing on the relativistic breakout regime ($\Gamma_{bo} \gtrsim 6$), we derive the properties of a shock breakout from a stellar wind based on a spherical explosion model,\footnote{Although a higly relativistic shock breakout is unlikely to be realized in a spherical explosion due to the large energies required, the spherical model remains applicable as long as the flow properties remain uniform within an angular scale of approximately $1/\Gamma$.} assuming that the shock-width scaling relation given in Equation (\ref{eq:SHwidth}) holds for highly relativistic shocks.

Following the prescription provided by GNL18, we consider a shock breaking out of a star characterized by a density profile of the form $\rho \propto (R_* - r)^n$, where $R_*$ is the stellar radius and $r$ is the distance from the stellar center. We further assume that the star is surrounded by an extended, wind-like medium with a density profile given by $\rho \propto r^{-2}$. For the case of $n = 3$, the ejecta, after breaking out from the stellar surface and subsequently accelerating, exhibit a mass profile described by $m_{}(\Gamma) \propto \Gamma^{-2.1}$, where $m_{}(\Gamma)$ denotes the mass of the ejecta with Lorentz factors greater than $\Gamma$ \citep{Johnson_Mckee1971, Pan_Sari2006}. Correspondingly, the energy of ejecta with Lorentz factors greater than $\Gamma$ can be expressed as $E(>\Gamma) \approx E_0 (\Gamma / \Gamma_0)^{-1.1}$, where $\Gamma_0$ and $E_0$ represent the Lorentz factor and the energy, respectively, of the breakout layer—defined as the leading edge of the shocked stellar matter that has a pair-unloaded optical depth of $\delta \tilde{\tau} \simeq 1$.
Following \citet{Nakar_Sari2012}, the Lorentz factor of this breakout layer is estimated as:
$\Gamma_0 \approx 45 E_{53}^{1.7} M_{ej,5}^{-1.2} R_{*,11}^{-0.95}$,
where $E_{ex} = 10^{53} E_{53}~ \mathrm{erg}$ is the explosion energy, $M_{ej} = 5 M_{ej,5} M_{\odot}$ is the stellar ejecta mass, and $R_{*} = 10^{11} R_{*,11}~ \mathrm{cm}$ is the stellar radius. The corresponding energy is given by:
$E_0 = 4\pi R_*^2 \Gamma_0 c^2 \kappa^{-1}$.
Here, we assume a constant opacity $\kappa$ for both the stellar and wind regions.

The stratified relativistic ejecta impinge upon the wind-like circumstellar medium, producing a shock that decelerates as it propagates outward. During propagation, energy is progressively injected into the shocked wind material because inner, more energetic but slower ejecta components eventually catch up with the decelerated forward shock region, thereby injecting additional energy into it.
The Lorentz factor of the forward shock, $\Gamma_{sh}$, at a given radius $r_{sh}$ can be roughly estimated by equating the energy of the shocked wind material with the energy of the ejecta that have caught up with the forward shock ($\Gamma \geq \Gamma_{sh}$):
$\Gamma_{sh}^2 m(r_{sh}) = E_0 ({\Gamma_{sh}}/{\Gamma_0})^{-1.1}$,
where $m(r_{sh}) \simeq 4\pi \tau_{w*} R_* \kappa^{-1} r_{sh}$ is the mass of the wind material swept up by the shock, and $\tau_{w*}$ represents the optical depth of the wind region from the stellar surface to infinity.

Next, we convert the shock radius to the optical depth to infinity, $\tau_w$, as:
$r_{sh} \simeq \tau_{w*} R_*/\tau_w$.
By equating this to the optical depth of the shock given in Equation (\ref{eq:SHwidth}), i.e., $\tau_w = \Delta \tilde{\tau}$, we obtain the shock Lorentz factor and radius as functions of the escape fraction $f_{esc}$:
\begin{eqnarray}
\label{eq:SHLor}
\Gamma_{sh} \approx 1.7 E_{53}^{0.73} M_{ej,5}^{-0.51} R_{*,11}^{-0.41} \tau_{w*}^{-0.41} f_{esc}^{-0.39} ,
\end{eqnarray}
\begin{eqnarray}
\label{eq:SHr}
r_{sh} \approx 6.2 \times 10^{13} E_{53}^{1.3} M_{ej,5}^{-0.93} R_{*,11}^{0.27} \tau_{w*}^{0.27} f_{esc}^{1.2} \quad \text{cm} .
\end{eqnarray}
Note that, since a small pair-unloaded optical depth ($\Delta \tilde{\tau} \ll 1$) is required to maintain the shock as radiation-mediated in highly relativistic shocks, an optical depth of $\tau_{w*} = 1$ is sufficient for the breakout to occur within the wind region.

From the above equations, a characteristic difference from the breakout estimates of GNL18 can be readily identified. Namely, in the present model, explosion parameters ($E_{\rm ex}$ and $M_{\rm ej}$) that lead to a larger breakout Lorentz factor $\Gamma_{\rm bo}$ (i.e., $\Gamma_{\rm sh}$ at $f_{\rm esc} \approx 1$) also result in a larger breakout radius $r_{\rm bo}$ (i.e., $r_{\rm sh}$ at $f_{\rm esc} \approx 1$).
This stands in stark contrast to the trend found in GNL18, where larger $\Gamma_{\rm bo}$---arising from higher explosion energy or lower ejecta mass---corresponds instead to a smaller breakout radius $r_{\rm bo}$.

The observables of the full breakout emission ($f_{esc} \approx 1$) can be estimated by substituting $f_{esc} = 1$ into Equations (\ref{eq:SHLor}) and (\ref{eq:SHr}). The total emitted energy $\approx E_{sh}$ at $f_{esc} \approx 1$) and the observed duration ($\approx r_{bo} / 2\Gamma_{bo}^{2} c$) are given by:
\begin{eqnarray}
\label{eq:Ebo}
E_{bo} \approx 9.6 \times 10^{47} \kappa_{0.2}^{-1} E_{53}^{2.8} M_{\rm ej,5}^{-2} R_{*,11}^{0.45} \tau_{w*}^{0.45}
 \quad \text{erg},
\end{eqnarray}
\begin{eqnarray}
\label{eq:Eto}
t_{bo} \approx 3.8 \times 10^{2} E_{53}^{-0.15} M_{\rm ej,5}^{0.1} R_{*,11}^{1.1} \tau_{w*}^{1.1}
\quad \text{s}.
\end{eqnarray}
As in Equation (\ref{eq:Tobs}), the observed temperature is estimated by assuming a comoving temperature of $m_e c^2 / 3$. The resulting observed temperature ($\approx \Gamma_{bo} m_e c^2/3$) is:
\begin{eqnarray}
\label{eq:Tobs2}
k T_{obs} \approx 280 E_{53}^{0.73} M_{ej,5}^{-0.51} R_{*,11}^{-0.41} \tau_{w*}^{-0.41}~{\rm keV}.
\end{eqnarray}
A particularly notable result is that the breakout duration, $t_{bo}$, is quite insensitive to the explosion parameters and instead predominantly determined by the properties of the ambient medium---specifically, the stellar radius $R_*$ and the wind optical depth $\tau_{w*}$.
This arises from the fact that, as mentioned above, a larger breakout Lorentz factor $\Gamma_{bo}$ also corresponds to a larger breakout radius $r_{bo}$, with the $r_{bo} \propto \Gamma_{bo}^{1.8}$ holding for fixed values of $R_*$ and $\tau_{w*}$.
This stands in contrast to the prediction by GNL18, in which the breakout duration exhibits strong sensitivity to explosion parameters:
$t_{bo} \approx 1.6 \times 10^{3} E_{53}^{-5.1} M_{\rm ej,5}^{3.6} R_{*,11}^{3.85} \tau_{w*}^{3.86}~\text{s}$.

To clearly demonstrate the difference between the previous analytical model and our current results at $\Gamma_{bo} \gtrsim 6$, let us consider a case where $E_{53} = 5$, with all other parameters fixed to their fiducial values ($R_{*,11} = M_{ej,5} = \tau_{w*} = 1$).
In this case, our model yields $\Gamma_{bo} \approx 5.5$ ($kT_{obs} \approx 0.9~{\rm MeV}$), $E_{bo} \approx 9 \times 10^{49}{\rm erg}$, and $t_{bo} \approx 300{\rm s}$.
In contrast, the analytical model predicts $\Gamma_{bo} \approx 19$ ($kT_{obs} \approx 3.3~{\rm MeV}$), $E_{bo} \approx 2 \times 10^{49}{\rm erg}$, and $t_{bo} \approx 0.42{\rm s}$.
These results indicate that for an explosion energetic enough to drive a highly relativistic shock in an extended wind region, the breakout duration is significantly longer in our model compared to previous estimates, while the total emitted energy and observed temperature remain of the same order.

We note that realizing a relativistic breakout by reducing $M_{\rm ej}$, instead of increasing $E_{ex}$, yields the same outcome: while our model predicts a stable breakout duration, the analytical model suggests a much shorter one.
It is also worth noting that the longer breakout duration predicted by the current study is favorable for observational prospects, enhancing detectability and enabling time-resolved spectral analysis.
However, our simulations currently cover only a limited portion of the parameter space, specifically $\Gamma_u = 6$, $10$, and $15$ and a finite range of $f_{esc}$.
To rigorously assess the robustness of the breakout estimates and validate the broader applicability of the scaling relation presented in Equation (\ref{eq:SHwidth}), a more systematic exploration across a wider range of $\Gamma_u$ and $f_{esc}$ is necessary.

During the breakout ($t_{obs} \lesssim t_{bo}$), the bolometric luminosity is expected to decline gradually with time:
\begin{eqnarray}
L_{bol} \approx \frac{f_{esc} E_{sh}}{t_{obs}} \propto t_{obs}^{-0.28}, 
\end{eqnarray}
with temperature also exhibiting a gradual decline with time:
\begin{eqnarray}
   T_{obs} \approx \frac{\Gamma_{sh} m_e c^2}{3} \propto t_{obs}^{-0.2} .
\end{eqnarray}
Hence, around the spectral peak energy $E_p \approx 3 k T_{obs}$, the luminosity is expected to show a very modest change. However, based on the spectral shape dependence on $f_{esc}$ found in the simulation (Figure \ref{fig:nufnu}), we expect to see a rapid increase in flux at energies well below the peak energy ($h\nu \ll E_p$), as well as an even more pronounced and sharp increase in flux above the peak energy ($h\nu > E_p$).

Regarding breakouts occurring at lower velocities, specifically in the mildly relativistic regime ($2 \lesssim \Gamma_{bo} < 6$), 
the dependence of the shock width on $\Gamma_u$ remains unclear and likely exhibits non-monotonic behavior.
This is primarily because the dominant deceleration mechanism changes within this range, as indicated by our intermediate $\Gamma_u=3.5$ simulations, which exhibit transitional properties.
  In parallel, the scaling with $f_{esc}$ also varies significantly—shifting from $\Delta \tilde{\tau} \propto f_{esc}^{-1.9}$ at $\Gamma_u \gtrsim 6$ to $\Delta \tilde{\tau} \propto f_{esc}^{-1}$ at $\Gamma_u \approx 2 - 3.5$, at least for $f_{esc} \gtrsim 0.1$.
  Consequently, it is challenging to derive consistent breakout estimates in this regime based on a single scaling relation, as was done above.
Nevertheless, our results suggest that the shock width tends to increase with decreasing $\Gamma_u$, at least down to some Lorentz factor between $\Gamma_u = 3.5$ and $2$ for finite shocks.
This again implies a trend in which explosions producing lower $\Gamma_{bo}$ tend to break out at smaller radii. However, this trend may reverse at some point between $\Gamma_{bo} = 3.5$ and $2$, given that the analytical model reproduces the shock structure at $\Gamma_u = 2$ quite well. 
A detailed and systematic study covering this intermediate $\Gamma_u$ range is therefore needed to identify the transition point, which is beyond the scope of the current work.

For breakouts occurring at $\Gamma_{bo} \approx 1$, as discussed in GNL18, the breakout likely takes place at sub-relativistic velocities ($\beta < 0.5$), where the pair opacity sharply drops. For a rough estimate in this regime, we refer the reader to GNL18.

\section{ {Limitations of the present analysis}}
\label{sec:limitations}
Despite the valuable insights provided by our simulations, several limitations should be acknowledged. Here we briefly summarize these limitations.
\subsection{On the assumption of steady-state and planar geometry}

The key assumptions in the present study are that the shock structure can be described by a steady-state and planar geometry.
For the two approximations to hold, an essentially identical requirement must be satisfied: the shock-crossing time, $t_{\rm cross}\sim \Gamma_u \Delta z/c$---the timescale for the shock to sweep up the mass contained in the shock transition layer of width $\Delta z$ measured in the shock rest frame—--must be much shorter than the dynamical timescale of the shock, which, in a wind-like environment, is given by $t_{\rm dyn}\sim r/c$; hence $t_{\rm cross}\ll t_{\rm dyn}$. While this condition is met well before the breakout,  its validity becomes marginal during the breakout phase, as breakout in a wind-like environment leads to the two timescale to be comparable $t_{\rm cross}\sim t_{\rm dyn}$.

In this regard, recent work by \citet{Wasserman2025}, based on dynamical simulations of sub-relativistic breakouts in a wind, argues that, contrary to the assumptions adopted in earlier studies \citep[][ILN20b]{Ioka2019}, the steady-state and planar approximations are likely to break down during breakout.
We note, however, that relativistic and sub-relativistic breakouts have markedly different natures. In sub-relativistic RMS breakouts, once breakout begins it is essentially completed within a single dynamical time, i.e., $f_{\rm esc}$ evolves from 0 to 1 over $\sim t_{\rm dyn}$. By contrast, as discussed in GNL18, relativistic breakout proceeds much more gradually, spanning many dynamical times during the evolution from $f_{\rm esc}=0$ to 1. 
This difference arises because, in sub-relativistic RMSs, the shock optical depth cannot fall significantly below the diffusion length, i.e., $\Delta \tilde{\tau} \sim \beta_u^{-1}$, thereby limiting the breakout duration to $\sim t_{\rm dyn}$. In relativistic RMSs, however, self-generated  opacity via pair production, allows the shock to propagate over several orders of magnitude in radius during breakout, as demontsrated in the present study.

More concretely, the scaling in Equation~(\ref{eq:SHwidth}) indicates that, $f_{esc}\gtrsim 0.1$, the pair-unloaded optical depth of the shock scales as $\Delta \tilde{\tau} \propto f_{esc}^{-1.9}$. Combined with the optical depth scaling in a wind, $\tilde{\tau} \propto r^{-1}$, this implies $f_{esc} \propto r^{1/1.9}$. As a result, $f_{esc}$ increases by only a factor of $\approx 2^{1/1.9}\sim 1.4$ within one dynamical time, leading to a gradual breakout in which $f_{esc}$ can be treated as an adiabatic parameter.
 Thus, although the steady-state and planar assumptions are marginal in both sub-relativistic and relativistic breakouts, the two cases should be clearly distinguished. Accordingly, the conclusion of \citet{Wasserman2025} does not directly carry over to the relativistic case, and the much more gradual evolution makes the steady-state approximation at least more justifiable. 
Nevertheless, validating these assumptions ultimately requires dynamical simulations of relativistic shock breakouts—an important but challenging direction for future research.

\subsection{On the effect of particle acceleration at the subshock}
\label{sec:paracc}

As demonstrated by our simulations, breakout from a wind is likely to be accompanied by a collisionless subshock whose strength increases as the escape fraction $f_{esc}$ rises. Because particle acceleration is expected to occur at the subshock, it may influence the high-energy spectrum; in particular, accelerated pairs could give rise to a non-thermal power-law extension in the emergent emission.
However, if the energy carried by non-thermal pairs constitutes only a small fraction of that of the thermally heated pairs, their effect on the breakout spectra presented here is expected to be modest. This expectation arises because (i) the lower energy content of the non-thermal pairs implies a correspondingly smaller contribution of Comptonized photons than that produced by thermal pairs, and (ii) the higher-energy photons generated are subject to stronger $\gamma-\gamma$ opacity owing to the increased number density of target photons above the pair-creation threshold, thereby further suppressing the escaping flux. As a result, although non-thermal pairs may extend the spectrum toward higher energies, the associated flux is expected to remain relatively small. Moreover, it is worth noting that, owing to the small energy content in the counterstreaming non-thermal component, Comptonization by accelerated pairs does not affect the deceleration dynamics of the shock either. 
 A quantitative prediction of the exact spectral shape would require a dedicated particle-acceleration model, which is beyond the scope of this work. At full breakout ($f_{\rm esc}=1$), when the entire shock has fully transitioned into a collisionless shock, non-thermal effects may become more relevant, but addressing this lies outside the scope of the present study. We defer a detailed investigation of this issue to future work.

\subsection{On the decoupling among plasma particles}

While we have adopted a single-fluid treatment of pairs and protons in the present study, the validity of this approximation remains uncertain.
 Specifically, as demonstrated by \citet{Levinson2020}, strong pair enrichment in RRMSs can lead to a decoupling between protons, electrons, and positrons, since the electric fields generated by small charge separations are no longer sufficient to maintain tight coupling. 
This decoupling may induce plasma instabilities and lead to microturbulence, as shown in \citet{Vanthieghem2022}. Such plasma effects, if present, could modify the escaping spectra through inverse Compton scattering by high-energy pairs produced by the plasma effects and may also alter the deceleration process by changing the efficiency of $\gamma$-$\gamma$ pair production within the shock.
Conversely, \citet{Mahlmann2023} showed that even a modest magnetic field, neglected here, may suffice to maintain tight coupling between species, potentially suppressing these instabilities. These issues warrant further investigation in future work.

\section{Summary and conlusions}\label{sec:summary}
In this study, we performed Monte Carlo simulations of photon starved relativistic RRMS, incorporating the effect of photon escape from the upstream region under the assumption of steady-state flow in a planar geometry. These simulations are applicable to RRMS propagating through shallowly decaying density profiles, such as those found in stellar winds. 
The calculations were carried out for shock Lorentz factors of $\Gamma_u = 2$, $3.5$, $6$,  $10$, and $15$ marking the
 first simulation to model relativitic shock breakout 
  emission from the first principles in the photon starved regime.\footnote{\citet{Lundman2021} conducted a first principles calculation of relativistic shock breakout in the “photon rich” regime, where photon production and pair production can be neglected.}
We explored a wide range of energy fractions carried by escaping photons, $f_{esc}$, and analyzed the resulting shock structures and the spectral properties of the escaping radiation.
We also conducted a detailed comparison between our numerical results and the shock structures predicted by an analytical model (GNL18), identifying deviations that suggest the need to reconsider the predicted observables of relativistic shock breakouts in a wind environment.

Our main findings are summarized below:

\begin{enumerate}

  \item  As found in the case of an infinite shock explored in ILN20a, vigorous pair production also acts as a thermostat in finite shocks, stabilizing the immediate downstream temperature at $\sim 100 - 200~{\rm keV}$, regardless of the escape fraction, $f_{esc}$. 
  In finite RRMS, its ability to self-generate opacity through pair production allows the pair-unloaded optical depth to decrease by several orders of magnitude as $f_{esc}$ increases.
  As indicated in the analytical model of GNL18,
  we find that a small escape fraction, $f_{esc} \sim 1/\Gamma_u^2$, is sufficient to significantly reduce the shock width. 
  As a result, in line with the prediction of the  analytical model, our reult suggets RRMS breakout in the wind persists over several decades in radius, in contrast to Newtonian RMS breakouts.

  \item  As found in infinite shocks (ILN20a), the presence of a subshock is a inherent feature of relativistic shocks, emerging near the end of the deceleration region. While the subshock is weak for $f_{\rm esc} = 0$, its strength increases with $f_{\rm esc}$, leading to more pronounced discontinuities in velocity and temperature. The post-subshock velocity asymptotically approaches that of the far downstream, whereas the elevated post-subshock temperature rapidly cools to the pair-regulated value of $\sim 100$-$200~{\rm keV}$, producing a distinct temperature spike.

  \item  While our results confirm the overall trends predicted by the analytical model, they also reveal systematic deviations. The model successfully reproduces the structure of infinite shocks for all tested Lorentz factors ($\Gamma_u = 2$, $3.5$, $6$, and $10$), and captures finite shock structures well for $\Gamma_u = 2$ across a wide range of $f_{\rm esc}$.
  However, for higher Lorentz factors ($\Gamma_u = 3.5$, $6$, $10$, and $15$), our simulations yield systematically narrower shock widths than predicted, particularly at large $f_{\rm esc}$. 
    Particularly, we find a scaling of $\Delta \tilde{\tau} \propto f_{\rm esc}^{-1.9}$, compared to the analytical prediction of $\Delta \tilde{\tau} \propto f_{\rm esc}^{-1}$.
  Furthermore, while the analytical model predicts a positive scaling with Lorentz factor ($\Delta \tilde{\tau} \propto \Gamma_u$), our simulations show an inverse trend, $\Delta \tilde{\tau} \propto \Gamma_u^{-1.8}$. 
  These discrepancies arise from a shift in the dominant deceleration mechanism at high escape fractions. In this regime, $\gamma$–$\gamma$ pair production between high-energy photons—upscattered by subshock-heated pairs—and the bulk population of counterstreaming photons becomes the main source of flow deceleration, overtaking the interaction between counterstreaming photons and the advected plasma originally assumed in the analytical model.

  \item Due to efficient temperature regulation in the downstream region, the spectra of escaping photons consistently peak at $E_{p} \approx 300$-$600~{\rm keV}$ in the shock frame, largely independent of $f_{\rm esc}$ and $\Gamma_u$.
  The spectral shape deviates significantly from a Wien distribution, exhibiting broadening at both low and high energies.
  At lower energies ($h\nu \ll E_p$), a flat spectral component with $f_\nu \propto \nu^0$ emerges due to free-free emission, and its prominence increases with larger $f_{esc}$.
  At higher energies ($h\nu > E_p$), our simulations reveal a spectral extension, particularly for $\Gamma_u = 6$, $10$, and $15$, originating from inverse Compton scattering by hot pairs in the post-subshock region.
  The strength and extent of this high-energy component increase with both $\Gamma_u$---reflecting higher post-subshock temperatures---and $f_{esc}$---due to reduced optical depth for $\gamma$–$\gamma$ pair production.
  These results indicate that the spectrum undergoes rapid broadening during breakout as $f_{\rm esc}$ increases.

  \item  Based on the deviations from the analytical model, we derive an updated closure relation 
  applicable to relativistic breakouts with $\Gamma_{bo} \gtrsim 6$. We also present revised estimates of breakout observables in wind environments using a spherical explosion model. Notably, we find that the breakout duration becomes largely insensitive to the explosion energy and ejecta mass, depending primarily on the progenitor radius and wind optical depth, following 
  $t_{bo} \sim 3.8 \times 10^{2} E_{53}^{-0.15} M_{\rm ej,5}^{0.1} R_{*,11}^{1.1} \tau_{w*}^{1.1}$. 
  This duration is significantly longer---by orders of magnitude---than that predicted by the analytical model for the same explosion parameters, reflecting the substantially reduced shock width in finite shocks.
  For example, a highly energetic explosion with isotropic equivalent energy $E_{ex} \sim 5 \times 10^{53} {\rm erg}$ from a compact progenitor ($R_* \sim 10^{11} {\rm cm}$) surrounded by a wind of optical depth $\tau_* \sim 1$ can produce a relativistic breakout with $\Gamma_{bo} \sim 6$. The resulting emission peaks at $\sim 3kT_{obs} \sim 3~{\rm MeV}$, with total radiated energy $E_{bo} \sim 10^{50}~{\rm erg}$ and a duration of approximately $300~{\rm s}$. In contrast, the analytical model predicts a sub-second duration for the same parameters.
  Such MeV-range emission may be detectable by current gamma-ray observatories such as Fermi-GBM up to distances of $\sim$100 Mpc. Future missions with enhanced sensitivity in the MeV band, including AMEGO-X, COSI, and e-ASTROGAM, could extend this detection horizon to several hundred Mpc, significantly increasing the prospects for observing relativistic shock breakouts.

\end{enumerate}

With the limitations summarized in Section~\ref{sec:limitations} in mind, the simulations presented here are, to the best of our knowledge, among the most sophisticated to date for RRMS breakout in wind-like environments. Our calculations, however, cover only a limited region of the parameter space in $\Gamma_u$ and $f_{\rm esc}$, leaving some uncertainty in our rough estimates of the breakout emission. Ultimately, fully dynamical simulations will be required to validate and extend these results. In this context, the present work should be regarded as a first step and a useful baseline for interpreting observations and for guiding future time-dependent studies.

\section*{Acknowledgements}
This work was supported by JSPS KAKENHI Grant Number JP24H01826,
JP23K25874, and JP25H00675. S.N. is supported by the JST ASPIRE Program "RIKEN-Berkeley mathematical quantum science initiative".
AL and EN acknowledge support by the Israel Science Foundation grant 1995/21, and a  BSF grant (2022314) and ERC grant JetNS (818899). 
Numerical computations and data analyses were carried out on the supercomputer HOKUSAI BigWaterfall2 (HBW2) at RIKEN, Cray XC50 and HPE Cray XD2000 at the Center for Computational Astrophysics, National Astronomical Observatory of Japan, and Yukawa-21 at the Yukawa Institute for Theoretical Physics, Kyoto University.

\section*{Data Availability}
The data underlying this article will be shared on reasonable request to the corresponding author.

\bibliographystyle{mnras}
\bibliography{reference}

\appendix
\section{Spectrum at low energies}
\label{App:fnu}

\begin{figure*}
  \centering

  \setlength{\columnsep}{0.1em}

  \setlength{\panelw}{\dimexpr0.5\textwidth - 0.5\columnsep\relax}

  \begin{minipage}[t]{\panelw}\centering
    \includegraphics[width=\linewidth,keepaspectratio]{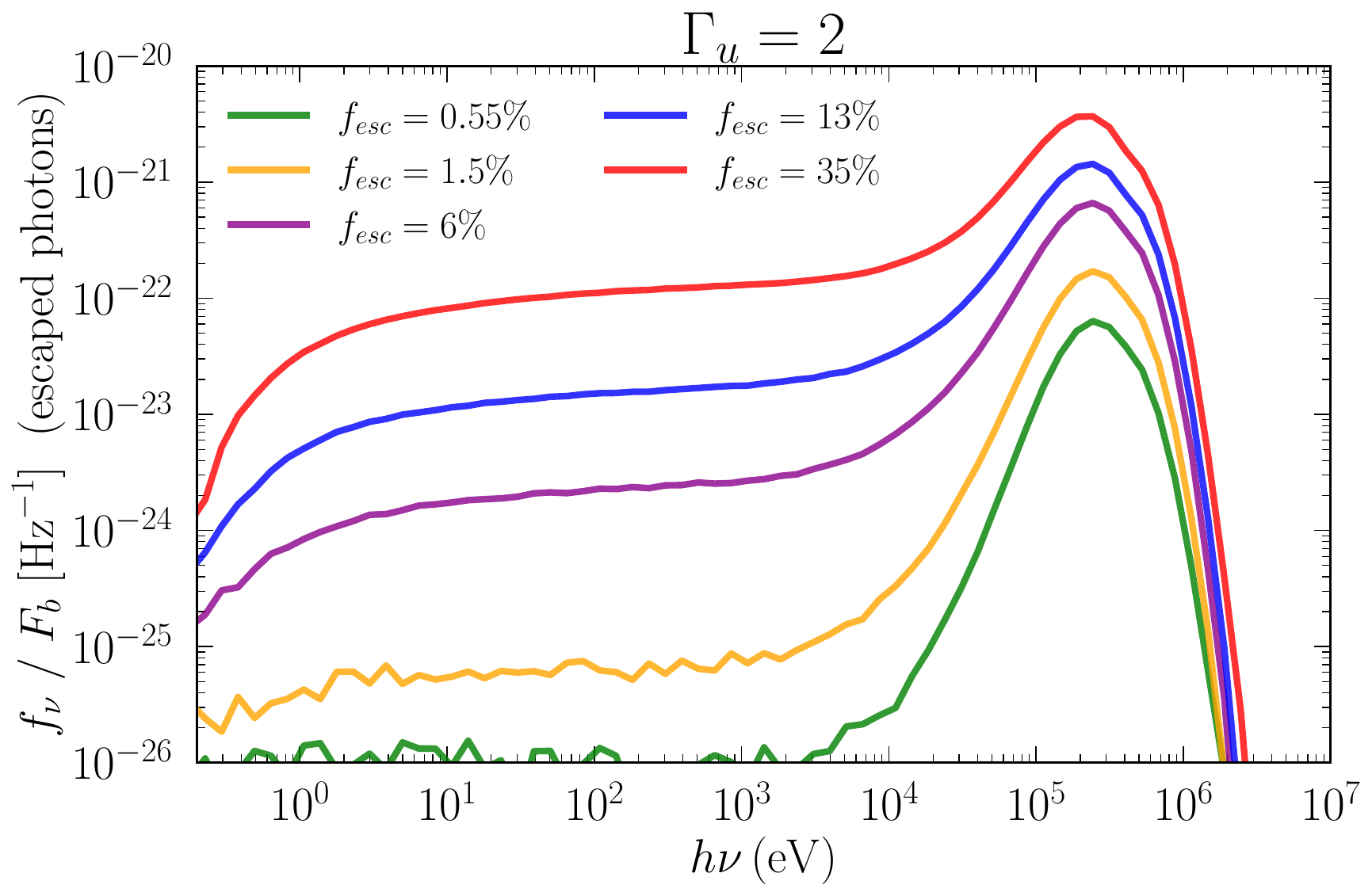}
  \end{minipage}\hspace{\columnsep}%
  \begin{minipage}[t]{\panelw}\centering
    \includegraphics[width=\linewidth,keepaspectratio]{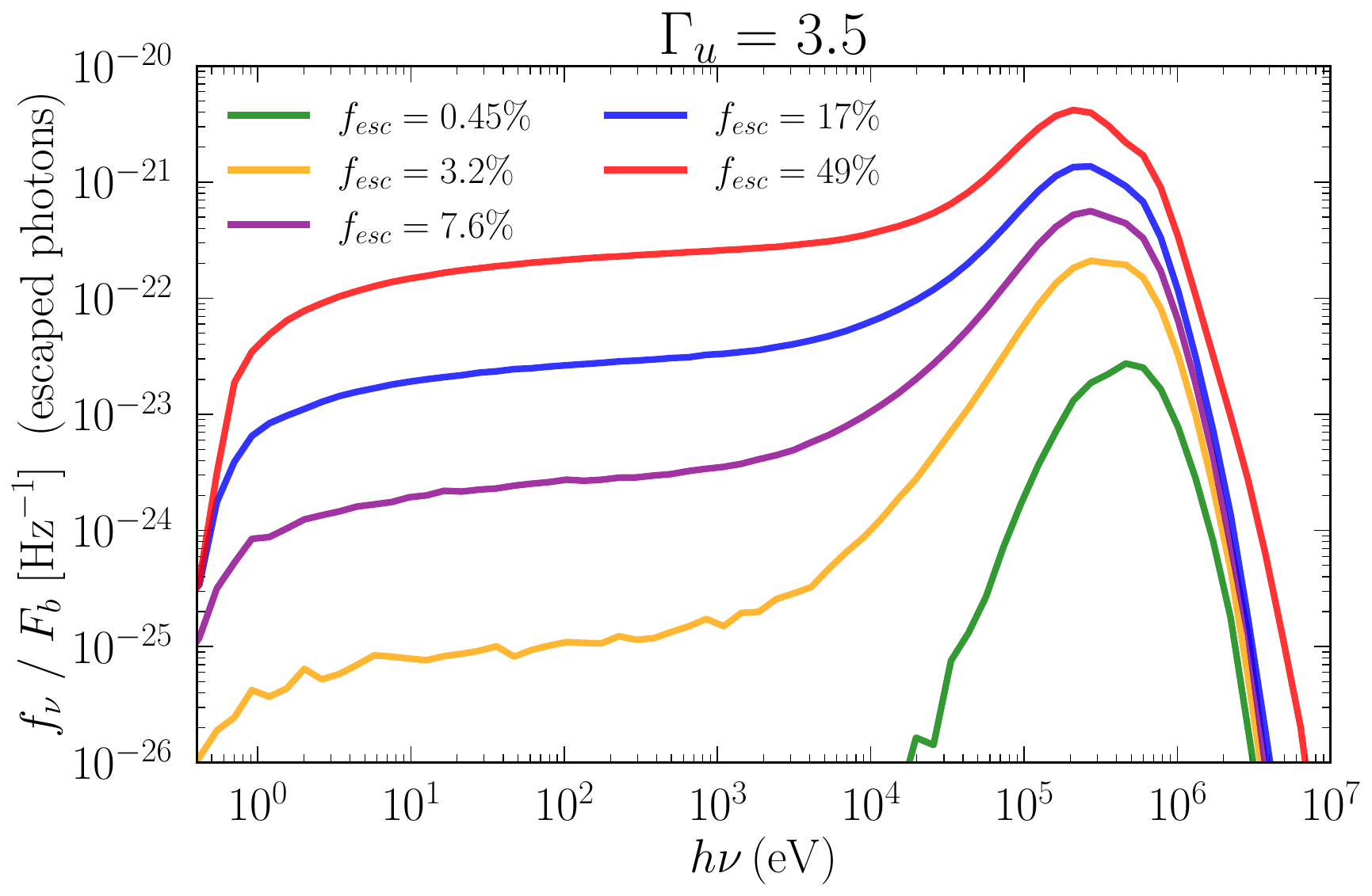}
  \end{minipage}

  \par\medskip %

   \begin{minipage}[t]{\panelw}\centering
    \includegraphics[width=\linewidth,keepaspectratio]{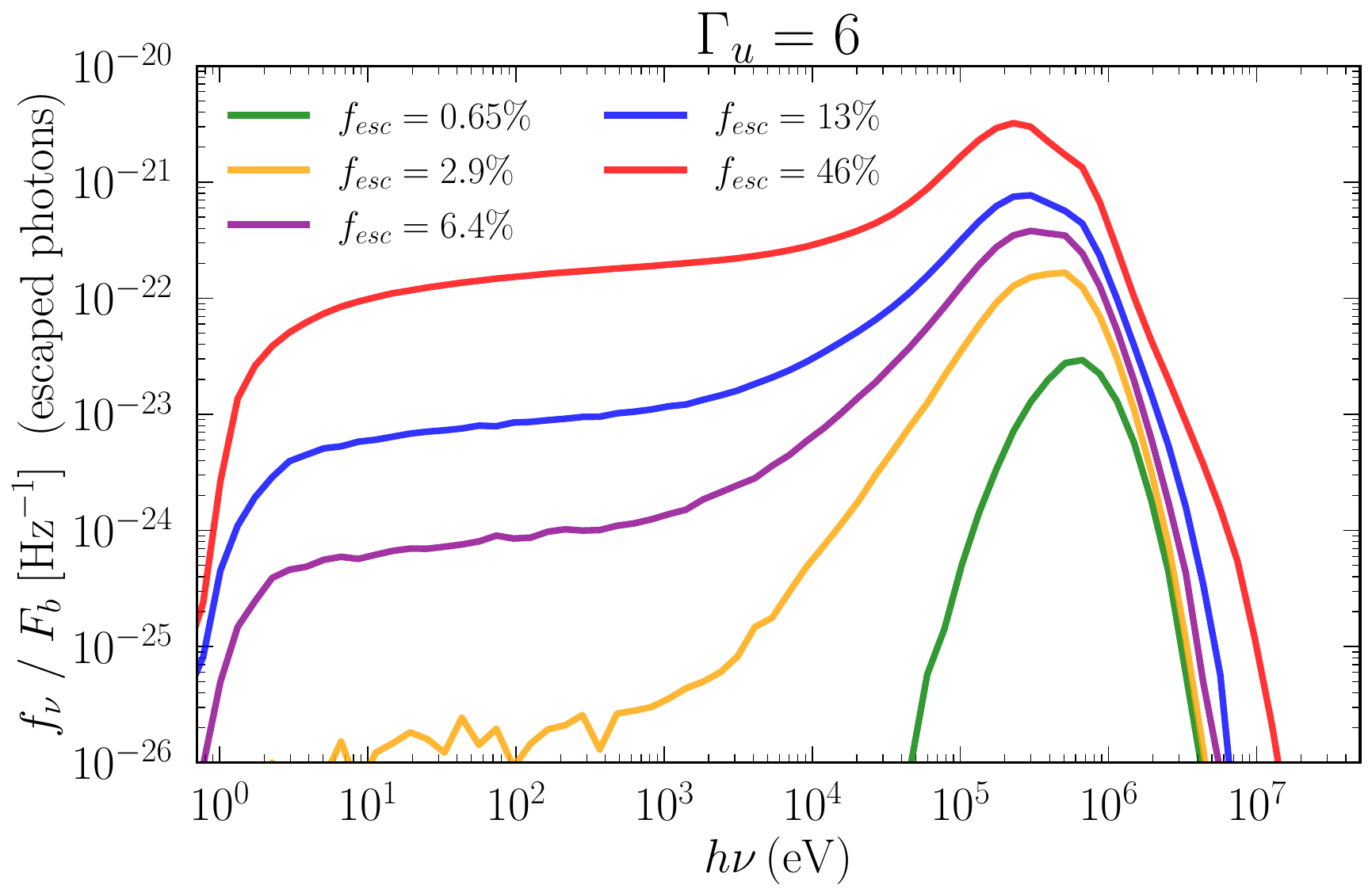}
  \end{minipage}\hspace{\columnsep}%
  \begin{minipage}[t]{\panelw}\centering
    \includegraphics[width=\linewidth,keepaspectratio]{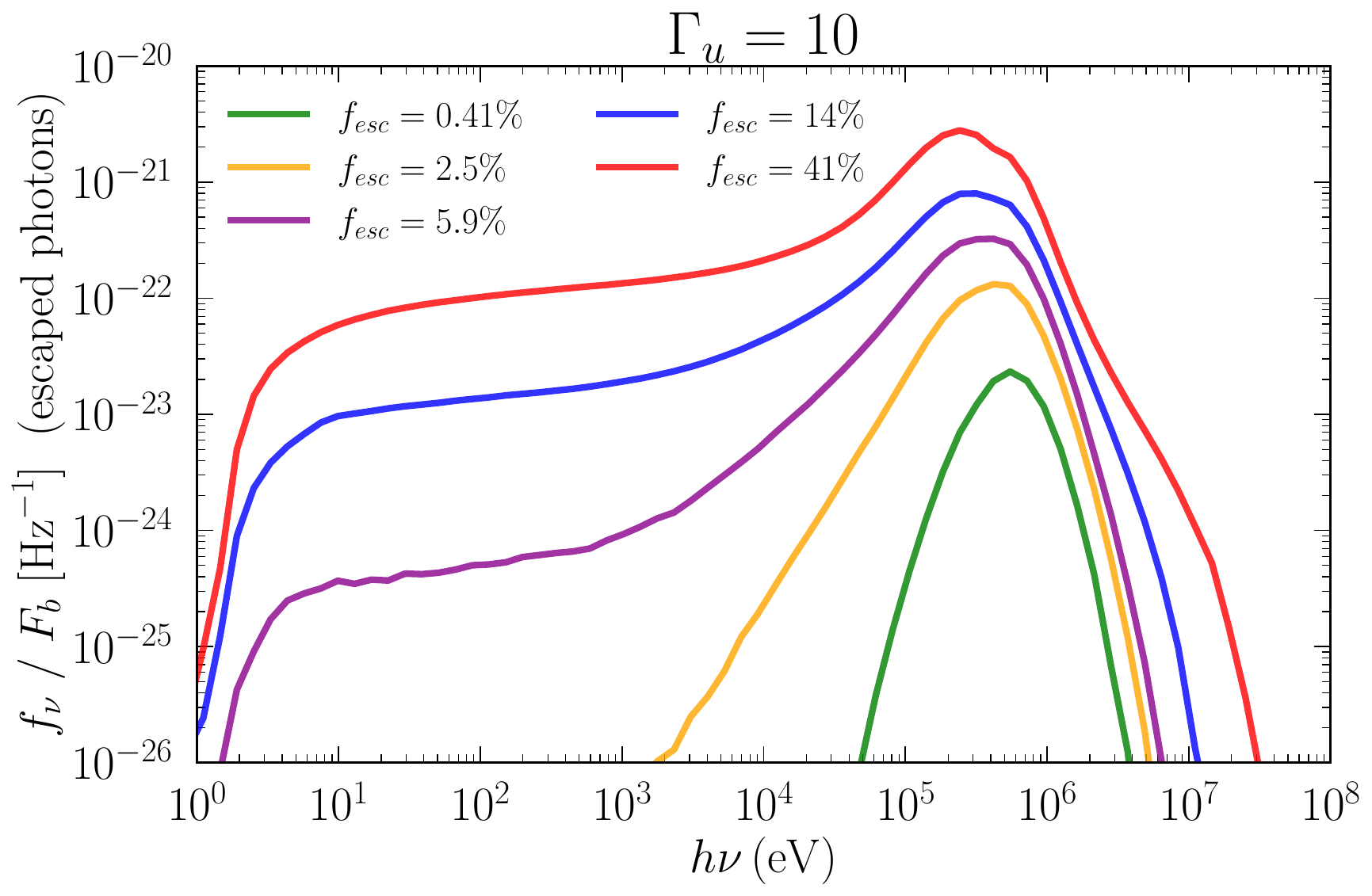}
  \end{minipage}

  \caption{Shock-frame, $f_\nu$ flux of escaping photons normalized by the total kinetic energy flux of baryons at the upstream boundary, $F_b = \Gamma_{u} (\Gamma_{u} - 1) n_{u} m_p c^3 \beta_{u}$. 
  Each panel display the results for shock velocities $\Gamma_{u} = 2$ (top left), $3.5$ (top right), $6$ (bottom left), and $10$ (bottom right), respectively. Within each panel, various lines represent different escape fractions, as specified in the legend.
  Note that the energy range of the spectra varies across different panels.}
  \label{fig:fnu}
\end{figure*}

In the main text, we presented the spectrum of photons escaping from the upstream boundary in the $\nu f_{\nu}$ form (Figure \ref{fig:nufnu}). Here, in Figure \ref{fig:fnu}, we illustrate the spectra in $f_{\nu}$ form to specifically highlight the characteristic spectral shape of $f_{\nu} \propto \nu^{0}$ below the spectral peak ($h\nu \ll E_p$). 
Note that the energy range covered in Figure \ref{fig:fnu} reaches much lower energies than those illustrated in Figure \ref{fig:nufnu}. Also, note that the jagged appearance of the SED at low energies in some lines is solely due to the lack of photon statistics.


The figure reveals that the broad spectral feature extends down to $h\nu \sim 1~{\rm eV}$, below which a significant break is observed. 
The break reflects the cutoff frequency for free-free emission, which in our simulations is set at a frequency where Coulomb screening suppresses the emission, as described in the Appendix of ILN20a:
\begin{eqnarray}
   \nu_{cut} = \frac{\gamma_{e, th}^2 \beta_{e, th} c}{2 \pi \lambda_D}, 
\label{eq:nucut}
\end{eqnarray}
where $\lambda_D = \sqrt{kT / 4 \pi e^2 (n + n_{\pm})}$ is the Debye length\footnote{There is a typo in the formula for the Debye length given in ILN20a; the factor $0.5$ before $n_{\pm}$ is mistakenly included.}, and $\gamma_{e, th} = 1 + 3/2 f(T) \Theta$ and $\beta_{e,th} = \sqrt{1 - \gamma_{e, th}^{-2}}$ represent the Lorentz factor and velocity of thermal motion of electrons, respectively. Here $f(T) = {\rm tanh}[({\rm ln}\Theta + 0.3) / 1.93] + 3/2$ is an analytical function of temperature defined in \citet{Budnik2010}, obtained from a fit to the exact equation of state of pairs.


\begin{figure*}
  \centering

  \setlength{\columnsep}{0.1em}

  \setlength{\panelw}{\dimexpr0.5\textwidth - 0.5\columnsep\relax}

  \begin{minipage}[t]{\panelw}\centering
    \includegraphics[width=\linewidth,keepaspectratio]{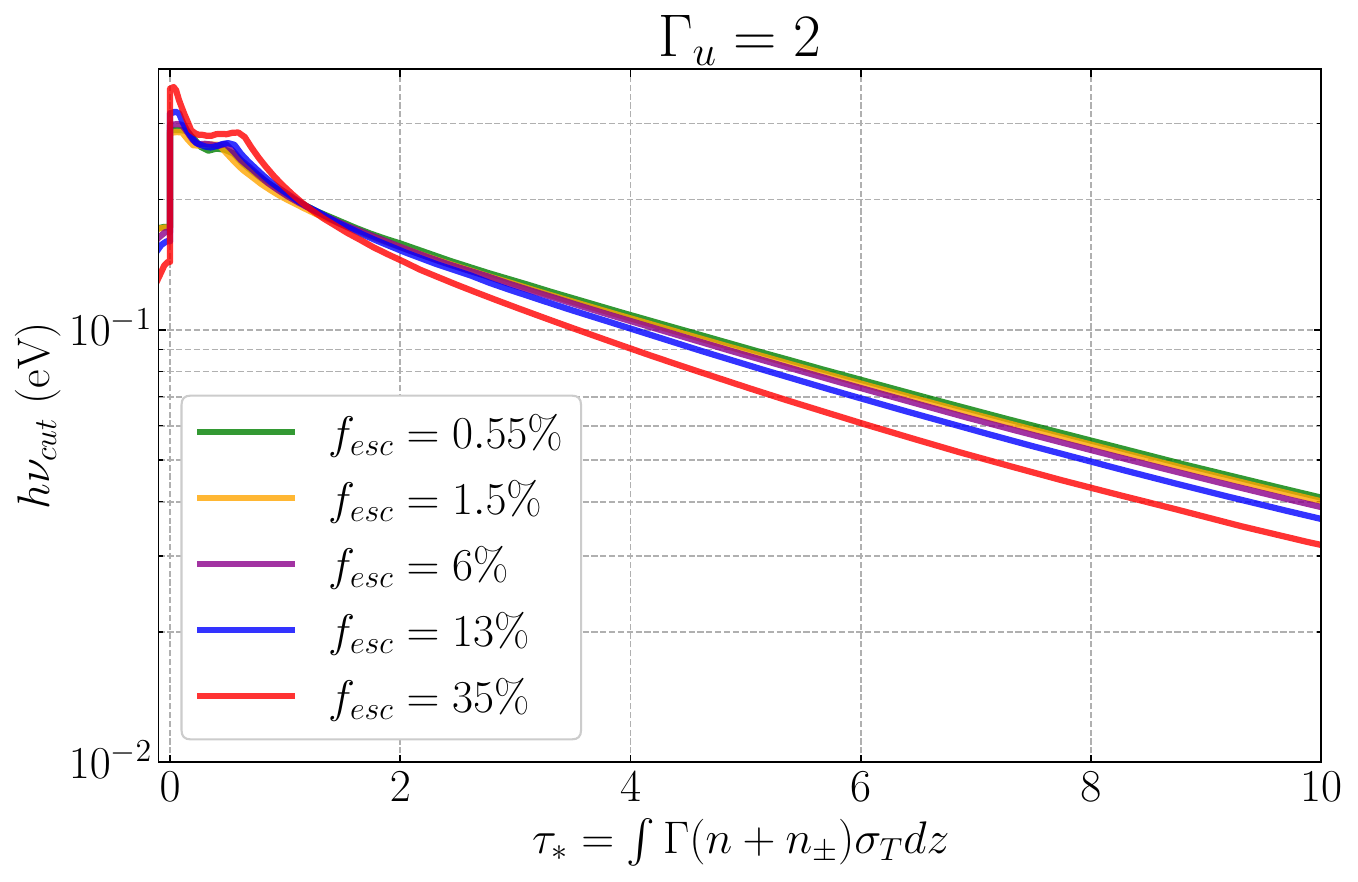}
  \end{minipage}\hspace{\columnsep}%
  \begin{minipage}[t]{\panelw}\centering
    \includegraphics[width=\linewidth,keepaspectratio]{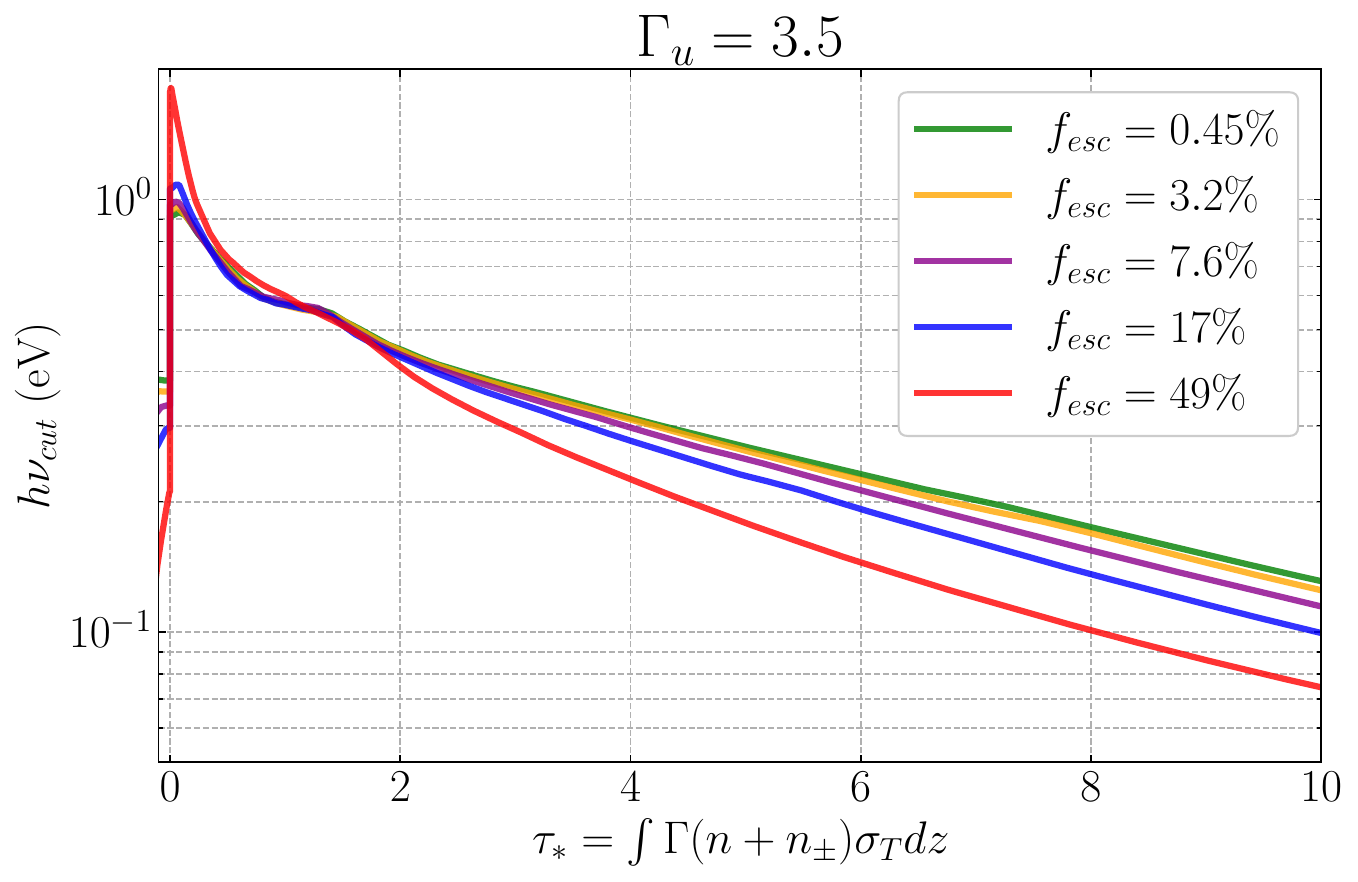}
  \end{minipage}

  \par\medskip %

   \begin{minipage}[t]{\panelw}\centering
    \includegraphics[width=\linewidth,keepaspectratio]{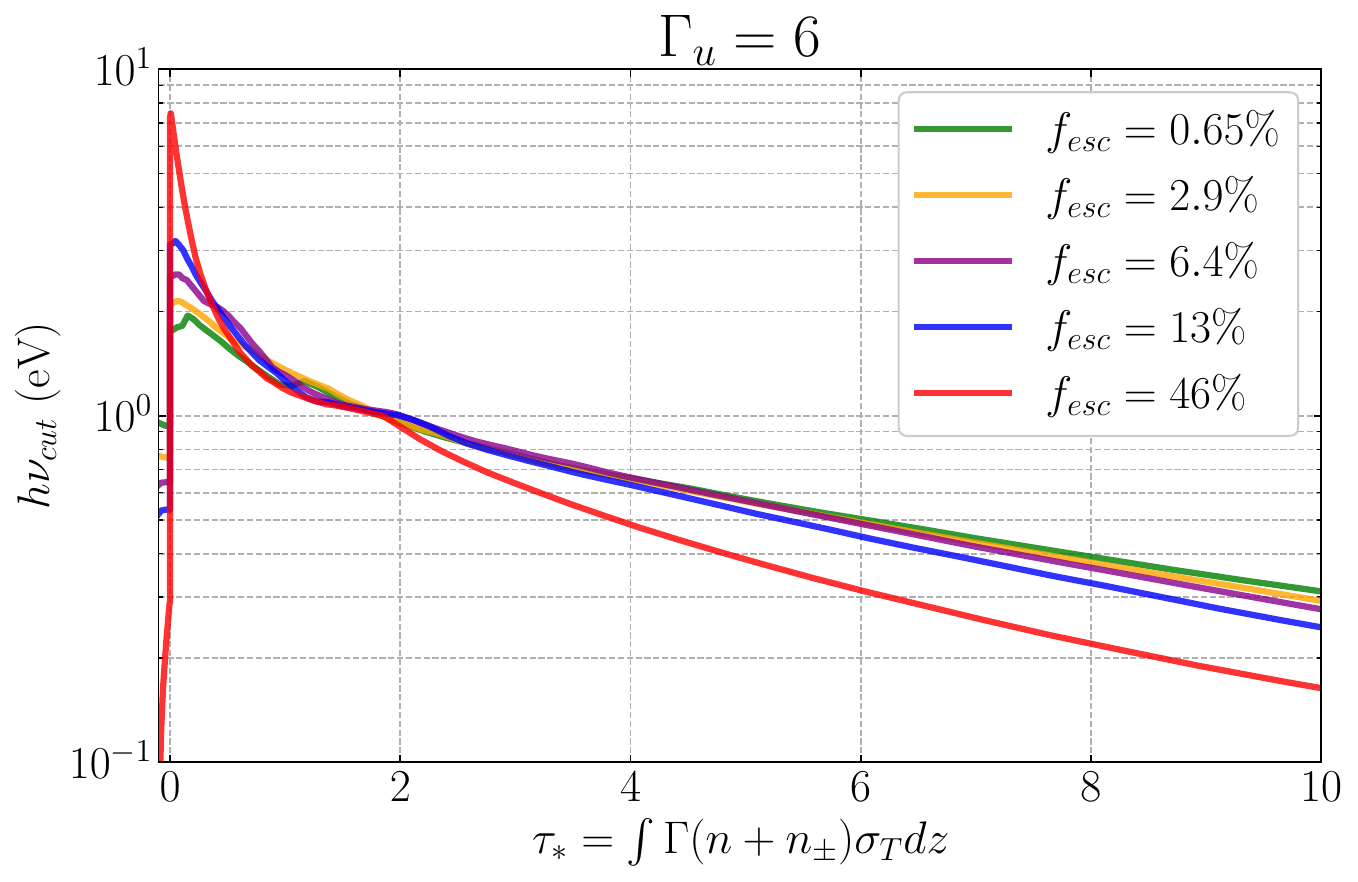}
  \end{minipage}\hspace{\columnsep}%
  \begin{minipage}[t]{\panelw}\centering
    \includegraphics[width=\linewidth,keepaspectratio]{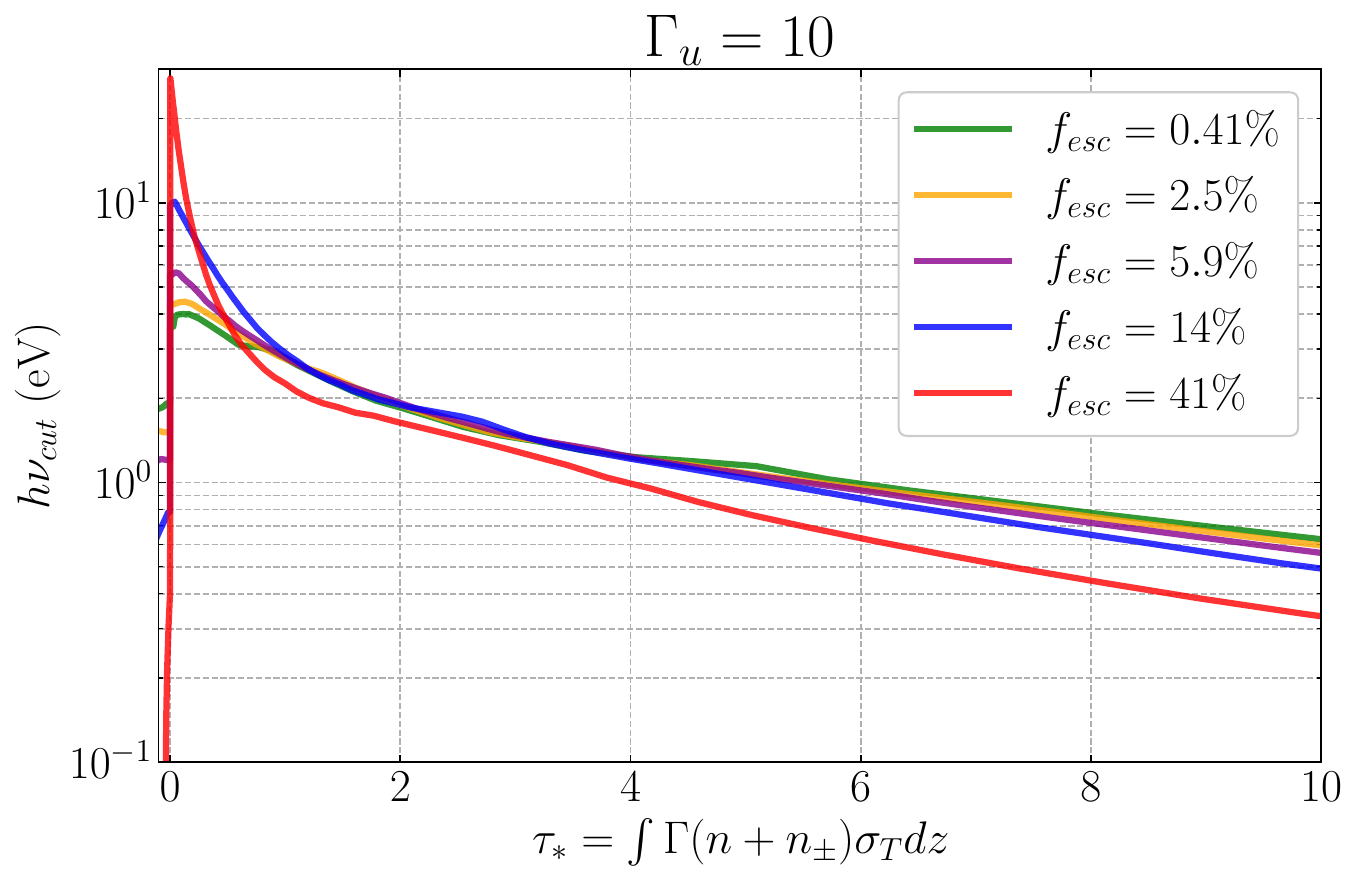}
  \end{minipage}

  \caption{The low cutoff energy for free-free emission, estimated from Equation (\ref{eq:nucut}), is shown for $\Gamma_u = 2$ (top left), $\Gamma_u = 3.5$ (top right), $\Gamma_u = 6$ (bottom left), and $\Gamma_u = 10$ (bottom right). Within each panel, various lines represent different escape fractions, as specified in the legend.}
  \label{fig:nuc}
\end{figure*}

As the equation indicates, $\nu_{cut}$ varies with position, depending on the local density of pairs and temperature. In the post-subshock region, where most photons are produced, the cutoff frequency progressively decreases in the downstream direction.
The spatial profile of $\nu_{cut}$ is displayed in Figure \ref{fig:nuc}. 
As seen in the figure, $h \nu_{cut}$  reaches maximum values of approximately $0.4$, $2$, $8$, and $30~{\rm eV}$ in simulations with the highest escape fractions for $\Gamma_u = 2$, $3.5$, $6$, and $10$, respectively, at the immediate post-subshock region.
These peak values are higher for higher escape fractions, reflecting elevated post-subshock temperatures.
Beyond the peak, $h \nu_{cut}$ decreases to about $0.03$, $0.7$, $0.2$, and $0.3~{\rm eV}$ at $\tau_{*} = 10$, which corresponds to roughly one diffusion length downstream from the subshock. 
While the maximum values somewhat depend on the escape fractions particularly for high $\Gamma_u$ shocks,  the profile beyond $\tau_{*} \gtrsim 1$ shows little variation across different escape fractions. This stability is attributed to the minimal variation in the temperature profile that follows the rapid cooling of subshock-heated pairs (Figure \ref{fig:profileGu2}-\ref{fig:profileGu10}).

It should be noted that while the roll-off feature observed at low energy in Figure \ref{fig:fnu} is robust, the spectral shape below this roll-off is not accurately computed in our simulations. This inaccuracy arises because we introduce a sharp cutoff at the frequency $\nu_c$; below this frequency, no photons are produced. In reality, although Coulomb screening should strongly suppress photon production, the efficiency is likely to decrease continuously at $\nu \lesssim \nu_c$.  Therefore, while the roll-off feature itself is robust, the spectrum at and below the roll-off should be considered a lower bound for $f_{\nu}$.

 This result differs from the spectra of mildly relativistic ($\beta_u = 0.5$) and fast Newtonian ($\beta_u = 0.25$ and $0.1$) shock breakouts computed in ILN20b. In these lower velocity shocks, the cutoff frequency $\nu_c$ is significantly lower due to the absence or reduced number of pairs and lower temperatures. Consequently, $\nu_c$ falls below the frequency at which the emission becomes optically thick from free-free absorption. Therefore, although a roll-off feature is also observed in the low-energy part of these spectra, it arises from absorption rather than the Coulomb screening effect. As a result, the spectral shape below the roll-off conforms to the Rayleigh-Jeans tail of the Planck function, $f_\nu \propto \nu^2$, ensuring accuracy. In contrast, the current simulations exhibit much harder spectra.

 \section{Description of  the Analytical Model}
 \label{App:Anamodel}

 Here, we provide a description of the analytical model of GNL18 and summarize the shock profile obtained from the model.
 The model describes the deceleration of a relativistic flow through interactions with counterstreaming photons emanating from the immediate downstream region. The shock structure is determined by the interplay between Compton scattering and pair production, which facilitates the conversion of kinetic energy into thermal energy.
 The counterstreaming photons are assumed to have energies of $\sim m_e c^2$,aligning with conditions found in RRMS. The model treats (i) the relativistic flow, composed of plasma (protons and pairs) and photons advected with the plasma, and (ii) counterstreaming photons as two colliding beams.
 Hereafter, we refer to (i) as the primary beam. Collisions occurring through Compton scattering and pair production inject new quanta (either backscattered photons or produced pairs) into the primary beam, leading to the conversion of kinetic energy into thermal energy.
 As a result, while the primary beam decelerates and thermalizes as it propagates toward the downstream region, the number of counterstreaming photons gradually decreases in the upstream direction.

 As a result of these interactions, the flow structure is derived as a function of $\tau$, the net optical depth of counterstreaming photons due to Compton scattering and pair production.  
 The derived flow structure determines the profiles of the following quantities:
 \begin{itemize}
  \item $x_l$: the number ratio of quanta (the sum of produced pairs and photons advected with the bulk plasma) to baryons, defined as  
  \begin{eqnarray}
      x_l = \frac{n_{\pm} + n_{\gamma \rightarrow d}}{n},
      \label{eq:xl}
  \end{eqnarray}
  where $n_{\gamma \rightarrow d}$ denotes the proper density of photons advected with the bulk plasma flow.
  \item $\hat{T} = \frac{k T}{m_e c^2}$: temperature normalized by the rest mass energy of an electron.     
     \item $\Gamma$: bulk Lorentz factor.     
\end{itemize}
Here, $\tau = 0$ corresponds to the immediate downstream region, where the deceleration has been completed ($\Gamma \approx 1$), and $\tau$ is measured from the immediate downstream region and is set to increase toward the upstream.\footnote{Note that the optical depth here is defined in the opposite direction from that used in our numerical simulations, where the optical depth is measured along the flow direction and increases toward the downstream.}

 Under the assumption that photons (pairs) in the primary beam originate solely from counterstreaming photons that collide with pairs (photons) in the primary beam, the governing equation for $x_l$ is written as  
 \begin{eqnarray}
   \frac{d x_l}{d\tau} = - (x_l + f_{esc} x_0), 
   \label{eq:xl}
 \end{eqnarray}
 where $x_0$ represents the value of $x_l$ at $\tau = 0$, and thus $f_{esc} x_0$ corresponds to the number density of downstream photons escaping at the upstream boundary, normalized by the baryon density.
 \footnote{For a more intuitive understanding of Equation (\ref{eq:xl}), note that the term $x_l + f_{esc} x_0$ represents the photon number flux of counterstreaming photons, where $f_{esc} x_0$ corresponds to the fraction of escaping photons, and $x_l$ represents the photon number flux of backscattered photons. Both quantities are evaluated in the shock frame and are normalized by the baryon density under the assumption of a relativistic flow ($\Gamma \gg 1$).  
 The reason why the counterstreaming photon flux is expressed as $x_l + f_{esc} x_0$ is due to the steady-state assumption: the local difference between the flux of backscattered photons and that of counterstreaming photons corresponds to the photon flux of escaping photons at the upstream boundary ($f_{esc} x_0$).  
 Thus, the equation expresses that the reduction in the counterstreaming photon flux (right-hand side of the equation) due to scattering is balanced by the increase in the flux of backscattered photons.}
 The above equation yields the solution:  
 \begin{eqnarray}
    x_l = x_0 [(1+f_{esc})e^{-\tau} - f_{esc}].
    \label{eq:xlsol}
 \end{eqnarray}  
 The value of $x_0$, which serves as the downstream boundary condition, is determined based on energy flux conservation and thermal energy deposition through two-beam collisions, as described below.

 The upstream boundary is defined as the position where  $x_l = 0$ , corresponding to the location beyond which scattered back photons or pairs no longer exist.
 From Equation (\ref{eq:xl}), it follows that in the case of an infinite shock ($f_{esc} = 0$), the solution extends to infinity ($\tau \rightarrow \infty$). On the other hand, for finite shocks ($f_{esc} \neq 0$), the upstream boundary is located at a finite optical depth given by $\tau_u = {\rm ln}( \frac{1+f_{esc}}{f_{esc}} )$.

 Assuming that the collision of counterstreaming photons leading to particle injection deposits thermal energy of approximately $\eta \Gamma m_e c^2$ into the primary beam, where $\eta$ is a free parameter expected to be of order unity, the temperature and Lorentz factor are derived to exhibit a universal dependence on $x_l$, independent of the escape fraction $f_{esc}$, as follows.  
 If the deposited thermal energy is evenly distributed among all particles, the temperature is given by  
 \begin{eqnarray}
   \hat{T} = \frac{\eta \Gamma (n_{\pm} + n_{\gamma \rightarrow d})}{2 n + n_{\pm} + n_{\gamma \rightarrow d}} =  \frac{\eta \Gamma x_l}{x_l + 2}.
   \label{eq:Tsol}
 \end{eqnarray}  
 This equation implies that once $x_l \gg 1$ is achieved, the temperature is given by $\hat{T} \approx \Gamma$.  
 Assuming that the contribution of counterstreaming photons to the net energy flux is small compared to that of the primary beam, which holds for $\Gamma \gg 1$ and when $f_{esc}$ is not too large,  
 energy flux conservation for the primary beam approximately holds and yields  
 \begin{eqnarray}
 \Gamma_u = \Gamma [1+(x_l + 1) \mu \hat{T}].
 \label{eq:conserve}
 \end{eqnarray}  
 Substituting Equation (\ref{eq:Tsol}) into Equation (\ref{eq:conserve}) leads to the Lorentz factor profile:  
 \begin{eqnarray}
   \Gamma = \frac{ \sqrt{1+ 16\mu \Gamma_u \eta \frac{x_l(x_l + 1)}{x_l + 2}} - 1 }{ 8\mu \eta \frac{x_l(x_l + 1)}{x_l + 2} }.
   \label{eq:Gasol}
 \end{eqnarray}  
 Note that this approximation breaks down as $f_{esc}$ approaches unity, where the counterstreaming photon energy flux becomes non-negligible.  
 
 The above equations readily determine the downstream value of $x_0$ as follows.  
 By assuming these equations extend to the immediate downstream region ($\Gamma \approx 1$), where $x_l \gg 1$, Equation (\ref{eq:Tsol}) gives $\hat{T} \approx \eta$.  
 Substituting this into Equation (\ref{eq:conserve}) with $\Gamma \approx 1$ leads to  
 \begin{eqnarray}
   x_0 = \frac{\Gamma_u}{4\mu \eta}.
   \label{eq:BC}
 \end{eqnarray}  
 
 In summary, the set of three equations---(\ref{eq:xlsol}), (\ref{eq:Tsol}), and (\ref{eq:Gasol})---together with the downstream boundary condition given by Equation (\ref{eq:BC}), describes the full shock structure as a function of $\tau$ for a given $\Gamma_u$ and $f_{esc}$, with $\eta$ as a free parameter.  
 It is important to note that, to convert the shock profile into a physical length scale, one must define the opacity, as described in the next section.  
 Additionally, the analytical solutions do not provide the relative abundance of photons and pairs in the primary beam.  
 Consequently, the relative contributions of Compton scattering and $\gamma$-$\gamma$ pair production to the opacity cannot be directly determined from this model.

\section{Detailed Comparison of Simulation Results with the Analytical Model for Infinite Shocks}
\label{App:CompINF}

\subsection{Conversion of Optical Depth}

 As shown in the previous section, the analytical model provides the shock profile as a function of the net optical depth, $\tau$, without explicitly specifying its relation to the spatial scale.  
 Consequently, converting $\tau$ into a physical length scale is necessary to facilitate the comparison between the simulation and the model.  
 In this study, we achieve this by translating $\tau$ to the pair unloaded Thomson optical depth $\tilde{\tau}$ ($= 2\Gamma_u \sigma_T z$), following the methodology outlined in GNL18:  
 \begin{eqnarray}
   d\tilde{\tau} = \frac{\sigma_T}{\sigma_{KN} (x_l + 1)}d\tau,
   \label{eq:tautau}
 \end{eqnarray}
 where  
 \begin{eqnarray}
   \sigma_{KN} = \frac{3}{8}\left(\frac{{\rm ln}2 X + 0.5}{X} \right)\sigma_T
   \label{eq:KN}
 \end{eqnarray}
 approximates the cross-section for Compton scattering in the Klein-Nishina (KN) regime as well as for $\gamma$-$\gamma$ pair production.  
 Here,  
 \begin{eqnarray}
   X = \Gamma (1 + a\hat{T}),
   \label{eq:X}
 \end{eqnarray}
 represents an approximate value for the typical photon energy of the counterstreaming photons observed in the rest frame of the pairs, normalized by $m_e c^2$ in the case of Compton scattering. At the same time, it also represents the product of the counterstreaming photon energy and the thermal photon energy advected with the flow, both evaluated in the comoving frame.\footnote{When $X \gg 1$ and the typical advected photon energy satisfies $kT \gg m_e c^2$, Equation (\ref{eq:KN}) provides a good approximation for the $\gamma$-$\gamma$ pair production cross-section.}  
 There is a slight modification to the cross-section given in Equation (\ref{eq:KN}) compared to GNL18, which does not include the additional $+0.5$ term in the numerator. This term is added solely to improve accuracy in the deep KN regime ($X \gg 1$), although it is not crucial, as no significant differences are observed regardless of which prescription is used.
 The parameter $a$, which is expected to be of order unity, is treated as an additional free parameter of the analytical model, along with $\eta$, both of which depend on the energy and angular distributions of pairs and photons.
 The above prescription approximates that each particle in the primary beam contributes equally to the opacity for the counterstreaming photons, regardless of whether the particle is a pair or a photon.

\subsection{Fitting to the simulation results}

Using the conversion from $\tau$ to $\tilde{\tau}$ given in Equation (\ref{eq:tautau}), the shock profile can be directly compared to the simulation results.  
Among the three outputs of the analytical solution, $\Gamma$, $x_l$, and $\hat{T}$, we chose the Lorentz factor profile $\Gamma$ in the upstream region of the subshock in the numerical results and fitted it to the analytical solution by optimizing the free parameters $a$ and $\eta$, as described in Section \ref{sec:comparison}.  
As also described in that section, since the analytical model does not account for the existence of the subshock, we shift the position of $\tilde{\tau} = 0$ in the analytical solution from its initial position, where $\Gamma \approx 1$, to a new position where $\Gamma = \Gamma_{u, sub}$, ensuring that the fitting starts from the same Lorentz factor $\Gamma_{u, sub}$ at $\tilde{\tau} = 0$.

\begin{figure}
  \begin{center}
  \includegraphics[width=1\columnwidth,keepaspectratio]{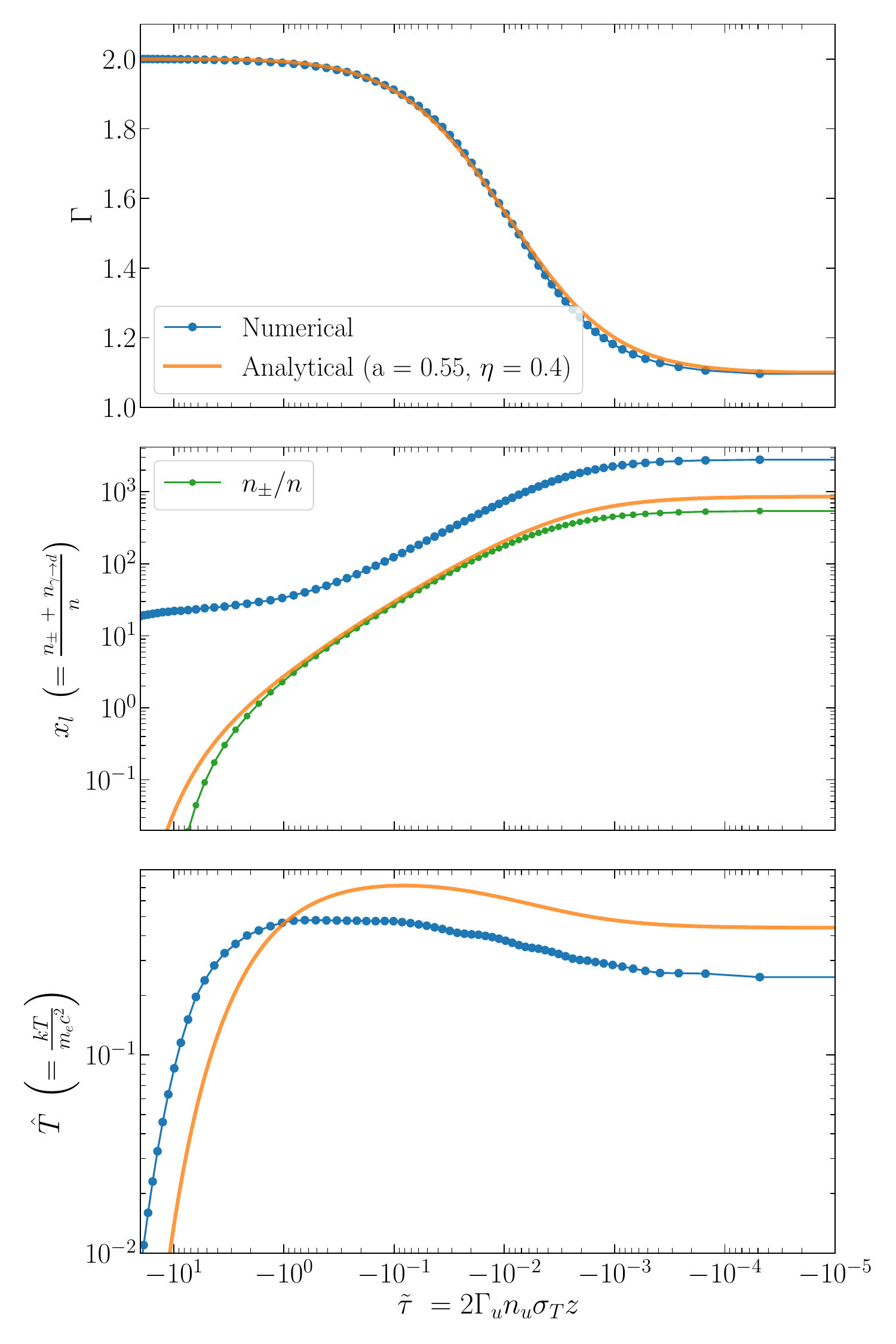}
  \end{center}
  \caption{
  Comparison of shock profiles as a function of pair-unloaded optical depth,
   $\tilde{\tau} = 2\Gamma_u n_u \sigma_T z$, with the analytical model of GNL18 for $\Gamma_u = 2$ and $f_{esc} = 0$ (infinite shock). The panels show the bulk Lorentz factor $\Gamma$ ({\it top}), the ratio of pair and advected photon density to baryon density, $x_l = (n_\pm + n_{\gamma \rightarrow d}) / n$ ({\it middle}), and the normalized temperature $kT / m_e c^2$ ({\it bottom}).
   The blue solid lines with dots represent the simulation results, while the orange solid lines show the best-fit analytical model. The green solid line with dots in the middle panel represents the pair-to-baryon density ratio, $n_{\pm}/n$, from the simulation. Note that the sign of the optical depth in the analytical model has been reversed to a negative value to match the definition employed in the simulation.}
  \label{fig:G2comp}
\end{figure}

\begin{figure}
  \begin{center}
  \includegraphics[width=1\columnwidth,keepaspectratio]{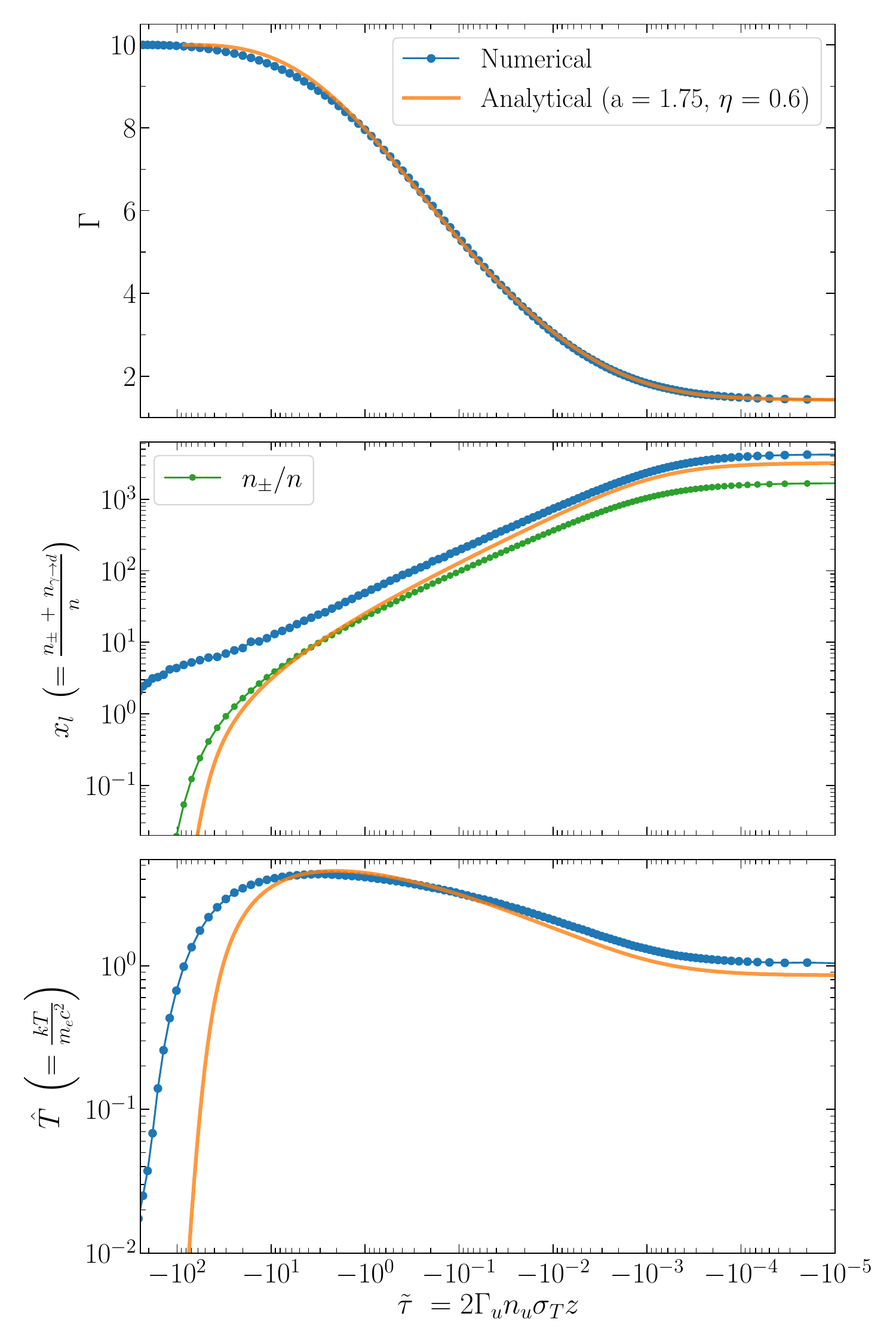}
  \end{center}
  \caption{Same as Figure \ref{fig:G2comp}, but for $\Gamma_u = 10$.}
  \label{fig:G10comp}
\end{figure}

The fitting is performed by scanning the free parameters $\eta$ in the range $[0.25, 1.0]$ and $a$ in the range $[0.5, 3]$ with an interval of $0.05$, minimizing the squared residuals between $\log(\tilde{\tau})$ from the simulation and the analytical solution.  
In this section, we present the fitted results, focusing on the case of infinite shocks, i.e., $f_{esc} = 0$.
 Figures \ref{fig:G2comp}-\ref{fig:G10comp} display the comparison of the simulation results with the best-fit analytical model. The figures show not only the $\Gamma$ profiles, which were used to optimize the fitting, but also the profiles of $\hat{T}$ and $x_l$.
  Note that $x_l$, a quantity introduced in the analytical model, includes contributions from both advected photons and produced pairs. However, the analytical model does not distinguish between these components and therefore does not provide their relative abundances.
  In contrast, in the simulations, $x_l$ is calculated by evaluating the individual components, $n_{\pm}$ and $n_{\gamma \rightarrow d}$, and substituting them into Equation (\ref{eq:xl}).
  Note that $n_{\gamma \rightarrow d}$ in the numerical simulations does not include only the backscattered photons which was originally counterstreaming photons before the scattering, 
  but also includes the photons produced by the thermal bremsstrahlung emission.

\begin{figure}
  \begin{center}
  \includegraphics[width=1\columnwidth,keepaspectratio]{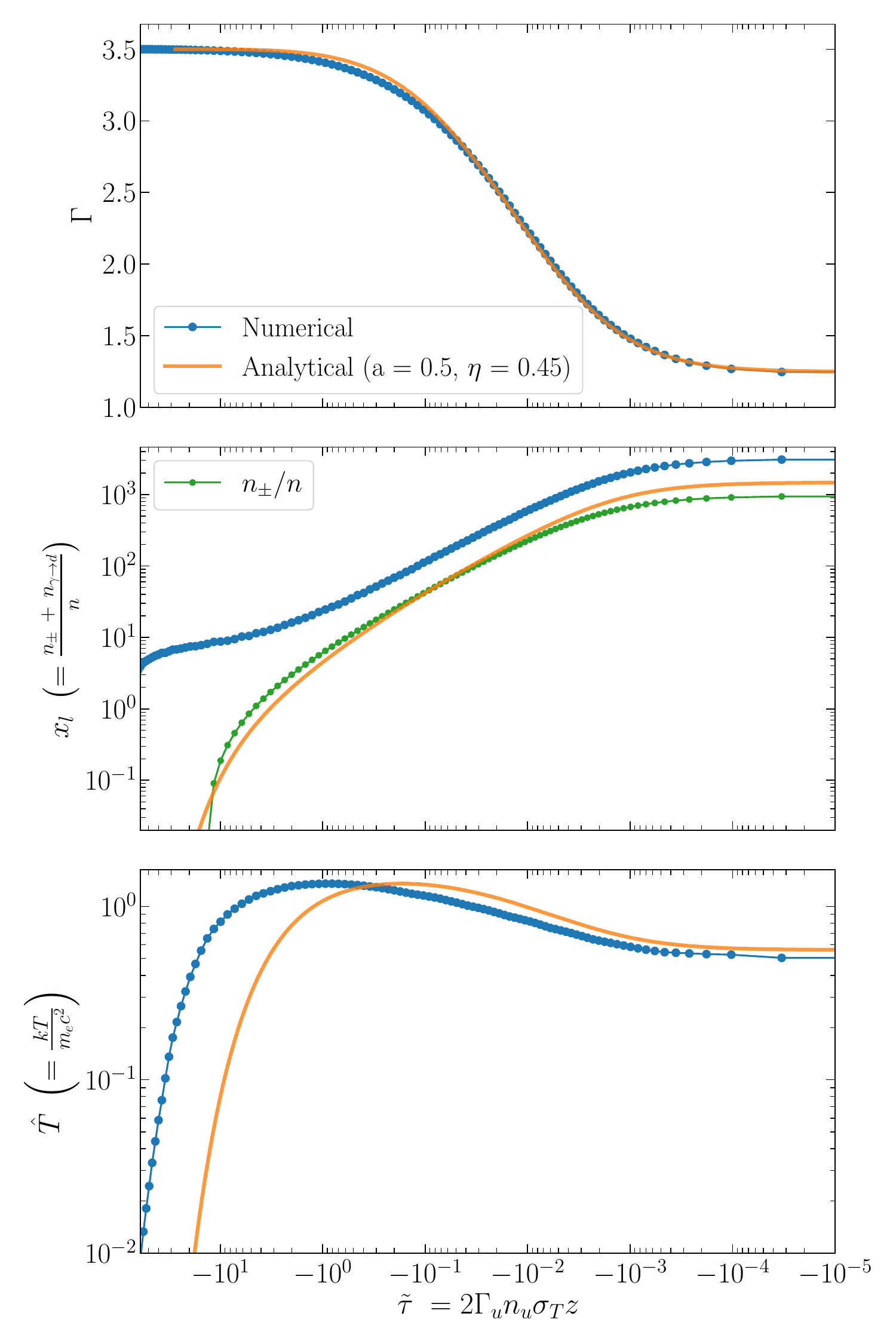}
  \end{center}
  \caption{
  Same as Figure \ref{fig:G2comp}, but for $\Gamma_u = 3.5$.}
  \label{fig:G6comp}
\end{figure}

\begin{figure}
  \begin{center}
  \includegraphics[width=1\columnwidth,keepaspectratio]{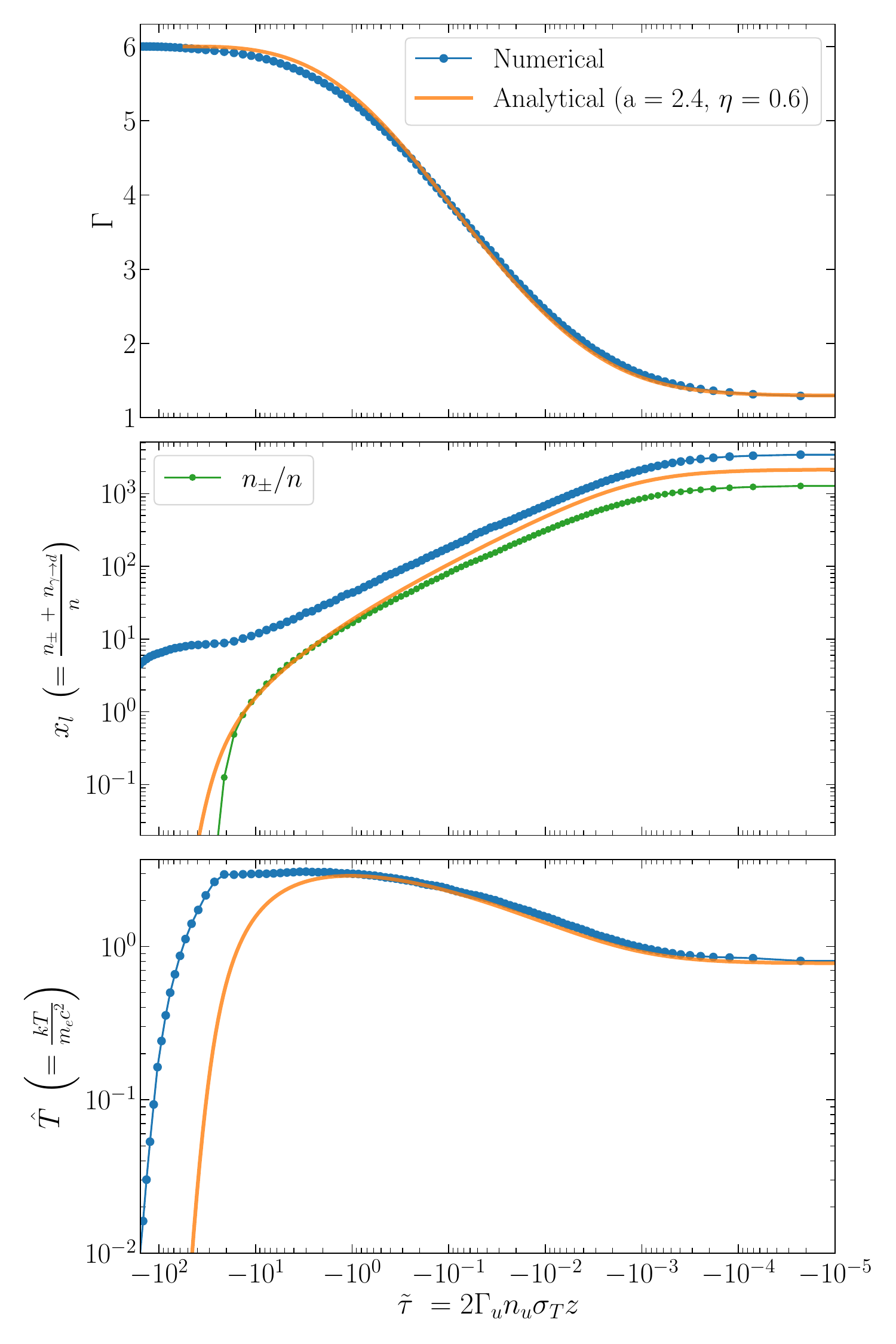}
  \end{center}
  \caption{
  Same as Figure \ref{fig:G2comp}, but for $\Gamma_u = 6$.}
  \label{fig:G6comp}
\end{figure}

\subsubsection{Comparison in the deceleration region}
\label{App:decreg}
As shown in the figures, the profiles of $\Gamma$ exhibit very good agreement between the analytical model and the simulation results.  
When excluding the far upstream region ($\Gamma \approx \Gamma_u$) where deceleration is negligible, broad agreement is also observed for the profiles of $\hat{T}$ and $x_l$.  
The discrepancies in $\hat{T}$ and $x_l$ between the analytical model and the simulation in the deceleration region remain within a factor of $\sim 2$ for $\Gamma_u = 3.5$, $6$, and $10$, and at most $\sim 3$ for $\Gamma_u = 2$.  
Given the simplified assumptions of the analytical model, this level of agreement highlights its ability to capture the essential features of shock dissipation physics.  
Notably, it is remarkable that even for the mildly relativistic shock with $\Gamma_u = 2$, the model exhibits reasonable agreement.  
This suggests that, although the model is derived under the assumption that the flow is highly relativistic ($\Gamma \gg 1$), it can be extended to the mildly relativistic regime, at least down to $\Gamma_u \sim 2$.  

On the other hand, a closer examination of the $x_l$ and $\hat{T}$ profiles for the $\Gamma_u = 2$ shock reveals relatively large discrepancies compared to higher Lorentz factor shocks, indicating the limitations of the model in the mildly relativistic regime.  
Focusing on the profile of $x_l$, the pair density ratio $n_{\pm}/n$ for the $\Gamma_u = 2$ shock shows much better agreement with the analytical model's $x_l$ than the computed $x_l$ itself.  
This can be understood from the relatively low temperature in the deceleration region ($\hat{T} \sim 0.2 - 0.3$), along with the relatively low energy of counterstreaming photons ($\lesssim \Gamma_u m_e c^2 = 2 m_e c^2$).
These factors significantly suppress the opacity for pair creation, as the energies of colliding photon pairs are close to or below the pair production threshold.
Therefore, the opacity is dominated by the scattering, and the assumption in the analytical model---that advected photons and pairs contribute equally to the opacity---breaks down in the simulation.  
As a result, the analytical solution's $x_l$ aligning well with the simulation's $n_{\pm}/n$ leads to a good match in the deceleration profile.

Compared with the $\Gamma_u=2$ case, shocks with higher Lorentz factors ($\Gamma_u=3.5$, $6$, and $10$) exhibit better agreement in the deceleration region, as mentioned earlier.
The temperature profile shows a particularly good match with the analytical model, as seen in the figures.
The quantity $x_l$ also shows good agreement between the simulations and the analytical solution, particularly for the high Lorentz factor shocks ($\Gamma_u=6$ and $10$). 
For $\Gamma_u=3.5$, as in the $\Gamma_u=2$ case, the simulated pair fraction $n_{\pm}/n$ is in closer agreement with the analytical $x_l$ than is the simulated $x_l$. Nonetheless, the simulated $x_l$ at $\Gamma_u=3.5$ exhibits slightly better agreement with the analytical $x_l$ than at $\Gamma_u=2$, indicating behavior intermediate between the low- and high-$\Gamma_u$ regimes.
For $\Gamma_u=6$ and $10$, the spatial profile of $x_l$ shows particularly good agreement between the simulations and the analytical solution, especially toward the downstream region.
However, a gap between the analytical model and the simulation gradually develops toward the upstream direction. This discrepancy arises because the advected photons in the simulation include not only scattered-back photons but also those generated by thermal bremsstrahlung. The thermal bremsstrahlung process produces numerous low-energy photons ($\lesssim m_e c^2$), which do not contribute to $\gamma$-$\gamma$ pair production. As a result, $x_l$ in the simulation accounts for a population of photons that do not contribute to the opacity.

It is worth noting that the qualitative difference between the mildly relativistic shocks ($\Gamma_u = 2$ and $3.5$) and highly relativistic shocks ($\Gamma_u = 6$ and $10$) is clearly reflected in a systematic difference in the values of the free parameters. Specifically, while the free parameters for highly relativistic shocks remain consistently within a certain range, the parameters for the mildly relativistic shock systematically exhibit lower values. Consequently, the shock width $\Delta \tilde{\tau}$ follows the expected scaling law, $\Delta \tilde{\tau} \propto \Gamma_u^3$, for $\Gamma_u \geq 6$, as reported in ILN20a, whereas at lower Lorentz factor $\Gamma_u \lesssim 3.5$, a noticeable deviation from this scaling emerges, as also pointed out in ILN20a.

\subsubsection{On the Discrepancy in the Far Upstream Region}
\label{App:farUS}

Examining the regions upstream of the deceleration zone, i.e., the far upstream region ($\Gamma \approx \Gamma_u$), a notable difference between the simulation results and the analytical solutions is observed across all models ($\Gamma_u = 2$, 6, and 10). One measurable deviation is the relatively extended temperature profile toward the upstream region found in the simulation.
This systematic deviation can be attributed to the fact that the counterstreaming photons are not monoenergetic. 
Higher-energy photons ($h\nu > m_e c^2$) are capable of penetrating deeper into the upstream region as a result of the reduced opacity caused by the Klein-Nishina effect, reaching larger optical depths compared to those predicted by the analytical solution.  Although the number of these surviving photons is too small to provide sufficient energy to decelerate the flow, they still carry enough energy to significantly heat the plasma, resulting in a substantial increase in temperature.
It is worth noting that GNL18 also reported a relatively extended temperature profile in the simulation results of \citet{Budnik2010} compared to the analytical solutions, presumably for the same reason.

This temperature rise also leads to a significant discrepancy in $x_l$ between the simulation and the analytical model in the far upstream region, as observed across all models ($\Gamma_u = 2$, 6, and 10). In this region, the simulation exhibits a significant rise in $x_l$ 
(taking place beyond the upstream boundary of the range displayed in Figures \ref{fig:G2comp} - \ref{fig:G10comp}), followed by a relatively flat profile. This profile reflects the fact that the photon population is dominated by photons produced through the thermal bremsstrahlung process.
In our simulation, the far-upstream condition is set as follows: the photons are in thermal equilibrium (blackbody distribution) with a very low temperature ($kT_u \ll m_e c^2$), and the photon-to-baryon ratio of the incoming flow from the upstream boundary is set to $n_{\gamma, u} / n_u = 10^{-2}$.\footnote{Specifically, $n_{\gamma, u} = 10^{-2}n_u = 10^{13}~{\rm cm}^{-3}$, which leads to
\[
k T_u =  k \left( \frac{n_{\gamma, u} c^3 h^3}{16 \pi \zeta(3) k^3} \right)^{1/3} \approx 0.68~{\rm eV},
\]
where $\zeta(3) \approx 1.202$ is the Riemann zeta function. However, note that the essential features of the shock structure do not depend on the details of the upstream condition as long as it remains sufficiently cold, ensuring that photons produced immediately downstream dominate over those from $n_{\gamma, u}$.}
This rapid heating drives the flow out of thermal equilibrium, leading to the rapid production of photons through thermal bremsstrahlung. While this process remains modest in the deceleration region, it has a noticeable effect in the far upstream region, significantly increasing the photon-to-baryon ratio $n_{\gamma}/n$ from its initial value.

A rough estimate of the increase in $n_{\gamma}/n$ at the far upstream region can be obtained as follows. For a pure proton-electron plasma, the effective photon generation rate due to thermal bremsstrahlung emission is given by
\begin{eqnarray}
 \dot{n}_{ff} = \alpha_e \sigma_T c n^2 \hat{T}^{-\frac{1}{2}} \Lambda_{ff},
  \label{eq:ndot}
\end{eqnarray}
where $\alpha_e$ is the fine-structure constant, and 
\begin{eqnarray}
  \Lambda_{ff} = E_1\left(\frac{h\nu_c}{ kT} \right) g_{ff},
\end{eqnarray}
with $E_1(x)$ being the exponential integral of $x$, $g_{ff}$ the Gaunt factor, and $\nu_c$ corresponding to the frequency of photons above which they are upscattered $\hat{T}$ times before being absorbed. This cutoff frequency is approximately given by
\begin{eqnarray}
  h\nu_c \approx 1.0 \times 10^{-3} n_{15}^{\frac{1}{2}} \hat{T}^{-\frac{5}{4}}~{\rm eV}
  \label{eq:nuc}
\end{eqnarray}
\citep{Chapline73,Weaver1976}, where $n_{15} = n/10^{15}~{\rm cm}^{-3}$.\footnote{Since the emissivity of thermal bremsstrahlung scales as $f_\nu \propto \nu^0$ at $h\nu \lesssim kT$, the total (frequency-integrated) photon generation rate diverges unless a low-energy cutoff is specified. An effective cutoff naturally arises in the system, as absorption below a certain frequency prevents the photon number from increasing. 
Note that this cutoff is automatically incorporated in the simulation, as free-free absorption is explicitly taken into account. Note also that, the cutoff due to Coulomb screening, given in Equation (\ref{eq:nucut}), is irrelevant in this region, as its value is significantly lower due to the lower temperature and density.}
To a good approximation, $\Lambda_{ff}$ can be expressed as
$
\Lambda_{ff} \approx \max\left[1, \frac{1}{2}\ln(y)\{1.6 + \ln(y)\}\right],
$
where $y = kT / h\nu_c$ \citep{Levinson_Nakar2020}. In the range $\hat{T} = 10^{-4}$ to $2 \times 10^{-2}$, $\Lambda_{ff}$ varies from $\sim 1$ to $\sim 74$ for $n_{15} = 1$.

Using the photon generation rate given in Equation~(\ref{eq:ndot}), the increase in $n_{\gamma}/n$ during the propagation over a distance $\Delta \tilde{\tau} = 2 \Gamma_u n_u \sigma_T \Delta z$ at the far upstream region is given by
\begin{eqnarray}
  \Delta \left(\frac{n_{\gamma}}{n} \right)  &\simeq  & \frac{\Delta n_{\gamma}}{n_u}  \simeq  \frac{\dot{n}_{ff}}{n_u} \frac{\Delta \tilde{\tau}}{2 \Gamma_u^2 n_u \sigma_T \beta_u c} \nonumber \\
   &\simeq & 0.5 \alpha_e \hat{T}^{-\frac{1}{2}} \Lambda_{ff} \Gamma_u^{-2} \Delta\tilde{\tau} \nonumber \\ 
   &\simeq &  1.8 \left( \frac{\hat{T}}{10^{-2}}\right)^{-\frac{1}{2}} \left( \frac{\Lambda_{ff}}{50} \right)  \left( \frac{\Gamma_u}{10} \right)^{-2}  \left( \frac{\Delta \tilde{\tau}}{100} \right) \,.
   \label{eq:Deltan}
\end{eqnarray}
The above equation shows that when the temperature is in the range $\hat{T} \sim 10^{-2}$, which corresponds to $\Lambda_{ff} \sim 50$ for $n_{15} = 1$, the photon generation becomes efficient enough to raise the photon-to-baryon ratio up to $\sim 2$ over a range of $\Delta \tilde{\tau} \sim 100$ for a pure proton-electron plasma propagating with a velocity corresponding to $\Gamma_u = 10$. This result is consistent with the value of $x_l$ found in the far upstream region for $\Gamma_u = 10$.
Once $n_{\gamma}/n$ reaches a certain value beyond which photon generation becomes ineffective within the flow, the increase in $x_l$ ceases. This occurs well before the temperature reaches its maximum value, as the photon generation becomes suppressed due to the high temperature and the limited spatial scale $\delta \tilde{\tau}$ remaining before the deceleration is completed.

The same equation also yields consistent results for the other two cases ($\Gamma_u = 2$ and $\Gamma_u = 6$). 
A lower Lorentz factor leads to a lower $\hat{T}$ in the far upstream region, which, combined with the reduced value of $\Gamma_u$, results in a relatively larger enhancement of $x_l$ in that region. For instance, for $\Gamma_u = 2$, $x_l$ rises from the boundary value of $10^{-2}$ to $\sim 20$ over a distance of $\Delta \tilde{\tau} \sim 40$. The typical temperature in this region is in the range $\hat{T} \sim 10^{-4} - 10^{-3}$, corresponding to $\Lambda_{ff} \sim 1 - 15$. Substituting these values into Equation~(\ref{eq:Deltan}) yields results consistent with the simulation, where $x_l \sim 20$ is observed.

To summarize, the discrepancy observed in the far upstream region is caused by a small number of high-energy counterstreaming photons that raise the plasma temperature without causing bulk deceleration. This drives the plasma out of thermal equilibrium, leading to significant photon production via thermal bremsstrahlung. This sets the value of $x_l$ in the upstream region of the deceleration zone; however, since the photons produced in this region are soft ($h\nu \ll m_e c^2$), they contribute negligibly to the opacity for counterstreaming photons.

\label{lastpage}

\end{document}